\documentclass[letterpaper,twocolumn,10pt]{article}

\usepackage{usenix2019_v3}

\usepackage{adjustbox}
\usepackage[ruled,vlined]{algorithm2e}
\usepackage{makecell}
\usepackage{algpseudocode}
\usepackage{amsthm}
\usepackage{amsmath} 
\usepackage{amssymb}
\usepackage{amsfonts}
\usepackage{appendix}
\usepackage{array}
\usepackage{comment}
\usepackage{float}
\usepackage{booktabs}
\usepackage{threeparttable}

\usepackage{caption} 
\usepackage{cite}
\usepackage{colortbl}
\usepackage{authblk} 
\usepackage{dsfont}

\usepackage{filecontents}

\usepackage{graphicx} 

\usepackage{listings}
\usepackage{longtable}

\usepackage{multicol}
\usepackage{multirow}

\usepackage{setspace}
\usepackage{stfloats}
\usepackage{subcaption}  

\usepackage{textcomp}
\usepackage{tikz} 
\usepackage{titlesec}

\usepackage[normalem]{ulem}

\usepackage{xcolor}

\usepackage{epstopdf}

\usepackage{tikz-cd}

\usepackage{pifont}

\usepackage{circledsteps}

\usepackage{tikz}
\usepackage{amsmath}

\usepackage{filecontents}

\def \xiaozhong #1{{\textcolor{blue}{[xiaozhong: #1]}}}

\newcommand{\toolname}{LLM-SPL }

\usepackage{xparse}
\newcommand{\bnm}{\begin{newmath}}
\newcommand{\enm}{\end{newmath}}

\newcommand{\bea}{\begin{eqnarray*}}%
\newcommand{\eea}{\end{eqnarray*}}%

\newcommand{\bne}{\begin{newequation}}
\newcommand{\ene}{\end{newequation}}

\newcommand{\bal}{\begin{newalign}}
\newcommand{\eal}{\end{newalign}}

\usepackage{xcolor} 
\usepackage[normalem]{ulem} 

\newcommand{\delete}[1]{\textcolor{blue}{\sout{#1}}}
\newcommand{\add}[1]{\textcolor{magenta}{#1}}

\newenvironment{newalign}{\begin{align}%
\setlength{\abovedisplayskip}{4pt}%
\setlength{\belowdisplayskip}{4pt}%
\setlength{\abovedisplayshortskip}{6pt}%
\setlength{\belowdisplayshortskip}{6pt} }{\end{align}}

\newenvironment{newmath}{\begin{displaymath}%
\setlength{\abovedisplayskip}{4pt}%
\setlength{\belowdisplayskip}{4pt}%
\setlength{\abovedisplayshortskip}{6pt}%
\setlength{\belowdisplayshortskip}{6pt} }{\end{displaymath}}

\newenvironment{newequation}{\begin{equation}%
\setlength{\abovedisplayskip}{4pt}%
\setlength{\belowdisplayskip}{4pt}%
\setlength{\abovedisplayshortskip}{6pt}%
\setlength{\belowdisplayshortskip}{6pt} }{\end{equation}}

\newcounter{ctr}

\newcounter{mytable}
\def\mytable{\begin{centering}\refstepcounter{mytable}}
\def\endmytable{\end{centering}}

\newcounter{myfig}
\def\myfig{\begin{centering}\refstepcounter{myfig}}
\def\endmyfig{\end{centering}}

\newlength{\saveparindent}
\newlength{\saveparskip}
\newcommand{\E}{{\rm I\kern-.3em E}}

\renewcommand{\eqref}[1]{\mbox{Equation~(\ref{#1})}}

\newcommand{\ignore}[1]{}

\newcommand\chenyi[1]{{\textcolor{magenta}{#1}}}

\newcommand\blue[1]{{\textcolor{blue}{#1}}}
\newcommand\magenta[1]{{\textcolor{magenta}{#1}}}
\newcommand\olive[1]{{\textcolor{olive}{#1}}}
\newcommand\teal[1]{\textcolor{teal}{#1}}

\def \part {part}

\renewcommand{\paragraph}[1]{\vspace*{6pt}\noindent\textbf{#1}\;}

\def \blackslug{\hbox{\hskip 1pt \vrule width 4pt height 8pt
    depth 1.5pt \hskip 1pt}}
\def \qed{\quad\blackslug\lower 8.5pt\null\par}

\newcounter{mynote}[section]

\newcounter{rcnote}[section]

\newcounter{mrnote}[section]

\newcounter{fknote}[section]

\newcounter{anote}[section]

\DeclareMathSymbol{\mlq}{\mathord}{operators}{``}
\DeclareMathSymbol{\mrq}{\mathord}{operators}{`'}

\newcommand{\rhf}[2]{R_{f, \gamma}}

\DeclareDocumentCommand{\edist}{o o}{
  \ensuremath{
    \IfNoValueTF{#1}{{d}}{{\sf d}(#1,#2)}
  }
}

\newcommand{\olrk}[1]{\ifx\nursymbol#1\else\!\!\mskip4.5mu plus 0.5mu\left(\mskip0.5mu plus0.5mu #1\mskip1.5mu plus0.5mu \right)\fi}

\NewDocumentCommand{\indseq}{ O{1} O{r} }{{#1}\ldots {#2}}

\begin{document}

\date{}

\title{\Large \bf LLM-Enhanced Software Patch Localization}



\author[1]{Jinhong Yu}
\author[2,3]{Yi Chen}
\author[2]{Di Tang}
\author[1]{Xiaozhong Liu}
\author[2]{XiaoFeng Wang}
\author[4]{Chen Wu}
\author[2]{Haixu Tang}

\affil[1]{Worcester Polytechnic Institute}
\affil[2]{Indiana University Bloomington}
\affil[3]{The University of Hong Kong}
\affil[4]{Microsoft}

\maketitle

\begin{abstract} 

Open source software (OSS) is integral to modern product development, and any vulnerability within it potentially compromises numerous products. While developers strive to apply security patches, pinpointing these patches among extensive OSS updates remains a challenge. Security patch localization (SPL) recommendation methods are leading approaches to address this. However, existing SPL models often falter when a commit lacks a clear association with its corresponding CVE, and do not consider a scenario that a vulnerability has multiple patches proposed over time before it has been fully resolved. To address these challenges, we introduce LLM-SPL, a recommendation-based SPL approach that leverages the capabilities of the Large Language Model (LLM) to locate the security patch commit for a given CVE. More specifically, we propose a joint learning framework, in which the outputs of LLM serves as additional features to aid our recommendation model in prioritizing security patches. Our evaluation on a dataset of 1,915 CVEs associated with 2,461 patches demonstrates that LLM-SPL excels in ranking patch commits, surpassing the state-of-the-art method in terms of Recall, while significantly reducing manual effort. Notably, for vulnerabilities requiring multiple patches, LLM-SPL significantly improves Recall by 22.83\%, NDCG by 19.41\%, and reduces manual effort by over 25\% when checking up to the top 10 rankings. The dataset and source code are available at \url{https://anonymous.4open.science/r/LLM-SPL-91F8}.

\end{abstract}
\vspace{-13pt}
\section{Introduction}
\label{section: introduction}
\vspace{-10pt}

Open source software (OSS) is now ubiquitous in product development. A recent report by Synopsys reveals that 96\% of the 1,700 commercial codebases examined across 17 industries incorporate open source components~\cite{synopsys}. Consequently, a single vulnerability in OSS can compromise the security of hundreds or even thousands of subsequent products.
A prominent example is the Log4J vulnerability in Apache~\cite{log4j}, which sent shockwaves across the global tech community, prompting numerous vendors using Apache to urgently deploy security patches.
To address such vulnerabilities from OSS, developers who integrate OSS into their products are often tasked with implementing security patches. 
However, they typically learn about OSS vulnerabilities from public sources such as the NVD/CVE vulnerability databases.
These sources usually do not specify the exact location of the security patches, as these patches are often buried within a bundle of OSS updates (e.g., commits on GitHub). 
As a result, when developers customize patches for their projects or craft hot-fixes based on the original patches, they might first have to navigate through numerous OSS modifications to identify the relevant security fixes, which is a labor-intensive and error-prone process.


\begin{figure}
    \centering
    \includegraphics[width=0.37\textwidth]{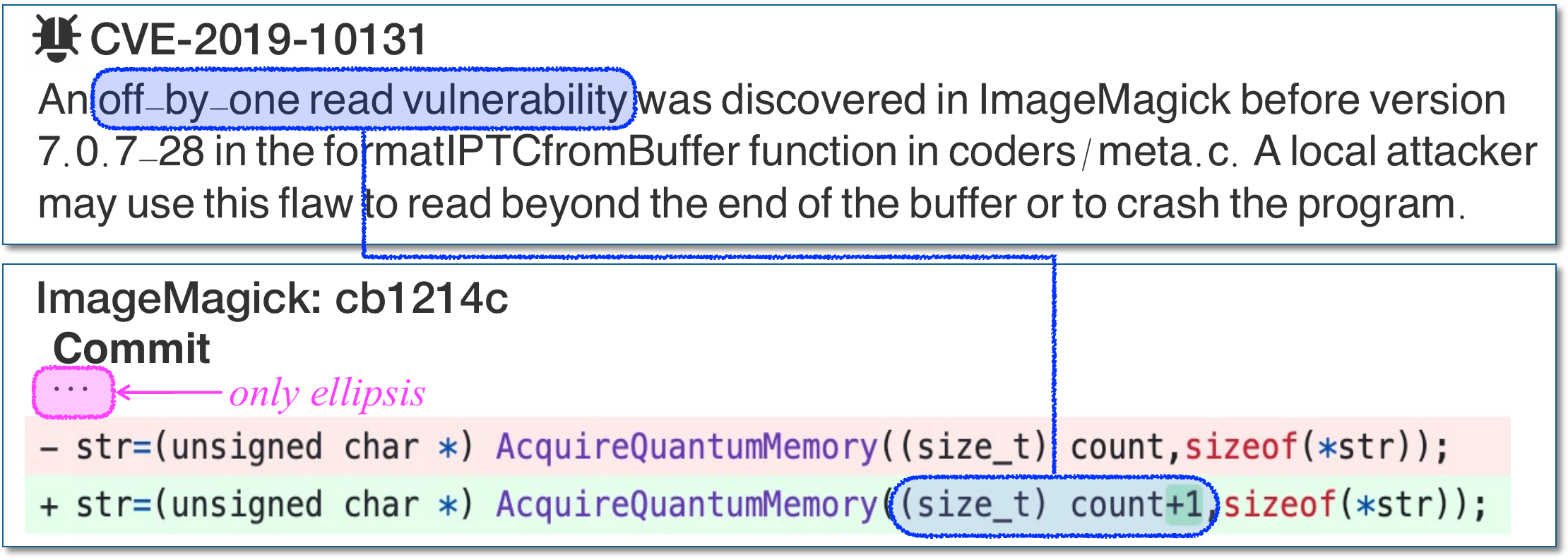}
    \vspace{-10pt}
    \caption{Unclear Association: CVE and Commit Example}
    \label{figure: Example of a commit as a patch for CVE without clear association.}
    \vspace{-18pt}
\end{figure}

\begin{figure*}
    \centering
    \centering \includegraphics[width=0.98\textwidth]{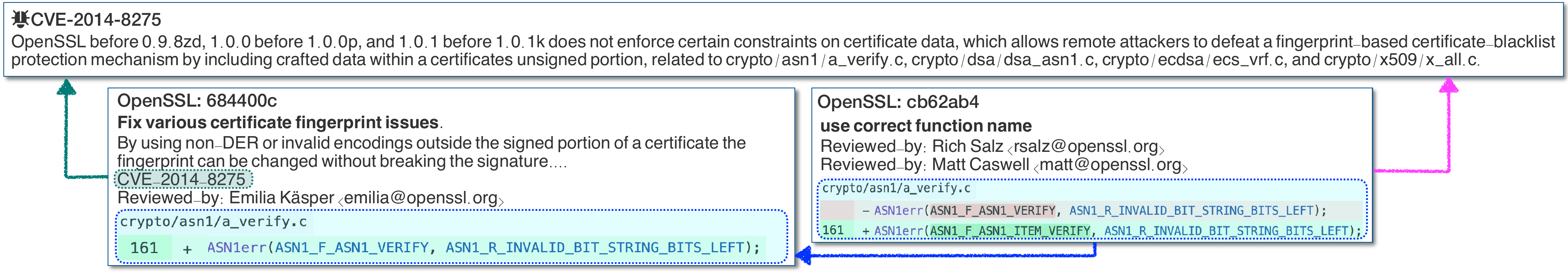}
    \vspace{-10pt}
    \caption{Example of a vulnerability fixed by multiple patches collaboratively.}
    \label{figure: Example of weak relationship between commit and CVE}
    \vspace{-15pt}
\end{figure*}

\vspace{-2pt}\noindent\textbf{Challenges in SPL}. To address this dilemma, automated techniques have been proposed to locate the security patch for a given CVE, such as using auxiliary information (e.g. CVE-ID) in vulnerability database~\cite{perl2015vccfinder,kim2017vuddy}, leveraging the external reference links in the CVE/NVD page~\cite{li2017large,perl2015vccfinder,wang2019detecting,wang2021patchdb,xu2022tracking}, and training a recommendation model based on CVEs and \chenyi{commits~\cite{wang2022vcmatch,tan2021locating}}. Among them, the security patch localization (SPL) recommendation methods notably stand out. 
%
However, since \ding{182} \textit{the content of CVEs and commits is often complex and intricate}\ignore{, and \ding{183} \textit{the available training data is rare}}, existing SPL recommendation models cannot handle cases if the commit has no clear association with the corresponding CVE.
Figure~\ref{figure: Example of a commit as a patch for CVE without clear association.} illustrates an example.
The commit cb1214c in OSS ImageMagick addresses the ``\textit{off-by-one read}'' vulnerability, referenced in CVE-2019-10131, by allocating an additional byte to memory (see code diffs in Figure~\ref{figure: Example of a commit as a patch for CVE without clear association.}). 
Existing SPL methods fail to identify this commit as it offers no descriptive content beyond the ellipsis, leaving no shared terms or similar semantics between the commit and CVE to be leveraged.
To recognize this commit as a patch for the vulnerability, one is expected to have an in-depth understanding of the CVE and commit, and possess security knowledge that associates adjusting memory allocation with a common remedy for the "\textit{off-by-one read}" issue.
Additionally, oftentimes, a vulnerability has multiple distinct patches proposed over time before it has been fully resolved, a scenario not considered by existing SPL solutions.
As shown in Figure~\ref{figure: Example of weak relationship between commit and CVE}, the commit cb62ab4 is a follow-up to the commit 684400c, which patch the OSS OpenSSL's vulnerability CVE-2014-8275, as can be easily identified from its commit description.
However, commit cb62ab4 has almost no direct association with the CVE in its content, thereby it would not be identified by existing SPL methods.
We believe this issue can be addressed by incorporating the relations between the two commits into the model training.
But, \textit{determining the inter-commit relationships poses another significant challenge (\ding{183}).}

\vspace{-1pt}\noindent\textbf{Our method}. 
Addressing these challenges in SPL requires extensive knowledge of software and security, especially vulnerabilities, as well as adept handling of semantic information from both natural language descriptions and programming code.
In some ways, Large Language Models (LLMs) have demonstrated owing the relevant capabilities~\cite{ fang2023large, noever2023chatbots, touvron2023llama, NEURIPS2020_1457c0d6, chen2021evaluating, falcon40b, vicuna2023, openai2023gpt35}.
Our experiments, in particular, revealed that LLM excels in comprehending CVEs and commits, notably by the high recall in recognizing their relationships.
However, directly applying LLM to solve SPL receives skepticism, evidenced by its high false positive rate ($>$85\%), which we will delve deeper into in Section~\ref{subsec: LLM Alone is Not Enough}.
Therefore, to adequately leverage LLM's capabilities, we proposed a joint learning approach, in which the outputs of LLM serve as additional features to aid the recommendation model in prioritizing security patches.
Specifically, our recommendation model leverages two LLM-based features: LLM's prediction of the relationship between CVEs and commits, and the LLM-endorsed inter-commit relationship graph of commits.
In the meanwhile, to make the cost of using LLM practical, we refined our recommendation algorithm by the LLM-feedback technique.
This technique necessitates seeking LLM feedback solely for commits positioned within the top-$k$ ranks, and successfully helped us drastically reduce costs from an initial estimate of 620,000 USD and a century-long processing time using the GPT-3.5 model, down to a mere 880 USD  with a processing duration of only 3 days.
We name this LLM-enhanced SPL approach as LLM-SPL.


\vspace{-1pt}\noindent\textbf{Results.}
We implemented LLM-SPL and evaluated its performance across a dataset of 1,915 CVEs with 2,461 associated patches. 
The results show that LLM-SPL effectively ranked the patches for 92.74\% CVEs within the top 10 positions, simultaneously delivering high-quality rankings as evidenced by the NDCG metric, which reached a high value of 87.33\%. This performance enables low manual effort, requiring the checking of only an average of 2.34 commits per CVE.
When compared to the state-of-the-art SPL method, VCMatch, LLM-SPL consistently outperformed in all metrics -- Recall, NDCG, and Manual Effort.
Particularly for CVEs requiring multiple collaborated patches, LLM-SPL significantly improved Recall from 60.30\% to 83.13\% (a 22.83\% increase), enhanced NDCG rom 60.99\% to 80.40\% (a 19.41\% increase), and reduced manual effort by over 25\% when checking up to the top 10 rankings in the practical application of LLM-SPL.
These results clearly demonstrate the effectiveness of our LLM-SPL approach.

\ignore{
We implemented LLM-SPL and evaluated it on a dataset including 1,915 CVEs associated with 2,461 patches.
Our evaluation shows that LLM-SPL can successfully rank the correct patch commit at the top for 70.54\% CVEs, place it within the top-5 for 90.43\%, top 8 for 92.73\%, and top 10 for 94.06\%.
Concurrently, LLM-SPL achieved an average NDCG of 84.16\% across metrics from $NDCG@1$ to $NDCG@10$, indicating the reliability of its rankings. 
When compared with the state-of-the-art SPL method, VCMatch~\cite{wang2022vcmatch}, LLM-SPL consistently outperformed in terms of both recalls and NDCGs.
Most notably, for vulnerabilities addressed by multiple collaborated patches, LLM-SPL exhibited marked improvements over VCMatch, enhancing the recall by 17.03\% and NDCG by 16.10\%.
%
These results highlight the capability of our LLM-SPL.
}

\vspace{0pt}\noindent\textbf{Contributions.} The contribution of this paper is a twofold:

\noindent$\bullet$\textit{~New Technique}.
We proposed an innovative SPL solution, LLM-SPL, which leverages the intelligence of large language models (LLMs) in comprehending CVEs, commits, and specialized knowledge in the field of software security. This solution efficiently and economically integrates insights from LLMs through a joint learning framework, utilizing them as additional features and feedback to refine the outcomes of a recommendation model for SPL.

\noindent$\bullet$\textit{~Performance Enhancement}. 
The evaluation results validate LLM-SPL's superior performance, consistently outperforming the state-of-the-art in all metrics—Recall, NDCG, and Manual Effort. Notably, for vulnerabilities requiring multiple patches, LLM-SPL significantly improves Recall by 22.83\%, NDCG by 19.41\%, and reduces manual effort by over 25\% when checking up to top 10 rankings.

\ignore{
\vspace{1pt}\noindent$\bullet$\textit{~New Technique}.
We propose an innovative and advanced SPL solution, LLM-SPL, which leverages the advantageous yet skeptical security intelligence of LLMs. 
It integrates the insights from LLM through a joint learning framework, utilizing them as additional features and feedback to refine the outcomes of a recommendation model for SPL. 

\vspace{1pt}\noindent$\bullet$\textit{~Performance Enhancement}. 
\chenyi{The evaluation results validate LLM-SPL's superior performance, consistently outperforming the state-of-the-art work in both recall and NDCG. Notably, for vulnerabilities needing multiple patches, LLM-SPL significantly improves recall by 17.03\% and NDCG by 16.10\%.}
}

\vspace{-15pt}
\section{Background}
\label{sec: background}
\vspace{-8pt}

\subsection{Vulnerability and Patch}
\label{subsec: Vulnerability and Patch}
\vspace{-8pt}

\vspace{2pt}\noindent\textbf{Common Vulnerabilities and Exposures.} 
CVE is a globally acknowledged catalog detailing known cybersecurity vulnerabilities.
Every entry in the CVE comprises details of a specific vulnerability, encompassing a unique identifier (CVE-ID), an in-depth description of the vulnerability, essential references that often cover bug reports and information about patches and so forth.
Figure~\ref{figure: CVE INFO} in Appendix illustrates an example.
For simplicity, we will use ``CVE'' to denote a CVE entry in the following paper.




\vspace{2pt}\noindent\textbf{Open source software patch.} 
Vulnerabilities in an OSS are addressed with patches, which involve changes to certain code segments. 
Typically, on GitHub~\cite{github}, these changes are made into commits that are subsequently pushed into the OSS's repository.
In a patch-related commit, there are \textit{code diffs} that highlight changes made on code and indicate the specific file being modified, together with the metadata including the author of this commit, the timestamp of this commit's creation, tags  representing the OSS's version, and the commit-id of the latest prior commit. Additionally, a commit contains a ``message'' section, referred as the \textit{commit description}, which might elucidate the specific vulnerability it seeks to resolve. We will use ``patch'' to denote a patch-related commit in following paper. An example is shown on Figure~\ref{figure: Commit INFO} in Appendix.



\vspace{-13pt}
\subsection{Security Patch Localization and VCMatch}
\label{subsec: Security Patch Localization}
\vspace{-8pt}

Security patch localization (SPL) is the task of pinpointing commits within an OSS's repository that serve as patches for addressing the vulnerability specified in a particular CVE.
Currently, the state-of-the-art (SOTA) SPL approach is VCMatch~\cite{wang2022vcmatch}, as supported by our experimental results in Section~\ref{sec: Evaluation}.
It forms the SPL as a recommendation problem, thus prioritizing patches with a higher ranking over non-patch commits. 
The recommendation algorithm employed by VCMatch utilizes five kinds of features: code behavior features, commit message identifiers, textual similarity features, security relevance features and temporal features. A comprehensive breakdown of these features can be found in Table~\ref{table: VCMatch Feature List} at Appendix.

\vspace{-13pt}
\subsection{Machine Learning}
\label{subsec: machine learning}
\vspace{-6pt}
\noindent\textbf{Large language model.} 
Large Language Models (LLMs) are advanced artificial intelligence systems capable of understanding and generating human-like text. 
These models not only exhibit exceptional proficiency in processing and producing both natural language and code~\cite{lample-etal-2016-ner,zukov2018-named,wang2021zero,wu2021openie,devlin2019bert,peters2018deep,chen2021evaluating,bubeck2023sparks}, but also possess  extensive domain knowledge, including expertise in software and security~\cite{noever2023chatbots,fang2023large,kang2023large}.

Such capabilities suggest that LLMs have the potential to comprehend the content of CVEs and commits, as well as discern their relationships, making them invaluable for supporting the SPL task.
However, LLMs are not without their limitations; they occasionally produce inaccuracies and hallucinations, which can compromise their reliability~\cite{huang2023survey}.
In Section~\ref{sec: A Potential Solver: Large Language Model}, we will discuss in detail both the potential and the limitations of LLMs within the SPL context. Subsequently, in Section~\ref{sec: design and implemenation}, we will introduce our methodology for leveraging LLMs to enhance SPL.

\vspace{0pt}\noindent\textbf{Relevance/Pseudo-relevance feedback for recommendation.}
In information retrieval, feedback mechanisms are crucial for refining search precision. Relevance feedback (RF) \cite{Radlinski2005query, rajaram2003classification} involves users in the retrieval process, enhancing results based on their input about relevant documents. This iterative process improves search results but at a high cost. Pseudo-relevance feedback (PRF)~\cite{Rocchio1971, li2018nprf} automates this process by assuming the top k results from an initial search are relevant. While this can enhance ranking performance, it may introduce noise.

\begin{figure}
    \centering
    \centering \includegraphics[width=0.42\textwidth]{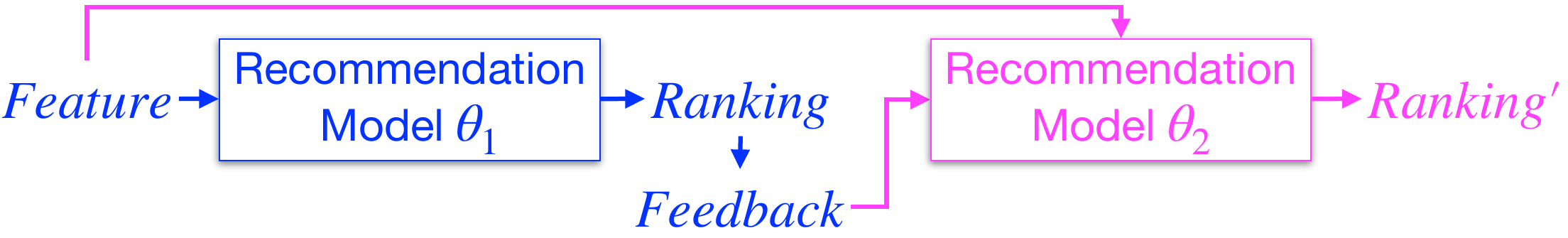}
    \vspace{-10pt}
    \caption{One-iteration feedback recommendation process.}
    \label{figure: Example of one-iteration feedback recommendation process}
    \vspace{-15pt}
\end{figure}

Figure~\ref{figure: Example of one-iteration feedback recommendation process} illustrates one-iteration of feedback recommendation process, where the feedback is from users in RF or assumption on top $k$ results in RPF.
Given the scarcity of domain experts and the high costs associated with their endeavor, employing RF to tackle the SPL challenge is nearly impractical. 
Meanwhile, relying on judgements from PRF poses a significant risk due to the prevalence of false positives in the top rankings produced by SOTA methods.
In our research, we proposed an approach, utilizing feedback from LLM, offers an invaluable, affordable and reliable solution for this precarious situation (see Section~\ref{subsec: Design}).




\vspace{-13pt}
\section{Challenges in Effective SPL}
\label{sec: challenges to SPL}
\vspace{-12pt}

Recent advances in SPL have utilized recommendation models, drawing on the content of CVEs and commits for model training, with the aim of identifying the commits that most likely address the vulnerabilities specified in given CVEs~\cite{wang2022vcmatch,tan2021locating}. 
The efficacy of these models is inherently tied to their ability to comprehend such content. 
However, the complexity and intricacy of CVE and commit content presents a significant challenge. 
Furthermore, certain vulnerabilities necessitate multiple distinct patches over time to achieve full resolution, a scenario inadequately addressed by current SPL solutions. 
We contend that incorporating the relationships among commits into the model training can effectively resolve this issue. Nevertheless, discerning these relationships is far from straightforward, presenting another challenge. 
Below, we will detail the two key challenges.

\vspace{-13pt}
\subsection{Challenge 1: Complexity of Content}
\label{subsec: challenge 1: complexity of content}
\vspace{-8pt}

Understanding or representing the content of CVEs and commits is crucial for success of SPL, as it not only facilitates the precise mapping of commits to their corresponding CVEs but is also essential for the effective feature extraction that is vital for training models. 
However, the content of both CVEs and commits is inherently complicated, demanding specialized knowledge for accurate comprehension.

\vspace{0pt}\noindent\textbf{CVE}.
As introduced in Section~\ref{subsec: Vulnerability and Patch}, a CVE is purely textual, detailing vulnerability information. To find what specific information CVEs actually provided, we carefully reviewed 100 randomly selected CVEs and summarized the types of information encountered. These findings, along with examples of how they are presented by CVEs, are shown in Table~\ref{table: information categories within CVE content} in the Appendix. The content within CVEs, as illustrated in the table, offers a wealth of information related to vulnerabilities, which can be categorized into four main areas: software information, vulnerability information, attack information, and patch information. Each category is further divided into subcategories. For example, the vulnerability information category may detail the type of vulnerability, the software's faulty functionality, and the names of files affected by the vulnerability, among others. Similarly, the attack information category might cover the payload of an attack packet, the method of attack, and the attack's impact.


However, understanding the information contained in a CVE is challenging. It demands not only knowledge of software security but also, in many cases, familiarity with the specific software affected by the vulnerability. 
For instance, without a background in security, one might not understand that \textit{integer overflow}, as documented in CVE-2012-2386, refers to a vulnerability type. 
Similarly, without knowledge of the Linux kernel, which carries a vulnerability documented in CVE-2020-11608, it becomes difficult to recognize that \textit{ov511\_mode\_init\_regs} and \textit{ov518\_mode\_init\_regs} are the names of the files affected by the vulnerability. 
Classical text learning methods, without dual knowledge in both security concepts and the affected software, can hardly generate high quality CVE/commit representations for effective recommendations.

\vspace{-1pt}\noindent\textbf{Commit}.
%
%
Compared to CVEs, commits feature a more complex content structure, including not only a textual description that notes the changes but also code diffs, which are the actual modifications made to the source code, as we introduced in Section~\ref{subsec: Vulnerability and Patch}.
To find the specific information offered by commit content -- similarly to how we approached CVE content -- we carefully reviewed 100 commits, randomly selected from 50 GitHub repositories. This review focused on both textual descriptions and code diffs. The diverse types of information encountered, along with examples, are detailed in Tables~\ref{table: information categories within commit description} and ~\ref{table: information categories within commit code diff} in the Appendix.
%
%

%
%
From Table~\ref{table: information categories within commit description}, we can see that the textual descriptions within commits closely align with the types of information found in CVE content, including but not limited to the vulnerability type, attack impact, and affected files.
Moreover, we observed that commit descriptions often provide information about fixes, such as fixing strategies or methods, as exemplified in Table~\ref{table: information categories within commit description}. 
This is logical, as their purpose is to describe or explain the modifications made.
Understanding commit descriptions, akin to understanding CVE content, requires not only knowledge of security concepts but also a deep familiarity with the affected software's source code, especially details related to fix information. 
This may present a challenge that, in some aspects, is even more substantial than understanding CVEs.
%
%

%
As illustrated in Table~\ref{table: information categories within commit code diff}, code diffs also convey diverse and rich information relevant to vulnerabilities.
While some code comments, written in natural language, clarify the purpose of modifications and functionalities of the modified code, the majority of code diffs comprise source code changes. These changes not only pinpoint the exact location of modifications but also provide clear insights into the functionalities of specific code segments. 
Indeed, comprehending source code is widely recognized as challenging~\cite{hansen2013makes}; however, understanding code diffs presents even greater difficulties.
This understanding demands not only programming knowledge and familiarity with the affected software's architecture and design but also a deep understanding of the context and rationale behind the changes. More specifically, it is essential to understand how a specific modification interacts with other code segments and addresses particular issues.
Acquiring such comprehensive knowledge is challenging, necessitating advanced programming skills and specialized training in software security.

\vspace{-13pt}
\subsection{Challenge 2: Inter-Commit Relations}
\label{subsec: Challenge 3: Establishment of Inter-commit Relationships}
\vspace{-8pt}


An analysis of all current CVEs reveals that 21.04\% of them require multiple patches for complete resolution.
This observation is understandable, as fully addressing an identified vulnerability sometimes necessitates several distinct patches over time.
For instance, additional or subsequent patches may be required in situations where an initial patch fails to completely address the vulnerability or introduces regression errors, or when more efficient or effective solutions are later developed. 
Moreover, in cases where a vulnerability impacts various components across a system, every affected component might require its own patch to ensure full protection.

%

Unfortunately, existing SPL approaches do not work well in these situations, where one vulnerability is associated with multiple patches, which we refer to as ``1-N'' in this paper. 
In our test set, nearly 40\% of 1-N CVEs\footnote{For clarity, a 1-N CVE refers to a CVE associated with multiple patches in the 1-N situation.} are not fully matched with their respective patches using the current SOTA SPL method~\cite{wang2022vcmatch}. 
We attribute this issue to the fact that existing SPL studies establish the association between a commit and a CVE based solely on the content of each commit and CVE, relying on a straightforward one-to-one relationship for determination. 
Consequently, patches that do not exhibit a clear relationship a CVE are often overlooked by these studies.
However, this phenomenon is particularly prevalent in 1-N situations, where we observed that many patches lack a clear relationship with their corresponding CVEs. 
Take, for example, the patch with commit ID cb62ab4 for OpenSSL, one of several patches collectively addressing the vulnerability CVE-2014-8275 (refer to Figure~\ref{figure: Example of weak relationship between commit and CVE}). 
This patch contains a brief description: \textit{``use correct function name ...''} and involves a one-line code modification, replacing the parameter \textit{``ASN1\_F\_ASN1\_VERIFY''} with \textit{``ASN1\_F\_ASN1\_ITEM\_VERIFY''}. 
Clearly, the content of this patch seems unrelated to the content within the CVE, making it challenging for even human experts to identify this commit as a patch for the CVE based solely on their individual content.

Upon revisiting commit cb62ab4, we discovered that to fix the vulnerability CVE-2014-8275, this commit works in conjunction with another commit, 684400c. The latter contains numerous clues indicating its role as a patch for CVE-2014-8275, even with the term ``CVE-2014-8275'' directly mentioned in its description, as shown in Figure~\ref{figure: Example of weak relationship between commit and CVE}.
Therefore, if relationships like the one between 684400c and cb62ab4 are identified (as highlighted in \blue{blue} in Figure~\ref{figure: Example of weak relationship between commit and CVE}), and such connections are considered during the model's learning process, it is likely that the model would not only recognize commit 684400c as a potential patch for CVE-2014-8275 based on their direct content association (as shown in \teal{green} in Figure~\ref{figure: Example of weak relationship between commit and CVE}) but also deduce a high probability of commit cb62ab4 being another patch for this CVE, taking into account the inter-commit relationships (as indicated in \magenta{red} in Figure~\ref{figure: Example of weak relationship between commit and CVE}).
Based on this insight, we believe that identifying the inter-commit relationships and incorporating them into the model's learning process would be a viable solution for addressing 1-N situations.

In our examination of the gathered 1-N data, we witnessed a strong linkage between the content of these commits, as evidenced by both the code diffs and the commit descriptions. 
Specifically, our investigation into the code revealed that these 1-N commits often modify the same sections of the codebase, focus on identical or closely related functionalities, or utilize similar modification patterns in various locations.
In the commit descriptions, we observed not only similarities in information related to vulnerabilities, attacks, and fixes, as categorized in Table~\ref{table: information categories within commit description}, but also additional details indicating connections among commits. For example, some commits share the same author or address the same issue ID, or one commit directly references another as a supplement or complement. 
However, understanding and identifying this information to build the inter-commit relationships demands extensive security knowledge and deep familiarity with the software, which, as previously discussed in Section ~\ref{subsec: challenge 1: complexity of content}\ignore{ and ~\ref{subsec: Challenge 2: Sparsity of Training Set}}, is significantly challenging.

\vspace{-20pt}
\section{A Potential Solution: Large Language Model}
\label{sec: A Potential Solver: Large Language Model}
\vspace{-5pt}




As discussed in Section~\ref{sec: challenges to SPL}, improving SPL recommendation methods fundamentally hinges on two challenges: understanding the content within CVEs and commits and recognizing the relationships among pieces of information in these contents. 
Our analysis, detailed in Section~\ref{sec: challenges to SPL}, reveals that overcoming these challenges necessitates a set of closely related comprehensive abilities: processing natural language, interpreting code, and possessing in-depth dual knowledge in the domains of security and software.
Emerging Large Language Model (LLM) technologies appear particularly well-suited to meet these demands. 
LLMs exhibit exceptional proficiency and accuracy in comprehending both natural language and code, as well as possessing extensive domain knowledge, including in security and software. 
Consequently, we believe that LLMs hold significant potential to address these challenges for enhancing SPL methods. 
In following sections, we will discuss both the potential and the limitations of LLMs in the SPL context.

\subsection{LLM Potential for Comprehension}
\label{subsec: Potential of LLM}
\vspace{-20pt}













The advancements in LLMs have demonstrated their near-human performance across a broad range of natural language processing (NLP) tasks, such as entity recognition~\cite{lample-etal-2016-ner, zukov2018-named}, information extraction~\cite{wang2021zero, wu2021openie}, and semantic understanding~\cite{devlin2019bert, peters2018deep}.
These capabilities are precisely what are needed to comprehend the textual descriptions within CVEs and commits, as shown in Table~\ref{table: information categories within CVE content} and ~\ref{table: information categories within commit description}.
For example, entity recognition enables the identification of key elements, such as software names, vulnerability types, and erroneous function names. 
Information extraction facilitates the gathering of detailed descriptions about faulty functionalities, attack methods, and fix strategies, among others. 
And semantic understanding allows for the comprehension of the inherent implications behind these texts.
When it comes to code, LLMs have also demonstrated remarkable proficiency in code-related tasks, such as code comprehension (understanding the logic and functionality of code segments), code summarization (creating concise descriptions of code functionality and its primary purpose), and code-to-comment alignment (associating code parts with their corresponding comments or descriptions)~\cite{chen2021evaluating, bubeck2023sparks}.
Such capabilities are crucial for a thorough understanding of code diffs, enabling comprehension of modified code functionalities and inference of the intentions behind these changes.
Particularly, as a generalized model, LLM possesses extensive domain knowledge including security and software.
This is showcased by LLMs' success in recognizing various types of vulnerabilities~\cite{noever2023chatbots}, identify malicious or suspicious behaviors within code~\cite{fang2023large}, and comprehend textual bug reports~\cite{kang2023large}.
Given these capabilities, we are convinced that LLMs hold significant potential for enhancing the comprehension of the content within CVEs and commits.

In addition, we conducted experiments using a real-world LLM model, specifically GPT-3.5, to empirically evaluate the LLM's capability in understanding CVEs and commits. 
We presented GPT-3.5 with 25 CVE and 25 commit samples, asking it to elucidate their contents. 
The results were compelling. 
The LLM not only demonstrated a deep and accurate understanding of the technical details within each sample but also successfully grasped their wider implications and context.
For instance, when we presented the CVE and commit illustrated in Figure~\ref{figure: Example of a commit as a patch for CVE without clear association.} to the LLM, it accurately interpreted the term \textit{``off\_by\_one read''} in CVE-2019-10131 as a vulnerability type, explaining it as related to a buffer issue: \textit{``In an off\_by\_one read vulnerability, the program incorrectly reads data from a buffer, by either accessing data beyond the buffer's bounds or ...''}. 
Regarding the commit cb1214c, the LLM effectively inferred the rationale behind the code modification to adjust memory allocation, suggesting that \textit{``Such a change likely indicates that there were some memory boundary issues in the original implementation.''}
These empirical results further strengthen our belief in the LLM's potential to comprehend the contents of CVEs and commits, thereby enhancing SPL.

\vspace{-13pt}
\subsection{LLM Potential for Relation Recognition}
\label{subsec: Potential of LLM}
\vspace{-5pt}

Furthermore, in recognizing relationships among pieces of information within CVEs and commits, we believe that LLMs have already mastered these connections, thanks to their inherent strengths in pattern recognition and their diverse training data.
Pattern recognition refers to identification of contextual connections within data, which is achieved by attention mechanism~\cite{vaswani2023attention} in LLMs.
This mechanism enables LLMs not only to recognize recurring structures but also uncover connections between seemingly unrelated pieces of information.
Moreover, the diversity of LLMs' training data, which includes both textual content and code~\cite{touvron2023llama}, empowers LLMs to forge relationships between text and code.
Thus, given the extensive training set of LLMs -- a well-established fact -- it is likely that LLMs have encountered and learned from patterns akin to those in CVEs and commits, perhaps not directly from these sources but in similar forms, multiple times during their training. 
Consequently, we believe that LLMs have already acquired the ability to discern patterns and establish these relationships among CVEs and commits.

To further validate our hypothesis, we conducted tests on LLM (specifically GPT-3.5) to assess its ability in recognizing the relationships between CVEs and commits, as well as among commits themselves. 
Specifically, we presented LLM with the content of 25 CVEs and their associated patches, asking whether the commits serve as patches for these CVEs; and we also provided LLM with 25 pairs of commits, inquiring if they collaboratively addressed the same vulnerability. 
The results were encouraging. 
LLM not only made correct judgments in the majority of cases 22 out of 25 for the first set, 19 out of 25 for the second set) but also provided reasonable and logical explanations for its decisions. 
Such experimental outcomes reinforce our belief that LLM has indeed learned these correlations, showcasing its potential as a valuable tool for establishing inter-commit relationships to enhance SPL recommendations.

\vspace{-1pt}

\vspace{-13pt}
\subsection{LLM Alone is Not Enough}
\label{subsec: LLM Alone is Not Enough}
\vspace{-5pt}





%
%

Given the LLM's potential to comprehend CVEs and commits content and its capability to identify relationships among these pieces of information, one might wonder why not directly apply LLMs to SPL by testing each commit of the affected software to see if it serves as a patch for a CVE.
To assess the effectiveness of this straightforward approach, we conducted experiments with GPT-3.5, using our collected dataset (for details of data collection, see Section~\ref{subsec: experimental setup}).
For the 1,915 CVEs in our dataset, we selected 50 commits for each, including the actual patches, from the affected software. We then formed a distinct CVE-commit pair for every individual commit, resulting in a total of 95,750 (1,915*50) pairs, which we used as input for the LLM.
The experimental results, as depicted in Table~\ref{table: Confusion matrix for LLM's identification} in the Appendix, reveal that directly applying LLMs to SPL is impractical: although the recall rate is reasonable at 87\%, the false positive rate is alarmingly high, exceeding 85\%.




This high false positive rate of 85\% indicates that for every 10 patches identified by the LLM, more than 8 are incorrect. In extreme cases, such as for CVE-2014-8545, the LLM identified as many as 38 commits as its patches, with 37 of them being incorrect, pushing the false positive rate above 97\%. This significant high false positive rate necessitates substantial manual effort from experts to sift through and identify the actual patches.
To understand why the LLM errs in identifying patches for CVEs so frequently, we analyzed 50 samples randomly selected from our dataset.
Our findings suggest that these false positives may stem from the LLM's excessive reliance on, or sensitivity to, specific patterns between a vulnerability and its patch.
For instance, commit d4fdceb was mistakenly identified as a patch for CVE-2016-2555, a SQL injection vulnerability in Atutor software.
This error likely occurred because the commit includes code for sanitizing a POST string, resembling an action to sanitize SQL input.
However, this change is unrelated to the vulnerability.



Additionally, the LLM missed 320 commits (false negatives). To investigate the underlying reasons, we analyzed 50 random samples from these overlooked commits.
What we observed was that the content of these commits lacks a clear association with the CVEs they actually patch, as demonstrated by the relationship between CVE-2014-8275 and its patch commit cdb2ab4, illustrated in Figure~\ref{figure: Example of weak relationship between commit and CVE}. 
However, we discovered that 26 of these commits were part of collaborative efforts with other commits to patch their associated CVEs. Notably, these collaborating commits had already been successfully identified by the LLM as patches for the CVEs. In the meanwhile, it was possible to establish the inter-commit relationships between these 26 commits and their collaborating commits through their content.
This observation aligns with the 1-N scenario we discussed in Section~\ref{subsec: Challenge 3: Establishment of Inter-commit Relationships}, reinforcing our idea that inter-commit relationships could help rectify these missed instances.
Given these insights, one may wonder why we cannot simply adjust the prompt to include all commits of the affected software along with the CVE, aiming to fully leverage LLMs for this purpose.
However, due to the input constraints of LLMs, such an approach is impractical.
Each interaction with an LLM is limited by input length and is independent of others~\cite{NEURIPS2020_1457c0d6}.
For example, the GPT-3.5 we used restricts a prompt to no more than 4,097 tokens~\cite{openai2023gpt35}.
Considering the substantial size of commits, which often include extensive code diffs, inputting multiple commits into the LLM simultaneously becomes challenging, if not impossible.




Our experiments and analysis indicate that directly using LLMs for SPL is inadequate, because the LLM has an excessively high false positive rate exceeding 85\%, along with the omission of certain commits that utilizing inter-commit relationships could potentially rectify.
Nevertheless, the LLM's recall rate of 87\% highlights its potential, specifically in identifying and establishing relationships between CVEs and commits.

\vspace{-13pt}
\section{LLM-SPL: Design and Implementation}
\label{sec: design and implemenation}
\vspace{-5pt}

\vspace{-5pt}
\subsection{Design}
\label{subsec: Design}

%
%
%
%

%
\vspace{-5pt}
As discussed in Section~\ref{sec: A Potential Solver: Large Language Model}, relying solely on LLMs to determine CVE-commit relationships proves inadequate and impractical. 
However, LLMs' impressive capabilities in comprehending CVE and commit content, along with identifying their complex relationships, align perfectly with the needs for addressing the challenges SPL currently faces.
Consequently, to leverage such capabilities for SPL, we proposed an LLM-enhanced approach, which, within a joint learning framework, augments the SPL recommendation method by considering LLM's judgement on CVE-commit relationships and LLM-endorsed inter-commit relationships.
Specifically, in training our model, besides using all the features from the SOTA SPL method (denoted as $F_0$) to preserve existing performance levels, we incorporate two additional sets of features endorsed by the LLM: \ding{182} the LLM's determination of whether a commit patches a CVE (denoted as $F_1$), and \ding{183} the inter-commit relationship graph established with the assistant of the LLM (denoted as $F_2$).

\begin{figure}
    \centering
    \includegraphics[width=0.44\textwidth]{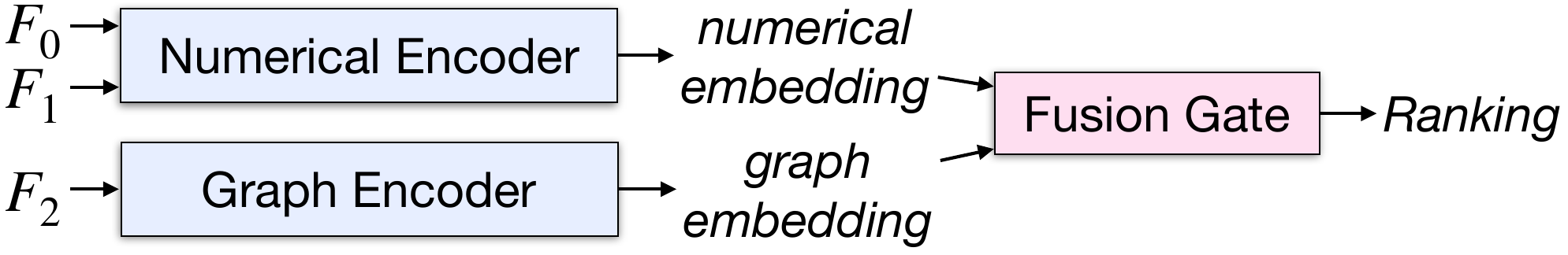}
    \vspace{-8pt}
    \caption{Design of joint learning framework.}
    \label{figure: Design of joint learning framework}
    \vspace{-15pt}
\end{figure}

\vspace{0pt}\noindent\textbf{Joint learning.}
Considering the diversity of features $F_0$, $F_1$, and $F_2$ -- with $F_0$ and $F_1$ presented in structured numerical formats and $F_2$ in an unstructured graph-based format -- and stemming from different data modalities (notably, $F_0$ captures the semantic meaning of texts, while $F_1$ and $F_2$ describe the relationships between objects) -- our design utilized joint learning. This method is well-suited for integration of varied feature representation and multi-modality features, ensuring our model effectively learns from these heterogeneous sets of features.
The primary idea behind joint learning for handling heterogeneous features is individually encoding each feature type and then strategically fusing them into a cohesive representation.
Based on this, we designed our joint learning framework, as illustrated in Figure~\ref{figure: Design of joint learning framework}. 
This framework includes two specialized encoders tailored to their respective feature types: a \texttt{Numerical Encoder} for $F_0$ and $F_1$, and a \texttt{Graph Encoder} for $F_2$.
The outputted embeddings are then blended by a \texttt{Fusion Gate} component, fusing the features into a singular score that represents the likelihood of a commit being a patch for a CVE. This score subsequently informs the ranking of commits.
Further details on the two encoders and the fusion gate are elaborated in Section~\ref{subsec: joint learning model}.

%

\vspace{0pt}\noindent\textbf{Challenges.}
However, during the model training with our dataset, we encountered prohibitive costs in both financial and temporal aspects when obtaining features $F_1$ and $F_2$ from the LLM. 
Financially, querying the LLM, GPT-3.5, that we used to determine whether a commit patches a CVE incurs an average cost of approximately 0.002 USD. 
For a dataset of $k$ software including $n$ CVEs and $m$ commits per software, the number of these queries is on the order of $O(nm)$, totaling 3 million in our dataset. 
Assessing whether two commits collaboratively address a vulnerability by GPT-3.5 requires a slightly higher cost of 0.003 USD per query, due to more tokens typically in commits than CVEs. 
These queries scale with $O(km^2)$, totaling 206 million in our dataset.
Cumulatively, the estimated total expense for our dataset exceeds 620,000 USD.
Regarding computational cost, each query, on average, takes 15 seconds to process in GPT-3.5, leading to time costs of $O(nm) + O(km^2)$. 
For our dataset, a lone user would take about \textit{``a century''} to complete all queries if working without any interruption\footnote{Other open-source LLM solutions, e.g., Llama, also have these computational cost challenges.}.
Notably, our dataset is relatively modest, only related to 200 software including 1,915 CVEs and a maximum of 1,500 commits for each software (as detailed in Section~\ref{subsec: experimental setup}). 
Yet, the estimated costs have already reached a prohibitive level.
Any further expansion of the dataset for enhanced performance would quadratically increase associated costs, making the application of LLMs untenable.

\begin{figure*}
    \centering
    \includegraphics[width=0.68\textwidth]{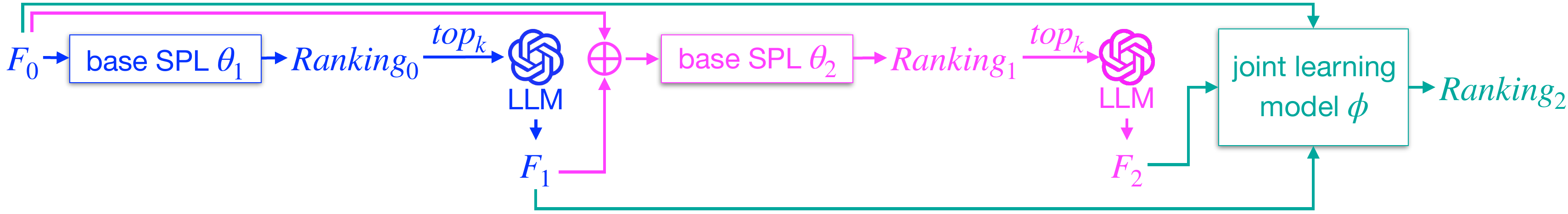}    
    \footnotesize
    \\
    \vspace{-2pt}
    \textit{Note: The detailed design of the joint learning model is illustrated in Figure~\ref{figure: Design of joint learning framework}.}
    \vspace{-8pt}
    \caption{Architecture of LLM-SPL.}
    \label{figure: Architecture of LLM-SPL}
    \vspace{-18pt}
\end{figure*}

\vspace{0pt}\noindent\textbf{Solution.}
To address this cost challenge of using LLMs, we referred to pseudo-relevance feedback recommendation, where the main concept is to annotate the top \textit{k} rankings from an initial model and then use these annotations as feedback to refine the model (see Section~\ref{subsec: machine learning}).
Inspired by this, we utilized $F_1$ and $F_2$ as LLM-feedback to refine the SPL recommendation model, but we limited LLM queries for $F_1$ and $F_2$ to only the top \textit{k} ranked commits from the initial SPL method, so that we can avoid a vast number of LLM queries for those ranked lower.
This strategy not only significantly reduces costs but also effectively leverages LLM insights to optimize the model's ranking.
Notably, regarding $F_2$, an inter-commit relationship graph, as discussed in Section~\ref{subsec: Challenge 3: Establishment of Inter-commit Relationships}, it is designed to elevate commits that might rank low individually but have strong ties to higher-ranked ones.
This means, for an inter-commit relationship to be impactful, at least one commit within it needs to be highly ranked.   
Thus, building a graph consisting solely of the top \textit{k} ranked commits is not just a cost-saving measure but also a practical solution which eliminates the effort of forming unnecessary connections in the graph.
In our experiments, instead of an estimated cost of 620,000 USD and a 100-year time span on our dataset, our method spent only 880 USD and just 3 days for a single user to complete the task.
\vspace{0pt}\noindent\textbf{Architecture.}
Figure~\ref{figure: Architecture of LLM-SPL} depicts the architecture of our method, named \toolname (\textbf{LLM}-Enhanced \textbf{S}ecurity \textbf{P}atch \textbf{L}ocalization).
The process is divided into three steps: 1) obtaining $F_1$ (indicated in \blue{blue}); 2) establishing $F_2$ (marked in \magenta{red}); and 3) training the joint learning model (highlighted in \teal{green}).
Specifically, in the first step, we use the LLM to label each of the top-ranked commits from an initial ranking ($Ranking_0$), which is generated by the base SPL recommendation model, as either patching a specific CVE or not, thereby obtaining the LLM-feedback $F_1$.
Notably, since a better initial ranking enhances the final outcome of the model in the pseudo-relevance feedback, in our implementation, we utilized the SOTA SPL model, VCMatch~\cite{wang2022vcmatch}, as our base SPL.
The $F_0$ in Figure~\ref{figure: Architecture of LLM-SPL} represents all features used by VCMatch (introduced in Section~\ref{subsec: Vulnerability and Patch}).
In the second step, for reasons similar to why we used the SOTA as the base SPL in Step 1, we hope the initial ranking used for establishing the graph $F_2$ is also as optimal as possible.
Considering that we already have feedback $F_1$, which can potentially enhance a model's ranking capabilities, thus, we take this feedback $F_1$ as the input with $F_0$ through concatenation operation to train the base SPL (VCMatch, again) to produce the initial ranking for this phase (denoted as $Ranking_1$).
After that, we establish inter-commit relationships specific to the top-ranked commits from $Ranking_1$, to finally construct the graph $F_2$.
Finally, in the third step, incorporating three features $F_0$, $F_1$ and $F_2$, we train the joint learning model, as illustrated in Figure~\ref{figure: Design of joint learning framework}, to deliver the ultimate commit ranking ($Ranking_2$) as our \toolname's output.

Notably, in our architecture, why we first obtain $F_1$ before $F_2$ is, once again, for saving costs.
This design is based on two primary considerations:
1) As just analyzed, procuring $F_2$ entails considerably greater costs than obtaining $F_1$;
2) Given the principle behind the effectiveness of $F_2$, as discussed in Section~\ref{subsec: Challenge 3: Establishment of Inter-commit Relationships}, optimizing the quality of the initial ranking (refer to $Ranking_1$ in Figure~\ref{figure: Architecture of LLM-SPL}) makes it possible to create a practically useful inter-commit relationship graph with fewer top-ranked commits that is just as effective as one using broader set of commits.
Therefore, our strategy is to first get $F_1$ and use it as a feedback to optimize the model, aiming to generate an improved ranking as the foundation for constructing $F_2$. Then, when LLM-SPL constructs the $F_2$ graph based on this ranking's top $k$ commits, the $k$ can be chosen as a relatively small number.
This strategy, while preserving the effectiveness of $F_2$, offers significant cost savings for building the graph.

\vspace{-13pt}
\subsection{Feature Generation based on LLM}
\label{subsec: feature generation based on LLM}
\vspace{-5pt}

As defined in Section~\ref{subsec: Design}, features $F_1$ and $F_2$ refer to the LLM's determination of whether a commit patches a CVE and whether two commits jointly address the same vulnerability.
In the following, we detail the prompts we used to generate $F_1$ and $F_2$ in \toolname, and elaborate on their generation during model learning.

\vspace{0pt}\noindent\textbf{Prompt templates.}

Figure~\ref{figure: template of prompt for F1} and~\ref{figure: template of prompt for F2} in Appendix illustrate the prompt templates we used to query the LLM for $F_1$ and $F_2$.
Each prompt consists of four parts.
Initially, Following established prompt engineering practices, we inform the LLM of its role as a software security analyst, highlighted in \olive{olive} in the figures, to leverage the LLM's expertise in software security.
Then, we provide the LLM with a CVE and a commit (for $F_1$) or two commits (for $F_2$), highlighted in \teal{green} in figures.
The task description, highlighted in \magenta{red}, instructs the LLM to analyze content and determine relationships. For $F_1$, the LLM accesses if the commit serves as a patch to the CVE.
For $F_2$, the LLM judges if two commits collaboratively address the same vulnerability. To ensure clear comprehension of the LLM's decision, our prompts (highlighted in \blue{blue}) direct the LLM to conclude with a single line stating YES, NO or Unknown. This structure allows us to efficiently extract the LLM's determination for $F_1$ and $F_2$ through simple string inspection.
Notably, when the LLM returns an UNKNOWN decision, considering its high recall rate and associated high false positive rate as discussed in Section~\ref{subsec: LLM Alone is Not Enough}, we treat such determinations as equivalent to NO.

\vspace{0pt}\noindent\textbf{Generation of $F_1$.}
As introduced in Section~\ref{subsec: Design}, the feature $F_1$, serving as LLM-feedback, represents the LLM's determination of whether each of the top $k$ commits in $Ranking_0$ patches a particular CVE.
In our model, we use a $k \times 1$ vector to represent it, where $k$ specifies the number of top-ranked commits considered.
Each entry in the vector is set to 0 if the LLM determines the commit does not patch the CVE, or 1 if it does.
To construct the $F_1$ vector, for every CVE in our dataset, we query the LLM with $k$ prompts (see Figure~\ref{figure: template of prompt for F1} in Appendix) regarding the top $k$ commits from $Ranking_0$, asking for the LLM's determination of their relevance to patching the CVE.
The responses from the LLM are used to populate the corresponding entries in the $F_1$ vector.

In our experiment, we set $k=100$.
To find the optimal value for $k$, we tested various numbers to see how often at least one patch for CVEs being ranked within top $k$ by the base SPL model.
Here, we considered only one patch because we believe $F_2$ would rectify any other commits that jointly address the CVE.
The experimental result is illustrated in Figure~\ref{figure: The recall of CVE with at least one patch ranked in top k on ranking 1}.
As it demonstrates, we observed an inflection point at $k=100$, where the recall of CVEs reached 98.12\%.
Increasing $k$ beyond this results in negligible improvements in detection.
Thus, we chose $k=100$ in our experiment to balance identification with overall costs.

\begin{figure*}[t]
   \noindent
   \begin{minipage}{0.22\textwidth}
     \centering
     \includegraphics[width=1\textwidth]{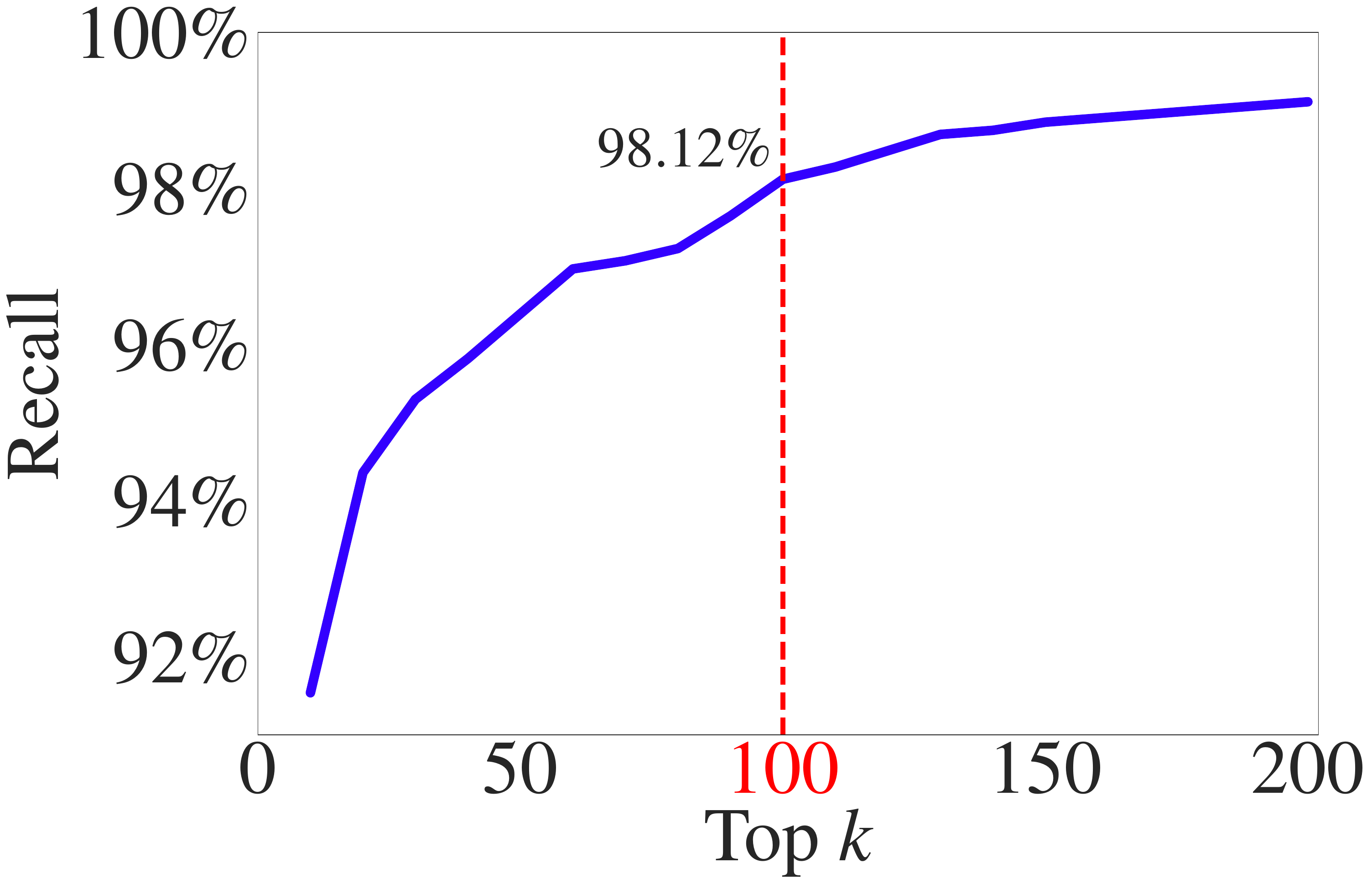}
     \vspace{-15pt}
     \captionsetup{skip=2pt}
     \caption{The recall of CVE with at least one patch ranked in top $k$ on $Ranking_0$.}
     \label{figure: The recall of CVE with at least one patch ranked in top k on ranking 1}
   \end{minipage}
   \hfill
    \noindent
   \begin{minipage}{0.23\textwidth}
     \centering
    \includegraphics[width=1\linewidth]{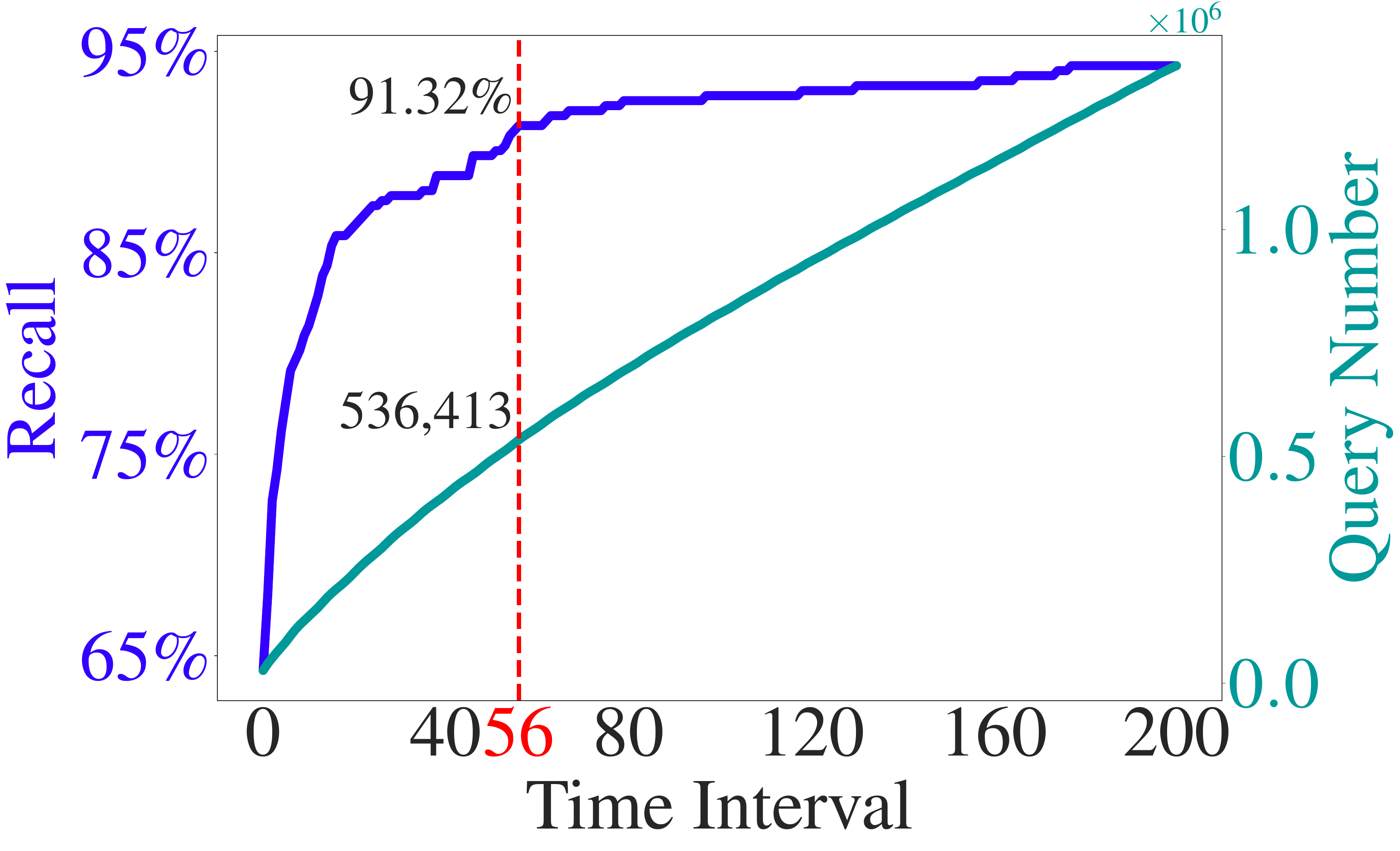}
    \vspace{-15pt}
    \caption{The recall of 1-N CVEs and the number of LLM queries over the time interval.}
    \label{figure: The recall of 1-N CVEs and the number of LLM queries over the time frame.}
    \end{minipage}
   \hfill
   \noindent
   \begin{minipage}{0.205\textwidth}
     \centering
    \includegraphics[width=1\linewidth]{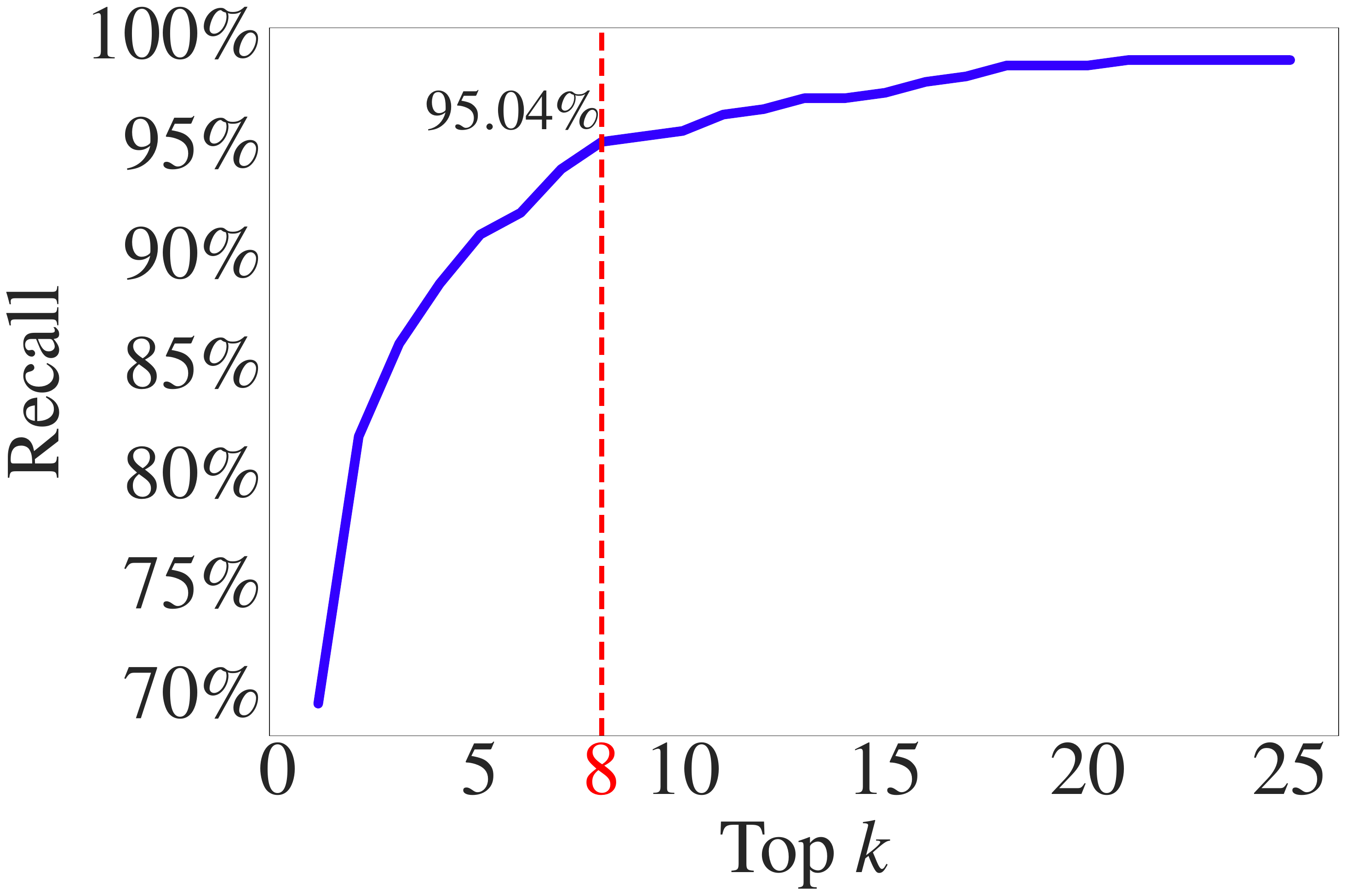}
    \vspace{-15pt}
    \caption{The recall of CVE with at least one patch ranked in top $k$ on $Ranking_1$.}
    \label{figure: The recall of CVE with at least one patch ranked in top k on ranking 2}
    \end{minipage}
    \hfill
    \noindent
    \begin{minipage}{0.28\textwidth}
    \centering
    \includegraphics[width=1\textwidth]{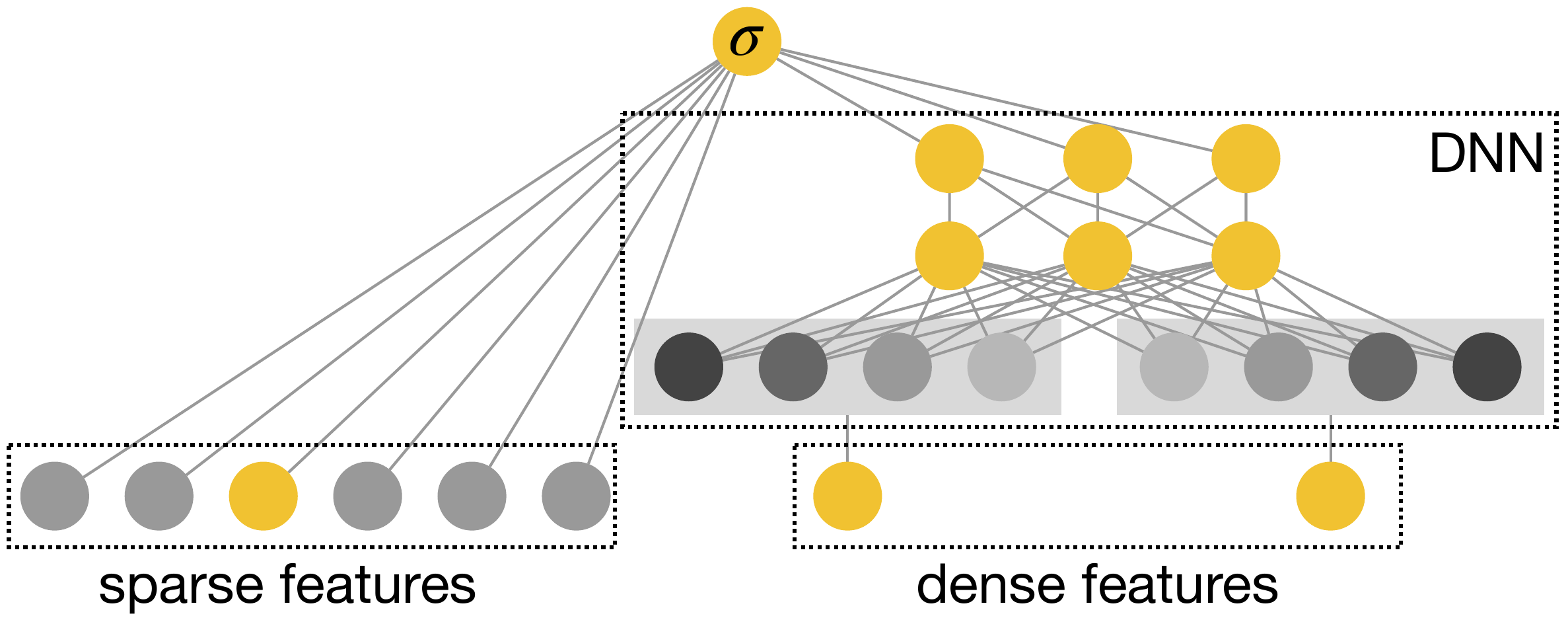}      
    \caption{Numerical encoder.}
    \label{figure: numerical encoder}
    \end{minipage}
    \vspace{-17pt}
\end{figure*}

\vspace{-3pt}\noindent\textbf{Generation of $F_2$.}
As introduced in Section~\ref{subsec: Design}, the feature $F_2$ serves as another LLM-feedback, representing the inter-commit relationship graph endorsed by the LLM, which pertains to the top $k$ commits in $Ranking_1$. 
In our study, 
$F_2$ is formally defined as a graph $G:\{V, E\}$, where $V$ represents a set of nodes, each node representing a commit, and $E$ represents a set of undirected edges, each edge representing a relationship between a pair of commits. 
For any given edge $e=\{u,v\}$ which connects nodes $u$ and $v$, the edge weight is assigned a value of 1 if the LLM determines that commits $u$ and $v$ collaboratively address the same vulnerability, or 0 otherwise.

\vspace{-2pt}
When constructing the graph $G$, a direct approach is to pair each top $k$ commit from $Ranking_1$ with every other commit in the software and query the LLM to determine if they collaboratively address the same vulnerability.
This method, for our dataset, would require approximately 9 million LLM queries, still leading to substantial costs (refer to Section~\ref{subsec: Design}).
To mitigate this, we proposed a strategy to further reduce the number of queries to the LLM.
Specifically, we only ask the LLM when a pair of commits satisfies the following conditions:
1) sharing the same author or committer; and 
2) being submitted within a specific time frame.
This strategy comes from our observations about the common practices of collaborated patches.
In our study, we observed that they are typically handled by the same individual, either the author, who designs the patch and crafts the commit, or the committer, who integrates the commit into the software.
Additionally, they are often consecutively applied in the software.
Thus, using this strategy, we query the LLM for the commit pairs that are more likely to address the same vulnerability, thereby achieving cost savings.
Notably, the information needed to determine if two commits meet these two conditions is presented in a standardized format within commit metadata (see Figure~\ref{figure: Commit INFO}), making it both easily accessible and straightforward for assessment.
Moreover, for the second condition, we set the time frame at 56 days based on our analysis.
As Figure~\ref{figure: The recall of 1-N CVEs and the number of LLM queries over the time frame.} illustrates, in our dataset, as the time frame is extended beyond 56 days, there is no significant increase in the coverage of collaborated commits, yet the number of queries continues to increase linearly.

More specifically, to establish the graph $G$ of $F2$, for each CVE, we enumerate the top $k$ commits in $Ranking_1$, querying the LLM using the number of $\sum_{i=1}^ks(V_{u_i})$ prompts (refer to Figure~\ref{figure: template of prompt for F2}), seeking its determination of whether commit $u$ and commit $v$ from $V_u$ collaboratively address the same vulnerability.
In this context, 
$u \in \{u_1, u_2, ..., u_k\}$, denoting the top $k$ commits in $Ranking_1$,
$V_u$ is a set of commits which satisfy the previously mentioned two conditions in relation to commit $u$,
$s(V_u)$ is the size of the set $V_u$.
Upon querying, if the response from the LLM is YES, indicating the two commits $u$ and $v$ jointly address a vulnerability, we assign a weight of 1 to the edge $e=\{u,v\}$; otherwise, it is set to 0.

Furthermore, during the graph building process for $F_2$, we set $k = 8$.
This choice comes from the similar analysis as we did to find the optimal top $k$ for obtaining $F_1$ (as referred to Figure~\ref{figure: The recall of CVE with at least one patch ranked in top k on ranking 1}).
Similarly, we tested with various values of $k$, assessing how many CVEs had at least one of their patches ranked within top $k$ in $Ranking_1$.
The result is illustrated in Figure~\ref{figure: The recall of CVE with at least one patch ranked in top k on ranking 2}. This figure reveals an inflection point at $k = 8$, achieving a recall of 95.04\% for the CVEs. Any further increase beyond $k = 8$ results in only marginal improvements in coverage. 
Thus, to balance the identification efficacy with overall costs, we selected $k=8$ in our experiment.

\vspace{-15pt}
\subsection{Joint Learning Model}
\label{subsec: joint learning model}
\vspace{-5pt}

Figure~\ref{figure: Design of joint learning framework} illustrates our design of the joint learning framework. We use a numerical encoder for $F_0$ and $F_1$, a graph encoder for $F_2$, and then a fusion gate to combine embeddings from these encoders. The resulting score represents the probability of a commit being a patch for a CVE, which informs LLM-SPL's recommendation ranking.


\vspace{0pt}\noindent\textbf{Numerical Encoder.} We utilize this encoder to merge shallow numerical features: $F_0$ and $F_1$, creating a consolidated embedding to enhance subsequent ranking outcomes. Specifically, $F_1$ is a sparse feature with values of either 0 or 1 (commit-relevance feedback from LLM), and $F_0$ encompasses 32 diverse features that could be dense or sparse. All features within $F_0$ and their respective types are detailed in Table~\ref{table: VCMatch Feature List}. For sparse features, cross-product transformations efficiently capture interpretable interactions. However,  this requires significant feature engineering for generalization. Deep neural networks (DNN) exhibit strong generalizations on dense features. Yet, DNNs tend to over-generalize and perform sub-optimally with sparse features. 
Therefore, we leverage the Wide \& Deep framework~\cite{cheng2016wide}, adept at balancing memorization and generalization, to efficiently manage both kinds of features concurrently.

To enrich the original $F_0$ and $F_1$, we incorporate dense feature categorization. Specifically, we derive an additional sparse feature by segmenting a dense feature into bins, which are determined based on its variance within our dataset. Formally, for a dense feature $x$, we derive a sparse one $x'$ by:
\vspace{-12pt}
\begin{equation} 
\notag
\begin{array}{r@{\ }l}
x' = \frac{x-x_{min}}{x_{intv}}, x_{intv}=\frac{x_{max}-x_{min}}{n_{bins}},
\end{array}
\vspace{-5pt}
\end{equation}
where $x_{max}$ and $x_{min}$ represent the maximum and minimum values of $x$ in our dataset, and $n_{bins}$ denotes the chosen number of bins. 


\ignore{
Among all sparse features (including $x'$), we manually create a set of cross-product transformations $\{\phi_k(x)\}$ based on their semantics using the following equation:
\begin{equation} 
\notag
\begin{array}{r@{\ }l}
\phi_k(\mathbf{x}) = \Pi x_i^{c_{ki}}, c_{ki} \in \{0,1\},
\end{array}
\end{equation}
where $c_{ki}$ is a boolean variable that is 1 if the i-th feature is part of the k-th transformation $\phi_k$, and 0 otherwise. This transformation captures the co-occurrence relationships among sparse features.
}

The overall structure of our numerical encoder is shown in Figure~\ref{figure: numerical encoder}. Formally, it produces embeddings according to:
\vspace{-5pt}
\begin{equation} 
\notag
\begin{array}{r@{\ }l}
\sigma(\mathbf{w}_{wide}^T \mathbf{x}_{sparse} + \mathbf{w}_{deep}^T f_{deep}(\mathbf{x}_{dense}) + b).
\end{array}
\vspace{-5pt}
\end{equation}
Here, $\mathbf{w}_{wide}$ and $\mathbf{w}_{deep}$ represent the weights for the wide and deep components, respectively. The term $\mathbf{x}_{sparse}$ denotes all sparse features including those from dense ones, and $f_{deep}(\mathbf{x}_{dense})$ refers to the embeddings from a DNN model $f_{deep}$ specific to the dense features $\mathbf{x}_{dense}$. Additionally, $b$ is the bias term, and $\sigma(\cdot)$ is a element-wide sigmoid function.


\vspace{0pt}\noindent\textbf{Graph Encoder.} 
We employ the graph encoder to generate a semantically-rich embedding for every node in the graph, elevating previously lower-ranked but actual patches of a CVE. While our graph, constructed based on LLM's feedback, includes edges that link high-ranked patches with their lower-ranked counterparts for positive impacts, it also has edges connecting non-patch commits that may introduce side effects. To ensure our model emphasizes the beneficial connections (edges) and minimizes the influence from the less desirable ones, we've integrated an attention mechanism into our model. 
Specifically, we utilized a Graph ATtention network (GAT)~\cite{velivckovic2017graph} to generate each node's embedding.

The embedding generation process of GAT is demonstrated as follows.
Initially, each node's embedding is initialized by using a representation of its corresponding commit, generated via a pre-trained language model (BERT). Subsequently, the embedding of node-$i$ undergoes iterative refinement by computing a weighted average of the feature vectors from its neighboring nodes:
$\mathbf{h}_i = \sigma(\sum_{k \in \mathcal{N}_i}\alpha_{ik} \mathbf{W}^T \mathbf{h}_k)$,
where $\alpha_{ik}$ signifies the attention coefficient, $\mathbf{h}_k$ denotes node-$k$'s embedding, and the feature vector of node-$k$ is $\mathbf{W}^T \mathbf{h}_k$, the matrix-vector product of the shared weight matrix $\mathbf{W}$ with its node embedding $\mathbf{h}_k$.

The attention coefficient, $\alpha_{ij}$, between nodes-$i$ and -$j$ is derived using the softmax function on the raw attention scores, $e_{ij}$, over all neighbors of node-$i$, represented by $\mathcal{N}_i$:
\vspace{-10pt}
\begin{equation} 
\notag
\begin{array}{r@{\ }l}
\alpha_{ij} = \frac{\exp(e_{ij})}{\Sigma_{k \in \mathcal{N}_i}\exp(e_{ik})}, 
e_{ij} = a(\mathbf{w}_e^T [\mathbf{W}^T \mathbf{h}_i, \mathbf{W}^T \mathbf{h}_j]).
\end{array}
\vspace{-6pt}
\end{equation}
 The raw attention score, $e_{ij}$, is obtained by applying the LeakyReLU activation function, $a(\cdot)$, to the dot product of the weight vector, $\mathbf{w}_e$, with the concatenated feature vectors of nodes, $[\mathbf{W}^T \mathbf{h}_i, \mathbf{W}^T \mathbf{h}_j]$. Notably, the weights $\mathbf{W}$ and $\mathbf{w}_e$ are consistent and used for all nodes in the graph.

\vspace{0pt}\noindent\textbf{Fusion Gate.}
We use it to combine two embeddings produced by encoders. Specifically, for a commit, it has a numerical embedding $E_N$ created by the numerical encoder and a graph embedding $E_G$ optimized by the graph encoder. The fusion gate combines them by:
\vspace{-8pt}
\begin{equation} 
\notag
\begin{array}{r@{\ }l}
E_N \frac{\exp(\mathbf{w}_E^T E_N)}{\exp(\mathbf{w}_E^T E_N)+\exp(\mathbf{w}_G^T E_G)} + E_G \frac{\exp(\mathbf{w}_G^T E_G)}{\exp(\mathbf{w}_E^T E_N)+\exp(\mathbf{w}_G^T E_G)},
\end{array}
\vspace{-8pt}
\end{equation}
where $\mathbf{w}_E$ and $\mathbf{w}_G$ are two weight vectors indicating the emphasis placed on each type of embedding.

\vspace{2pt}\noindent\textbf{Loss Function.}
By integrating a linear classifier after the fusion gate, we perform binary classification to determine if a commit acts as a patch for a designated CVE. Given the skewness in our dataset, where patch commits are vastly outnumbered by non-patch commits, we use the Focal loss~\cite{lin2017focal}, as employed in SOTA SPL~\cite{wang2022vcmatch}, to mitigate this imbalance. Essentially, the Focal loss diminishes the relative loss for correctly classified instances and prioritizes challenging, misclassified cases. Furthermore, the Focal loss introduces a parameter to balance the weights of positive and negative samples, which, according to our dataset, is set to be $\frac{1}{1500}$. 

After adequately training our joint learning model, we obtain a score from the final linear classifier indicating the likelihood of a commit being a patch for a specific CVE. By sorting these scores for each commit, we obtain the final ranking of them, $Ranking_2$.

\vspace{-14pt}
\section{Evaluation}
\label{sec: Evaluation}
\vspace{-10pt}

\subsection{Experimental Setup}
\label{subsec: experimental setup}
\vspace{-10pt}

\vspace{5pt}\noindent\textbf{Data}.
Our data set is collected from the National Vulnerability Database (NVD)~\cite{NVD} and Open Source Software (OSS) repositories on GitHub~\cite{github}. We gathered all the commits from 200 OSS repositories on GitHub, along with their associated CVEs that are listed in the NVD. We selected these 200 OSS repositories based on the most recent CVEs reported to the NVD.
For a CVE, we identify a commit as its patch only if the commit is explicitly mentioned in the CVE's report (as illustrated in Figure~\ref{figure: CVE INFO}).
Any other commits related to the associated OSS are not considered as patches for this CVE.
It's important to note that the information of a CVE we use for labeling commits is not included in our to-be-analyzed content of CVE.
In total, from 200 OSS repositories, we collected 1,915 CVEs and 2,461 associated patches across all their commits.

When constructing our dataset, we define a positive sample as a CVE paired with its patch, and a negative sample as a CVE paired with a commit that is not its patch.
To ensure that positive samples are not overshadowed by negative ones, we follow previous SPL works~\cite{wang2022vcmatch,tan2021locating} by restricting the number of non-patch commits used to create negative samples for each CVE to an experimental maximum of 1,500.
In total, our dataset contains 2,461 positive samples and 3,120,898 negative samples.

To comprehensively evaluate LLM-SPL, we established three datasets: Full, 1-1, and 1-N.
The Full dataset includes every sample we labeled.
The 1-1 dataset consists of samples corresponding to 1,512 CVEs that can be fully resolved with just one patch.
The 1-N dataset includes samples related to 403 CVEs each requiring multiple patches to fix jointly.

\vspace{-1pt}\noindent\textbf{Implementation.}
In the first two steps of LLM-SPL, we utilized the SOTA VCMatch framework as the base SPL and followed its configuration. Step 1 used feature set $F_0$ (model parameter $\theta_1$,  while step 2 concatenated $F_0$ and $F_1$ (parameter $\theta_2$). Both models were trained on the same labeled dataset. 
The third step's joint learning model used Adam optimizer (learning rate 0.001, L2 regularization 1e-5, dropout 0.4), trained over 100 epochs with batch size 10240. Numerical Encoder using 10 bins for discretization, followed by a three-layer MLP(256 dimensions per layer.) The Graph Encoder employed a two-layer GAT with 4 attention heads and 256 hidden sizes per layer.


\vspace{0pt}\noindent\textbf{Experimental environment.} We used an Ubuntu 20.04.6 LTS 64-bit machine (with 503 GB memory, an AMD EPYC 7543 32-Core Processor and 1 NVIDIA A100 GPU) for model training.

\vspace{0pt}\noindent\textbf{Evaluation method.}
To evaluate LLM-SPL's performance, we adopt a five-fold cross-validation, consistent with the SOTA VCMatch, against which we compared in our evaluation. The results presented are the average values of each metric obtained over the five iterations.
\vspace{-13pt}
\subsection{Effectiveness}
\label{subsec: effectiveness}
\vspace{-10pt}


\vspace{2pt}\noindent\textbf{Research questions.}
To evaluate the effectiveness of LLM-SPL, we conducted experiments focusing on the following two questions:
\ding{182} \textit{How thoroughly can LLM-SPL retrieve all patches for a CVE?}
This assessment is crucial as it directly correlates with LLM-SPL's practical utility. Specifically, the ability to uncover all patches in 1-N scenarios is vital to ensure comprehensive vulnerability remediation.
\ding{183} \textit{How effectively can LLM-SPL prioritize patches by ranking them above non-patch commits?}
This question assesses the quality of the ranking produced by LLM-SPL, which is crucial for enabling users to quickly identify a CVE's patch(es).

\vspace{0pt}\noindent\textbf{Metrics.}
To answer these questions, our metrics are:

\vspace{-1pt}\noindent$\bullet$\textit{~Recall.} 
To answer the first questoin, we utilize the metric Recall at top \textit{k} (\textit{R@k}). This metric measures the proportion of CVEs for which LLM-SPL successfully locates all corresponding patches within the top \textit{k} results, with \textit{k} being a parameter configurable by users. Specifically, \textit{R@k} is the ratio of the number of CVEs whose patches are all located within the top \textit{k} rankings to the total number of CVEs analyzed.

\vspace{0pt}\noindent$\bullet$\textit{~NDCG.}
For the second question, we utilize the metric Normalized Discounted Cumulative Gain (NDCG)~\cite{WikiNDCG} at top \textit{k} (denoted as \textit{N@k}). \textit{N@k} is a widely recognized metric for evaluating ranking quality; it not only considers the position of relevant items within the top \textit{k} rankings but also assigns greater weight to those items positioned higher in the list. 
An \textit{N@k} score, ranging from 0 to 1, indicates that a higher score reflects a more favorable ranking of patches over non-patch commits within the top \textit{k} results, thus reflecting a more effective recommendation system. 

\vspace{0pt}\noindent$\bullet$\textit{~Manual Effort.}
Besides using the metric NDCG to measure the quality of rankings, we calculate the Manual Effort at top \textit{k} (denoted as \textit{M@k}), following precedents established in previous SPL recommendation studies~\cite{wang2022vcmatch,tan2021locating}.
\textit{M@k} quantifies the average number of commits that users need to manually review in the list of top \textit{k} items from top to bottom before locating all the patches for a CVE.
Specifically, for a given CVE, if the lowest ranked patch is placed at the r-th position by LLM-SPL and users choose to check the top \textit{k} commits, the manual efforts needed for this CVE is calculated as $min(r,k)$.
For a dataset of \textit{n} CVEs, the average manual effort, $M@k$, is calculated using this formula: $M@k=\frac{\sum_{i=1}^{n}min(r_i,k)}{n}$, where $r_i$ is the ranking position of the lowest patch for the i-th CVE within the top \textit{k} results.

\begin{table*}[t]
    \centering
    \captionsetup{skip=2pt}
    \caption{Recall ($R@k$) of PatchScout, FixFinder, VCMatch and LLM-SPL.}    
    \label{table: recall of models}
    \vspace{-2pt}
    \begin{adjustbox}{width=0.87\textwidth}

\begin{tabular}{|c|cccc|cccc|cccc|}
\hline
\multirow{2}{*}{\textit{k}} 
    & \multicolumn{4}{c|}{\textbf{Full}}      
    & \multicolumn{4}{c|}{\textbf{1-1}}       
    & \multicolumn{4}{c|}{\textbf{1-N}}  \\ \cline{2-13} 
    & \multicolumn{1}{c|}{\textbf{PatchScout}} 
    & \multicolumn{1}{c|}{\textbf{FixFinder}} 
    & \multicolumn{1}{c|}{\textbf{VCMatch}} 
    & \multicolumn{1}{c|}{\textbf{LLM-SPL}}                        
    & \multicolumn{1}{c|}{\textbf{PatchScout}} 
    & \multicolumn{1}{c|}{\textbf{FixFinder}} 
    & \multicolumn{1}{c|}{\textbf{VCMatch}} 
    & \multicolumn{1}{c|}{\textbf{LLM-SPL}}                        
    & \multicolumn{1}{c|}{\textbf{PatchScout}} 
    & \multicolumn{1}{c|}{\textbf{FixFinder}} 
    & \multicolumn{1}{c|}{\textbf{VCMatch}} 
    & \multicolumn{1}{c|}{\textbf{LLM-SPL}}          \\ \hline

\textbf{1}  & \multicolumn{1}{c|}{45.69\%}  
            & \multicolumn{1}{c|}{44.28\%} 
            & \multicolumn{1}{c|}{62.14\%}  
            & \multicolumn{1}{c|}{64.86\% \magenta{$\uparrow$}}  
            
            & \multicolumn{1}{c|}{57.87\%}  
            & \multicolumn{1}{c|}{56.08\%} 
            & \multicolumn{1}{c|}{78.70\%}  
            & \multicolumn{1}{c|}{82.14\% \magenta{$\uparrow$}}  
            
            & \multicolumn{1}{c|}{0.00\%}   
            & \multicolumn{1}{c|}{0.00\% }           
            & \multicolumn{1}{c|}{0.00\%}  
            & \multicolumn{1}{c|}{0.00\% \teal{$=$}}  \\ \hline

\textbf{2}  & \multicolumn{1}{c|}{57.75\%}  
            & \multicolumn{1}{c|}{55.77\% } 
            & \multicolumn{1}{c|}{74.99\%}  
            & \multicolumn{1}{c|}{79.22\% \magenta{$\uparrow$}} 
            
            & \multicolumn{1}{c|}{69.11\%}  
            & \multicolumn{1}{c|}{67.59\% } 
            & \multicolumn{1}{c|}{87.57\%}  
            & \multicolumn{1}{c|}{89.22\% \magenta{$\uparrow$}}
            
            & \multicolumn{1}{c|}{15.14\%}  
            & \multicolumn{1}{c|}{11.41\%} 
            & \multicolumn{1}{c|}{27.79\%}  
            & \multicolumn{1}{c|}{41.69\% \magenta{$\uparrow$}}  \\ \hline

\textbf{3}  & \multicolumn{1}{c|}{61.67\%}  
            & \multicolumn{1}{c|}{61.57\% } 
            & \multicolumn{1}{c|}{78.49\%}  
            & \multicolumn{1}{c|}{84.18\% \magenta{$\uparrow$}}  
            
            & \multicolumn{1}{c|}{73.21\%}  
            & \multicolumn{1}{c|}{73.74\%} 
            & \multicolumn{1}{c|}{89.55\%}  
            & \multicolumn{1}{c|}{91.60\% \magenta{$\uparrow$}}
            
            & \multicolumn{1}{c|}{18.36\%}  
            & \multicolumn{1}{c|}{15.88\% } 
            & \multicolumn{1}{c|}{36.97\%}  
            & \multicolumn{1}{c|}{56.33\% \magenta{$\uparrow$}}  \\ \hline

\textbf{4}  & \multicolumn{1}{c|}{64.02\%}  
            & \multicolumn{1}{c|}{64.54\%} 
            & \multicolumn{1}{c|}{80.68\%}  
            & \multicolumn{1}{c|}{86.58\% \magenta{$\uparrow$}}
            
            & \multicolumn{1}{c|}{75.66\%}  
            & \multicolumn{1}{c|}{76.85\% } 
            & \multicolumn{1}{c|}{90.67\%}  
            & \multicolumn{1}{c|}{92.72\% \magenta{$\uparrow$}}
            
            & \multicolumn{1}{c|}{20.35\%} 
            & \multicolumn{1}{c|}{18.36\%} 
            & \multicolumn{1}{c|}{43.18\%}  
            & \multicolumn{1}{c|}{63.52\% \magenta{$\uparrow$}}  \\ \hline

\textbf{5}  & \multicolumn{1}{c|}{65.33\%}  
            & \multicolumn{1}{c|}{66.21\%} 
            & \multicolumn{1}{c|}{82.45\%}  
            & \multicolumn{1}{c|}{88.56\% \magenta{$\uparrow$}}
            
            & \multicolumn{1}{c|}{77.05\%}  
            & \multicolumn{1}{c|}{78.57\%} 
            & \multicolumn{1}{c|}{91.67\%} 
            & \multicolumn{1}{c|}{93.65\% \magenta{$\uparrow$}}
            
            & \multicolumn{1}{c|}{21.34\%}
            & \multicolumn{1}{c|}{19.85\%}
            & \multicolumn{1}{c|}{47.89\%}  
            & \multicolumn{1}{c|}{69.48\% \magenta{$\uparrow$}} \\ \hline

\textbf{6}  & \multicolumn{1}{c|}{66.37\%}  
            & \multicolumn{1}{c|}{67.57\%} 
            & \multicolumn{1}{c|}{83.66\%} 
            & \multicolumn{1}{c|}{89.45\% \magenta{$\uparrow$}} 
            
            & \multicolumn{1}{c|}{78.17\%}  
            & \multicolumn{1}{c|}{80.03\% }
            & \multicolumn{1}{c|}{92.39\%}  
            & \multicolumn{1}{c|}{94.18\% \magenta{$\uparrow$}} 
            
            & \multicolumn{1}{c|}{22.08\%}  
            & \multicolumn{1}{c|}{20.84\%} 
            & \multicolumn{1}{c|}{50.87\%}  
            & \multicolumn{1}{c|}{71.71\% \magenta{$\uparrow$}}  \\ \hline

\textbf{7}  & \multicolumn{1}{c|}{67.10\%}  
            & \multicolumn{1}{c|}{68.41\%}
            & \multicolumn{1}{c|}{84.75\%}  
            & \multicolumn{1}{c|}{90.34\% \magenta{$\uparrow$}}  
            
            & \multicolumn{1}{c|}{79.03\%}  
            & \multicolumn{1}{c|}{80.82\%}
            & \multicolumn{1}{c|}{93.12\%} 
            & \multicolumn{1}{c|}{94.71\% \magenta{$\uparrow$}}
            
            & \multicolumn{1}{c|}{22.33\%} 
            & \multicolumn{1}{c|}{21.84\%}
            & \multicolumn{1}{c|}{53.35\%}  
            & \multicolumn{1}{c|}{73.95\% \magenta{$\uparrow$}}  \\ \hline

\textbf{8}  & \multicolumn{1}{c|}{67.68\%}  
            & \multicolumn{1}{c|}{69.56\% }
            & \multicolumn{1}{c|}{85.43\%}  
            & \multicolumn{1}{c|}{90.97\% \magenta{$\uparrow$}}
            
            & \multicolumn{1}{c|}{79.63\%} 
            & \multicolumn{1}{c|}{81.81\%} 
            & \multicolumn{1}{c|}{93.58\%}  
            & \multicolumn{1}{c|}{94.84\% \magenta{$\uparrow$}}  
            
            & \multicolumn{1}{c|}{22.83\%} 
            & \multicolumn{1}{c|}{23.57\%} 
            & \multicolumn{1}{c|}{54.84\%} 
            & \multicolumn{1}{c|}{76.43\% \magenta{$\uparrow$}}  \\ \hline

\textbf{9}  & \multicolumn{1}{c|}{67.99\%}  
            & \multicolumn{1}{c|}{70.29\%} 
            & \multicolumn{1}{c|}{86.27\%}  
            & \multicolumn{1}{c|}{91.64\% \magenta{$\uparrow$}}
            
            & \multicolumn{1}{c|}{79.96\%}  
            & \multicolumn{1}{c|}{82.61\% } 
            & \multicolumn{1}{c|}{93.98\%}  
            & \multicolumn{1}{c|}{94.84\% \magenta{$\uparrow$}}
            
            & \multicolumn{1}{c|}{23.08\%}  
            & \multicolumn{1}{c|}{24.07\%} 
            & \multicolumn{1}{c|}{57.32\%}  
            & \multicolumn{1}{c|}{79.65\% \magenta{$\uparrow$}}  \\ \hline

\textbf{10} & \multicolumn{1}{c|}{68.56\%}  
            & \multicolumn{1}{c|}{71.17\%} 
            & \multicolumn{1}{c|}{87.10\%}  
            & \multicolumn{1}{c|}{92.74\% \magenta{$\uparrow$}}
            
            & \multicolumn{1}{c|}{80.22\%}  
            & \multicolumn{1}{c|}{83.60\% } 
            & \multicolumn{1}{c|}{94.25\%}  
            & \multicolumn{1}{c|}{95.30\% \magenta{$\uparrow$}} 
            
            & \multicolumn{1}{c|}{24.81\%}  
            & \multicolumn{1}{c|}{24.57\%} 
            & \multicolumn{1}{c|}{60.30\%}  
            & \multicolumn{1}{c|}{83.13\% \magenta{$\uparrow$}}  \\ \hline

\end{tabular}
    
    \end{adjustbox}
\vspace{-10pt}    
\end{table*}

\begin{table*}[t]
    \centering
    \captionsetup{skip=2pt}
    \caption{NDCG ($N@k$) of PatchScout, FixFinder, VCMatch, and LLM-SPL.}    
    \label{table: NDCG of models}
    \vspace{-2pt}
    \begin{adjustbox}{width=0.87\textwidth}

\begin{tabular}{|c|cccc|cccc|cccc|}
\hline
\multirow{2}{*}{\textit{k}} 
    & \multicolumn{4}{c|}{\textbf{Full}}                                                   
    & \multicolumn{4}{c|}{\textbf{1-1}}                                                       & \multicolumn{4}{c|}{\textbf{1-N}}                                                    \\ \cline{2-13} 
                            
    & \multicolumn{1}{c|}{\textbf{PatchScout}} 
    & \multicolumn{1}{c|}{\textbf{FixFinder}} 
    & \multicolumn{1}{c|}{\textbf{VCMatch}} 
    & \multicolumn{1}{c|}{\textbf{LLM-SPL}} 
    
    & \multicolumn{1}{c|}{\textbf{PatchScout}} 
    & \multicolumn{1}{c|}{\textbf{FixFinder}} 
    & \multicolumn{1}{c|}{\textbf{VCMatch}} 
    & \multicolumn{1}{c|}{\textbf{LLM-SPL}}
    
    & \multicolumn{1}{c|}{\textbf{PatchScout}} 
    & \multicolumn{1}{c|}{\textbf{FixFinder}} 
    & \multicolumn{1}{c|}{\textbf{VCMatch}} 
    & \multicolumn{1}{c|}{\textbf{LLM-SPL}} \\ \hline
    
\textbf{1}                  
    & \multicolumn{1}{c|}{51.36\%}          
    & \multicolumn{1}{c|}{50.39\%} 
    & \multicolumn{1}{c|}{73.63\%}          
    & \multicolumn{1}{c|}{80.10\% \magenta{$\uparrow$}} 
    
    & \multicolumn{1}{c|}{57.90\%} 
    & \multicolumn{1}{c|}{56.08\%} 
    & \multicolumn{1}{c|}{78.70\%}          
    & \multicolumn{1}{c|}{82.14\% \magenta{$\uparrow$}}    
    
    & \multicolumn{1}{c|}{26.80\%}           
    & \multicolumn{1}{c|}{29.03\%} 
    & \multicolumn{1}{c|}{54.59\%}           
    & \multicolumn{1}{c|}{72.46\% \magenta{$\uparrow$}}               \\ \hline

\textbf{2}                  
    & \multicolumn{1}{c|}{56.48\%}          
    & \multicolumn{1}{c|}{55.38\%} 
    & \multicolumn{1}{c|}{77.20\%}    
    & \multicolumn{1}{c|}{82.92\% \magenta{$\uparrow$}} 
    
    & \multicolumn{1}{c|}{64.98\%} 
    & \multicolumn{1}{c|}{63.35\%} 
    & \multicolumn{1}{c|}{84.30\%          }
    & \multicolumn{1}{c|}{86.61\% \magenta{$\uparrow$}}
    
    & \multicolumn{1}{c|}{24.59\%}           
    & \multicolumn{1}{c|}{25.48\%} 
    & \multicolumn{1}{c|}{50.56\%}           
    & \multicolumn{1}{c|}{69.10\% \magenta{$\uparrow$}} \\ \hline
    
\textbf{3}                  
    & \multicolumn{1}{c|}{58.38\%}          
    & \multicolumn{1}{c|}{58.33\%} 
    & \multicolumn{1}{c|}{78.58\%}
    & \multicolumn{1}{c|}{84.62\% \magenta{$\uparrow$}} 
    
    & \multicolumn{1}{c|}{67.03\%} 
    & \multicolumn{1}{c|}{66.42\%} 
    & \multicolumn{1}{c|}{85.29\%}
    & \multicolumn{1}{c|}{87.80\% \magenta{$\uparrow$}}
    
    & \multicolumn{1}{c|}{25.95\%}           
    & \multicolumn{1}{c|}{27.97\%} 
    & \multicolumn{1}{c|}{53.43\%}
    & \multicolumn{1}{c|}{72.71\% \magenta{$\uparrow$}} \\ \hline
    
\textbf{4}                  
    & \multicolumn{1}{c|}{59.39\%}          
    & \multicolumn{1}{c|}{59.65\%} 
    & \multicolumn{1}{c|}{79.35\%}
    & \multicolumn{1}{c|}{85.48\% \magenta{$\uparrow$}}
    
    & \multicolumn{1}{c|}{68.07\%}
    & \multicolumn{1}{c|}{67.76\%} 
    & \multicolumn{1}{c|}{85.77\%}          
    & \multicolumn{1}{c|}{88.28\% \magenta{$\uparrow$}}
    
    & \multicolumn{1}{c|}{26.84\%}       
    & \multicolumn{1}{c|}{29.24\%} 
    & \multicolumn{1}{c|}{55.27\%}
    & \multicolumn{1}{c|}{74.95\% \magenta{$\uparrow$}} \\ \hline

\textbf{5}                
    & \multicolumn{1}{c|}{59.95\%}   
    & \multicolumn{1}{c|}{60.32\%} 
    & \multicolumn{1}{c|}{79.94\%}
    & \multicolumn{1}{c|}{86.10\% \magenta{$\uparrow$}} 
    
    & \multicolumn{1}{c|}{68.60\%} 
    & \multicolumn{1}{c|}{68.42\%} 
    & \multicolumn{1}{c|}{86.16\% }        
    & \multicolumn{1}{c|}{88.64\% \magenta{$\uparrow$}}
    
    & \multicolumn{1}{c|}{27.47\%}           
    & \multicolumn{1}{c|}{29.92\%} 
    & \multicolumn{1}{c|}{56.62\%}       
    & \multicolumn{1}{c|}{76.58\% \magenta{$\uparrow$}} \\ \hline
    
\textbf{6}                  
    & \multicolumn{1}{c|}{60.37\%}      
    & \multicolumn{1}{c|}{60.85\%} 
    & \multicolumn{1}{c|}{80.36\%}       
    & \multicolumn{1}{c|}{86.44\% \magenta{$\uparrow$}}
    
    & \multicolumn{1}{c|}{69.03\%} 
    & \multicolumn{1}{c|}{68.94\%} 
    & \multicolumn{1}{c|}{86.41\%}      
    & \multicolumn{1}{c|}{88.83\% \magenta{$\uparrow$}}
    
    & \multicolumn{1}{c|}{27.89\%}  
    & \multicolumn{1}{c|}{30.50\%} 
    & \multicolumn{1}{c|}{57.62\%}        
    & \multicolumn{1}{c|}{77.47\% \magenta{$\uparrow$}} \\ \hline
    
\textbf{7}
    & \multicolumn{1}{c|}{60.66\%}          
    & \multicolumn{1}{c|}{61.19\%} 
    & \multicolumn{1}{c|}{80.74\%  }      
    & \multicolumn{1}{c|}{86.73\% \magenta{$\uparrow$}} 
    
    & \multicolumn{1}{c|}{69.30\%} 
    & \multicolumn{1}{c|}{69.21\%} 
    & \multicolumn{1}{c|}{86.66\%}         
    & \multicolumn{1}{c|}{89.01\% \magenta{$\uparrow$}}
    
    & \multicolumn{1}{c|}{28.23\%}       
    & \multicolumn{1}{c|}{31.11\%} 
    & \multicolumn{1}{c|}{58.53\%}
    & \multicolumn{1}{c|}{78.17\% \magenta{$\uparrow$}} \\ \hline
    
\textbf{8}                  
    & \multicolumn{1}{c|}{60.86\%}      
    & \multicolumn{1}{c|}{61.55\%} 
    & \multicolumn{1}{c|}{81.01\%}         
    & \multicolumn{1}{c|}{86.93\% \magenta{$\uparrow$}} 
    
    & \multicolumn{1}{c|}{69.49\%} 
    & \multicolumn{1}{c|}{69.52\%} 
    & \multicolumn{1}{c|}{86.80\%}         
    & \multicolumn{1}{c|}{89.05\% \magenta{$\uparrow$}}
    
    & \multicolumn{1}{c|}{28.49\%}         
    & \multicolumn{1}{c|}{31.64\%}
    & \multicolumn{1}{c|}{59.29\%}          
    & \multicolumn{1}{c|}{78.99\% \magenta{$\uparrow$}} \\ \hline

\textbf{9}                  
    & \multicolumn{1}{c|}{60.97\%}
    & \multicolumn{1}{c|}{61.82\%} 
    & \multicolumn{1}{c|}{81.31\% }       
    & \multicolumn{1}{c|}{87.09\% \magenta{$\uparrow$}} 
    
    & \multicolumn{1}{c|}{69.59\%} 
    & \multicolumn{1}{c|}{69.76\%} 
    & \multicolumn{1}{c|}{86.92\%}          
    & \multicolumn{1}{c|}{89.05\% \magenta{$\uparrow$}} 
    
    & \multicolumn{1}{c|}{28.64\%}       
    & \multicolumn{1}{c|}{32.06\%} 
    & \multicolumn{1}{c|}{60.25\%}         
    & \multicolumn{1}{c|}{79.74\% \magenta{$\uparrow$}} \\ \hline
    
\textbf{10}                

    & \multicolumn{1}{c|}{61.15\%}     
    & \multicolumn{1}{c|}{62.12\%}
    & \multicolumn{1}{c|}{81.52\%}
    & \multicolumn{1}{c|}{87.33\% \magenta{$\uparrow$}}
    
    & \multicolumn{1}{c|}{69.67\%} 
    & \multicolumn{1}{c|}{70.05\%} 
    & \multicolumn{1}{c|}{87.00\%}          
    & \multicolumn{1}{c|}{89.18\% \magenta{$\uparrow$}}   
    
    & \multicolumn{1}{c|}{29.21\%}       
    & \multicolumn{1}{c|}{32.38\%} 
    & \multicolumn{1}{c|}{60.99\%}           
    & \multicolumn{1}{c|}{80.40\% \magenta{$\uparrow$}} \\ \hline
\end{tabular}
    
    \end{adjustbox}
    \vspace{-15pt}
\end{table*}

\begin{table*}[t]
    \centering
    \captionsetup{skip=2pt}
    \caption{Manual Effort ($M@k$) of PatchScout, FixFinder, VCMatch, and LLM-SPL.}    
    \label{table: Manual Effort of models}
    \vspace{-2pt}
    \begin{adjustbox}{width=0.87\textwidth}

\begin{tabular}{|c|cccc|cccc|cccc|}
\hline
\multirow{2}{*}{\textit{k}} 
    & \multicolumn{4}{c|}{\textbf{Full}}  
    & \multicolumn{4}{c|}{\textbf{1-1}}    
    & \multicolumn{4}{c|}{\textbf{1-N}}                                                    \\ \cline{2-13} 
                            
    & \multicolumn{1}{c|}{\textbf{PatchScout}} 
    & \multicolumn{1}{c|}{\textbf{FixFinder}} 
    & \multicolumn{1}{c|}{\textbf{VCMatch}} 
    & \multicolumn{1}{c|}{\textbf{LLM-SPL}}
    
    & \multicolumn{1}{c|}{\textbf{PatchScout}} 
    & \multicolumn{1}{c|}{\textbf{FixFinder}} 
    & \multicolumn{1}{c|}{\textbf{VCMatch}} 
    & \multicolumn{1}{c|}{\textbf{LLM-SPL}}
    
    & \multicolumn{1}{c|}{\textbf{PatchScout}} 
    & \multicolumn{1}{c|}{\textbf{FixFinder}} 
    & \multicolumn{1}{c|}{\textbf{VCMatch}} 
    & \multicolumn{1}{c|}{\textbf{LLM-SPL}} \\ \hline
    
\textbf{1}                  
    & \multicolumn{1}{c|}{1.00} 
    & \multicolumn{1}{c|}{1.00} 
    & \multicolumn{1}{c|}{1.00} 
    & \multicolumn{1}{c|}{1.00 \teal{$=$}} 
    
    & \multicolumn{1}{c|}{1.00}
    & \multicolumn{1}{c|}{1.00}
    & \multicolumn{1}{c|}{1.00} 
    & \multicolumn{1}{c|}{1.00 \teal{$=$}} 
    
    & \multicolumn{1}{c|}{1.00}
    & \multicolumn{1}{c|}{1.00}
    & \multicolumn{1}{c|}{1.00} 
    & \multicolumn{1}{c|}{1.00 \teal{$=$}} \\ \hline
    
\textbf{2}                  
    & \multicolumn{1}{c|}{1.54} 
    & \multicolumn{1}{c|}{1.56} 
    & \multicolumn{1}{c|}{1.38} 
    & \multicolumn{1}{c|}{1.35 \blue{$\downarrow$}} 
    
    & \multicolumn{1}{c|}{1.42} 
    & \multicolumn{1}{c|}{1.44} 
    & \multicolumn{1}{c|}{1.21} 
    & \multicolumn{1}{c|}{1.18 \blue{$\downarrow$}} 
    
    & \multicolumn{1}{c|}{2.00} 
    & \multicolumn{1}{c|}{2.00} 
    & \multicolumn{1}{c|}{2.00} 
    & \multicolumn{1}{c|}{2.00 \teal{$=$}} \\ \hline
    
\textbf{3}                  

    & \multicolumn{1}{c|}{1.97} 
    & \multicolumn{1}{c|}{2.00} 
    & \multicolumn{1}{c|}{1.63} 
    & \multicolumn{1}{c|}{1.56 \blue{$\downarrow$}}
    
    & \multicolumn{1}{c|}{1.73} 
    & \multicolumn{1}{c|}{1.76} 
    & \multicolumn{1}{c|}{1.34} 
    & \multicolumn{1}{c|}{1.29 \blue{$\downarrow$}} 
    
    & \multicolumn{1}{c|}{2.85} 
    & \multicolumn{1}{c|}{2.89} 
    & \multicolumn{1}{c|}{2.72} 
    & \multicolumn{1}{c|}{2.58 \blue{$\downarrow$}} \\ \hline
    
\textbf{4}                  
    & \multicolumn{1}{c|}{2.35}
    & \multicolumn{1}{c|}{2.38} 
    & \multicolumn{1}{c|}{1.84} 
    & \multicolumn{1}{c|}{1.72 \blue{$\downarrow$}} 
    
    & \multicolumn{1}{c|}{2.00} 
    & \multicolumn{1}{c|}{2.03} 
    & \multicolumn{1}{c|}{1.44} 
    & \multicolumn{1}{c|}{1.37 \blue{$\downarrow$}}
    
    & \multicolumn{1}{c|}{3.67} 
    & \multicolumn{1}{c|}{3.73} 
    & \multicolumn{1}{c|}{3.35} 
    & \multicolumn{1}{c|}{3.02 \blue{$\downarrow$}} \\ \hline
    
\textbf{5}                  
    & \multicolumn{1}{c|}{2.71} 
    & \multicolumn{1}{c|}{2.74} 
    & \multicolumn{1}{c|}{2.04} 
    & \multicolumn{1}{c|}{1.85 \blue{$\downarrow$}} 
    
    & \multicolumn{1}{c|}{2.24} 
    & \multicolumn{1}{c|}{2.26} 
    & \multicolumn{1}{c|}{1.54} 
    & \multicolumn{1}{c|}{1.44 \blue{$\downarrow$}} 
    
    & \multicolumn{1}{c|}{4.46} 
    & \multicolumn{1}{c|}{4.54}
    & \multicolumn{1}{c|}{3.92} 
    & \multicolumn{1}{c|}{3.38 \blue{$\downarrow$}} \\ \hline
    
\textbf{6}                  
    & \multicolumn{1}{c|}{3.06} 
    & \multicolumn{1}{c|}{3.08}
    & \multicolumn{1}{c|}{2.21} 
    & \multicolumn{1}{c|}{1.97 \blue{$\downarrow$}} 
    
    & \multicolumn{1}{c|}{2.47} 
    & \multicolumn{1}{c|}{2.47}
    & \multicolumn{1}{c|}{1.62} 
    & \multicolumn{1}{c|}{1.51 \blue{$\downarrow$}}
    
    & \multicolumn{1}{c|}{5.25} 
    & \multicolumn{1}{c|}{5.34}
    & \multicolumn{1}{c|}{4.44} 
    & \multicolumn{1}{c|}{3.69 \blue{$\downarrow$}} \\ \hline
    
\textbf{7}                  
    & \multicolumn{1}{c|}{3.39} 
    & \multicolumn{1}{c|}{3.40} 
    & \multicolumn{1}{c|}{2.38} 
    & \multicolumn{1}{c|}{2.07 \blue{$\downarrow$}} 
    
    & \multicolumn{1}{c|}{2.69} 
    & \multicolumn{1}{c|}{2.67} 
    & \multicolumn{1}{c|}{1.69} 
    & \multicolumn{1}{c|}{1.56 \blue{$\downarrow$}}
    
    & \multicolumn{1}{c|}{6.03}
    & \multicolumn{1}{c|}{6.14}
    & \multicolumn{1}{c|}{4.93} 
    & \multicolumn{1}{c|}{3.97 \blue{$\downarrow$}} \\ \hline
    
\textbf{8}                  
    & \multicolumn{1}{c|}{3.72} 
    & \multicolumn{1}{c|}{3.72} 
    & \multicolumn{1}{c|}{2.53} 
    & \multicolumn{1}{c|}{2.17 \blue{$\downarrow$}} 
    
    & \multicolumn{1}{c|}{2.90}
    & \multicolumn{1}{c|}{2.86} 
    & \multicolumn{1}{c|}{1.76} 
    & \multicolumn{1}{c|}{1.62 \blue{$\downarrow$}} 
    
    & \multicolumn{1}{c|}{6.80} 
    & \multicolumn{1}{c|}{6.92}
    & \multicolumn{1}{c|}{5.40} 
    & \multicolumn{1}{c|}{4.23 \blue{$\downarrow$}} \\ \hline
    
\textbf{9}                  
    & \multicolumn{1}{c|}{4.04} 
    & \multicolumn{1}{c|}{4.02} 
    & \multicolumn{1}{c|}{2.67} 
    & \multicolumn{1}{c|}{2.26 \blue{$\downarrow$}} 
    
    & \multicolumn{1}{c|}{3.10} 
    & \multicolumn{1}{c|}{3.04} 
    & \multicolumn{1}{c|}{1.83} 
    & \multicolumn{1}{c|}{1.67 \blue{$\downarrow$}} 
    
    & \multicolumn{1}{c|}{7.58} 
    & \multicolumn{1}{c|}{7.68} 
    & \multicolumn{1}{c|}{5.85} 
    & \multicolumn{1}{c|}{4.47 \blue{$\downarrow$}} \\ \hline
    
\textbf{10}                 
    & \multicolumn{1}{c|}{4.36}
    & \multicolumn{1}{c|}{4.32} 
    & \multicolumn{1}{c|}{2.81} 
    & \multicolumn{1}{c|}{2.34 \blue{$\downarrow$}}
    
    & \multicolumn{1}{c|}{3.30} 
    & \multicolumn{1}{c|}{3.22} 
    & \multicolumn{1}{c|}{1.89} 
    & \multicolumn{1}{c|}{1.72 \blue{$\downarrow$}} 
    
    & \multicolumn{1}{c|}{8.34} 
    & \multicolumn{1}{c|}{8.44} 
    & \multicolumn{1}{c|}{6.28} 
    & \multicolumn{1}{c|}{4.67 \blue{$\downarrow$}} \\ \hline
\end{tabular}                                               
    
    \end{adjustbox}
    \vspace{-15pt}
    
\end{table*}

\vspace{-1pt}\noindent\textbf{Overall results.}
Tables~\ref{table: recall of models}, ~\ref{table: NDCG of models}, and~\ref{table: Manual Effort of models} showcase the performance results of LLM-SPL across three datasets: 1-1, 1-N, and Full, as detailed in Section~\ref{subsec: experimental setup}. 
For the 1-1 dataset, LLM-SPL ranked patches in the first position for 82.14\% of CVEs and within top 10 for 95.30\%. High ranking quality ($N@10$ = 89.18\%) allow locating all patches for 95.30\% by checking only 1.72 commits on average ($M@10$ = 1.72)
In the 1-N dataset, LLM-SPL successfully identified and ranked all the collaborated patches within top 10 for 83.13\% of 1-N CVEs. Moreover, the high quality of these rankings, as evidenced by the NDCG value of $N@10$=80.40\%, enables the examination of only 4.67 commits ($M@10$=4.67) on average.
In the Full dataset, LLM-SPL achieved high recall values (over 90\% when $k>=7$) and consistently high NDCG values (above 80\% for every \textit{k} from 1 to 10).
These results clearly demonstrate LLM-SPL's effectiveness in locating patches for vulnerabilities with minimal manual effort.


To better understand the effectiveness of LLM-SPL, we compared its performance against prior SPL works, including VCMatch~\cite{wang2022vcmatch}, PatchScout~\cite{tan2021locating}, and FixFinder~\cite{hommersom2021automated}. The comparative results, as illustrated in Tables~\ref{table: recall of models},~\ref{table: NDCG of models}, ~\ref{table: Manual Effort of models}, and Figure~\ref{figure: Increase in Recall, NDCG and Manual Effort for LLM-SPL over VCMatch.} in Appendix, indicate that LLM-SPL significantly outperforms all prior works, across all datasets (Full, 1-1, and 1-N) for all \textit{k} values (from 1 to 10) and across all metrics (Recall, NDCG, and Manual Effort).
Notably, our experimental results indicate that VCMatch is the state-of-the-art (SOTA) SPL approach. In the following, we detail the specific improvements our method offers over VCMatch.

For Recall, while VCMatch achieves high recall on the 1-1 dataset (e.g., $R@4$ = 90.67\%, $R@10$ = 94.25\%), LLM-SPL increases average recall by 1.77\% across all \textit{k} values (1 to 10). Notably, it improves $R@1$ by 3.44\%, demonstrating the enhanced capability to rank patches in the topmost position compared to VCMatch. 
In 1-N scenarios, improvements are substantial: $R@5$ increases from 47.89\% to 69.48\% and $R@10$ from 60.30\% to 83.13\%, gains of 21.59\% and 22.83\%, respectively. In our 403-CVE 1-N dataset, this allows the complete remediation of 86 additional CVEs in the top 5 and 92 in the top 10 rankings.
Regarding NDCG (~\ref{table: NDCG of models} and Figure~\ref{figure: Increase in Recall, NDCG and Manual Effort for LLM-SPL over VCMatch.}(b)), LLM-SPL shows consistent improvements, with average increases of 6.01\% (Full dataset), 2.46\% (1-1) and 19.34\% (1-N). This improved ranking quality significantly reduces manual effort(~\ref{table: Manual Effort of models} and Figure~\ref{figure: Increase in Recall, NDCG and Manual Effort for LLM-SPL over VCMatch.}(c)). Compared to VCMatch, Users experience an average 10\% effort reduction in the Full dataset, with the 1-N dataset showing about 20\% reduction at $M@7$ and over 25\% at $M@10$.
Overall, LLM-SPL not only identifies all necessary patches for CVEs but also ranks them higher, substantially reducing practical manual effort compared to VCMatch.


\vspace{0pt}\noindent\textbf{Declined rankings in LLM-SPL against VCMatch.}
In our analysis, we observed that among 2,461 security patches in our dataset, 97 (3.94\%) were ranked lower by our LLM-SPL compared to VCMatch. 
Upon a detailed manual analysis of these 97 commits, we discovered that the primary reason for their lower rankings was due to false determinations made by the LLM (GPT-3.5 used in our research), which included both false positives (FPs) and false negatives (FNs) in generating features $F_1$ and $F_2$.
Specifically, with the feature $F_1$, when the LLM failed to recognize an actual CVE patch (FN) and simultaneously misidentified a non-patch commit as the patch for this CVE (FP), our model tended to assign a higher ranking to the FP commit, which consequently resulted in the genuine patch being ranked lower.
For the feature $F_2$, particularly in 1-N scenario, if the LLM did not detect a collaborated relationship between actual patches (FN) and incorrectly linked a non-patch with a high-ranked patch (FP), our model likely elevated the ranking of the FP non-patch, thereby causing the actual patches to be ranked lower.
Among the 97 commits analyzed, 41 were affected by issues related to $F_1$, 30 by $F_2$, and the remaining 26 were influenced by a combination of errors in both features.

\vspace{-8pt}
As discussed in Section~\ref{sec: A Potential Solver: Large Language Model}, the LLM indeed exhibits a high false positive rate. However, it also maintains a high recall rate, indicating that false negatives are comparatively infrequent. Thus, we further analyzed the causes of false negatives in features $F_1$ and $F_2$, ultimately identifying two primary reasons for these occurrences.
Firstly, insufficient information within CVEs and commits significantly hampers the LLM's performance.
For example, CVE-2020-7772 is briefly described as \textit{``This affects the package doc-path before 2.1.2.''}, severely lacking details about the vulnerability (refer to Table~\ref{table: information categories within CVE content} for typical details included in CVE descriptions).
Similarly, the security patch for CVE-2020-12607, commit 7b64e3e, contains merely the message \textit{``Update docs to v2.1.2''} along with a one-line code change to update release information, offering minimal insight into its relevance to the CVE or its connections with other patches.
Such sparse information results in LLM's failures to identify a commit that actually patches a CVE (resulting in a FN for $F_1$) and to recognize jointly working patches (leading to a FN for $F_2$).
Another factor contributing to false negatives is the extensive length of some commit diffs. For instance, the security patch commit 2430929 for CVE-2021-23365 contains 50,434 tokens, far exceeding the maximum token limit of a single GPT-3.5 prompt, which is 4,097. In such cases, we truncated the extensive code diffs during prompt construction. Unfortunately, this truncation often resulted in the loss of critical information, consequently impeding the LLM's ability to recognize the relationships necessary for $F_1$ and $F_2$, leading to false negatives.

However, such a decrease was minimal in our results. Among these 97 commits, the average drop was 4.77 positions, with 36.08\% declining by only one position, 57.73\% falling within three positions, and 75.26\% dropping within five positions.

\vspace{-15pt}
\subsection{Ablation Study}
\vspace{-7pt}

As introduced in Section~\ref{sec: design and implemenation}, our LLM-SPL incorporates two key features, $F_1$ and $F_2$, endorsed by the LLM to enhance the SPL recommendations.
The substantial enhancements resulting from the entire design have been demonstrated in Section~\ref{subsec: effectiveness}.
This section investigates the individual contributions of $F_1$ and $F_2$ to the model's performance, respectively.
To this end, we conducted ablation studies using models with different feature combinations:
1) $F_0$ alone, which corresponds to VCMatch,
2) $F_0$ and $F_1$ only, excluding $F_2$, and
3) all features ($F_0$, $F_1$, and $F_2$), which constitute our LLM-SPL.
The results of these experiments are illustrated in Tables~\ref{table: result of ablation study on recall},~\ref{table: result of ablation study on NDCG}, and~\ref{table: result of ablation study on manual effort} in Appendix.
%

To evaluate the impact of feature $F_1$, we compared models using only $F_0$ against both $F_0$ and $F_1$. Results (as detailed in Tables~\ref{table: result of ablation study on recall},~\ref{table: result of ablation study on NDCG}, and~\ref{table: result of ablation study on manual effort} and illustrated in Figures~\ref{figure: ablation study - recall.},~\ref{figure: ablation study - NDCG.}), show consistent improvement across all metrics ($R@k$, $N@k$, and $M@k$ for $k=1$ to 10) across all dataset(Full, 1-1 and 1-N). At $k=5$, including $F_1$ improved Recall ($R@5$) by 5.01\%, 1.59\%, and 17.87\% for Full, 1-1, and 1-N datasets respectively. NDCG ($N@5$) increased by 5.32\%, 2.05\%, and 17.57\%, while Manual Effort ($M@5$) reduced by 8.05\%, 5.47\%, and 11.84\%. These improvements account for approximately 85\% of LLM-SPL's overall enhancement (Figures 15-17), confirming $F_1$'s effectiveness in leveraging LLM capabilities for CVE patch identification and prioritization. 

To assess the contribution of $F_2$, we compared the performance of LLM-SPL, which incorporates all features ($F_0$, $F_1$, and $F_2$), against a model that uses only $F_0$ and $F_1$.
Given that $F_2$ is specifically designed for the 1-N scenario, our analysis focused on the 1-N dataset.
As illustrated in Tables~\ref{table: result of ablation study on recall},~\ref{table: result of ablation study on NDCG}, and~\ref{table: result of ablation study on manual effort} and Figures~\ref{figure: ablation study - recall.}(c),~\ref{figure: ablation study - NDCG.}(c), and~\ref{figure: ablation study - manual effort.}(c) in Appendix, $F_2$ provided comprehensive improvements across all metrics (Recall, NDCG and Manual Effort) in the 1-N dataset.
For instance, at $k=10$, $F_2$ increased Recall($R@10$) from 76.18\% to 83.13\%, a boost of 6.95\%, and enhanced NDCG($N@10$) by 3.10\%, which in turn led to a reduction in manual effort by 4.23\% ($M@10$).
This means that in our 1-N dataset of 403 CVEs, on top of the reduction from manually checking 540 commits achieved by $F_1$, $F_2$ further reduced the need to check an additional 100+ commits.
Overall, F2 contributed approximately 15\% to the improvement in each performance metric for LLM-SPL, as shown in Figures Figures~\ref{figure: ablation study - recall.}(c),~\ref{figure: ablation study - NDCG.}(c), and~\ref{figure: ablation study - manual effort.}(c).
These results demonstrate the efficacy of $F_2$, indicating its ability to effectively address 1-N scenarios.
Furthermore, in the 1-1 dataset, $F_2$ had minimal impact, which aligns with our expectations.

\vspace{-10pt}
\section{Related Work}
\vspace{-12pt}

\vspace{2pt}\noindent\textbf{Security patch localization.}
Initial approaches~\cite{kim2017vuddy,li2017large,perl2015vccfinder,wang2019detecting,wang2021patchdb,xu2022tracking} relied heavily on direct references between CVE descriptions and patches. However, the effectiveness of these methods was limited by the availability of explicit information. 
To address this limitation, subsequent studies like FixFinder~\cite{hommersom2021automated}, PatchScout~\cite{tan2021locating} modeled SPL as a recommendation problem. They extracted rule-based features to describe CVEs-commit relationships, applying predefined feature weights or various machine learning models to rank the commits most likely associated with a given CVE.
Building on these efforts, VCMatch~\cite{wang2022vcmatch}, the current SOTA, 
extended these features by introducing embeddings to represent CVEs and commits, enhancing the semantic understanding of vulnerability and patches. Our work, LLM-SPL, not only adopts these features but also leverages LLMs to comprehend CVEs, commits, and specialized software security knowledge, significantly improving SPL recommendations. Moreover, unlike previous works focused on 1-1 scenarios (single patch per vulnerability), our approach effectively addresses complex 1-N scenarios where multiple patches collaboratively resolve a vulnerability.

\vspace{0pt}\noindent\textbf{Recommendation algorithm.}
Our approach incorporates LLM feedback into recommendation algorithms, combining the advantages of Relevance Feedback (RF) and Pseudo-Relevance Feedback (PRF). It mitigates RF's high user involvement cost ~\cite{Radlinski2005query, rajaram2003classification} and PRF's potential biases from relying solely on top $k$ results ~\cite{Rocchio1971, li2018nprf}. Additionally, our recommendation algorithm leverages information from graphs. Graph-based recommendations have proven effective, especially for candidates with inherent relationships: pharmaceutical recommendations use medical knowledge graphs ~\cite{bhoi2021personalizing}, travel recommendations employ map graphs ~\cite{turno2023graph}, social media leverages social networks ~\cite{li2017survey, jamali2010matrix, walter2008model} and citation systems utilize scholarly graphs ~\cite{liu2014meta, ma2020review, jeong2020context}. Unlike these systems using pre-existing graphs built on accumulated knowledge, our SPL research requires constructing a domain-specific security graph from scratch. We utilize LLMs to assist in building this graph, demonstrating the potential of LLM-based graph generation for enhancing SPL.

\vspace{-15pt}
\section{Conclusion}
\vspace{-10pt}

In this study, we proposed LLM-SPL, a novel SPL (Security Patch Localization) recommendation model leveraging Large Language Models (LLMs) to address the challenges inherent in SPL. Our approach extends beyond mere utilization of LLM outputs; through a joint learning framework, we incorporate the determinations of LLMs on CVE-commit and commit-commit relationships as additional features/feedback to refine the recommendation model to improve its performance. Our evaluation demonstrates the effectiveness of LLM-SPL. Compared to the state-of-the-art work, VCMatch, our LLM-SPL consistently outperformed it in terms of Recall while also achieving a notable reduction in manual effort. Notably, for scenarios where one vulnerability requires multiple jointly working patches, LLM-SPL demonstrates significant improvements in all metrics. These results underscore the potential of LLM-SPL as a valuable SPL approach in real-world applications.

\ignore{


In this study, we propose an innovative and advanced SPL solution, LLM-SPL, which leverages the advantageous yet skeptical security intelligence of LLMs. Our solution goes beyond using the LLM output as is, instead encapsulating the LLM end products in a SPL supervised learning architecture that shines a light on the complex correlations between commits and their compatible CVEs.
%
The evaluation results demonstrate the superiority of LLM-SPL. 
When compared against the state-of-the-art SPL method, VCMatch, LLM-SPL consistently exhibits enhanced performance in both recall and NDCG metrics, with particularly notable improvements for the 1-N scenario.
}

\bibliographystyle{plain}
\bibliography{ref}

\appendix
\vspace{-10pt}

\begin{figure}[H]
    \centering
    \includegraphics[width=0.45\textwidth]{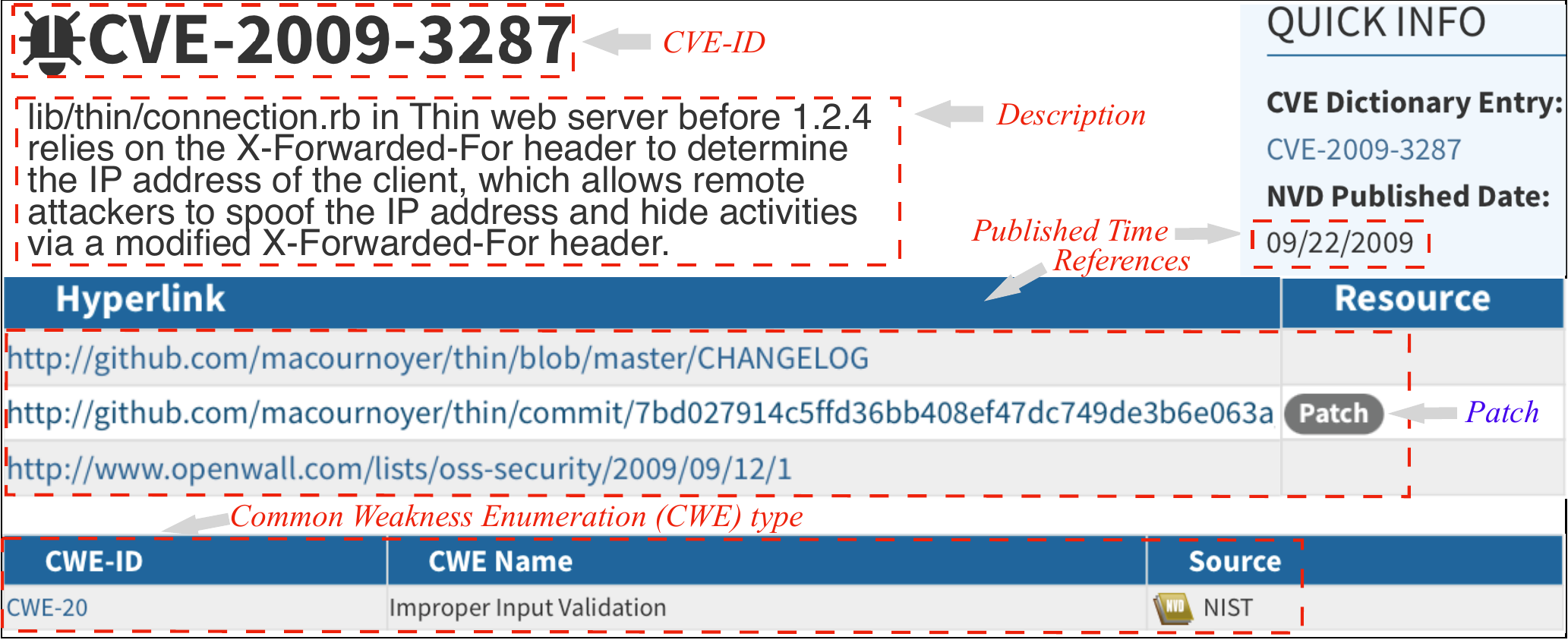}  
    \vspace{-3pt}
    \caption{Example of CVE information.}
    \label{figure: CVE INFO}
\end{figure}

\vspace{-5pt}

\begin{figure}[H]
    \centering
    \includegraphics[width=0.45\textwidth]{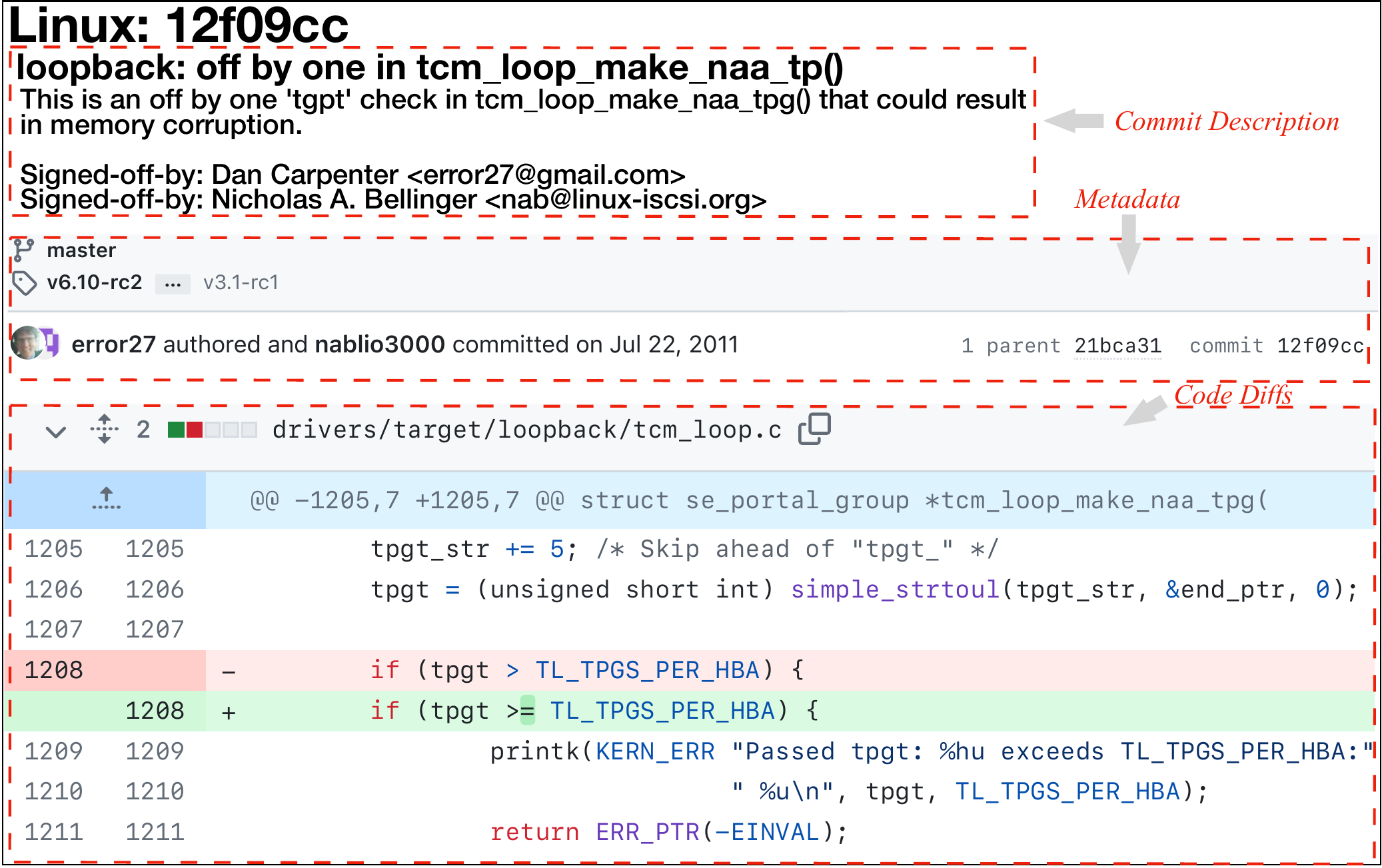}
    \vspace{-3pt}
    \caption{Example of commit information.}
    \label{figure: Commit INFO}
\end{figure}

\vspace{-5pt}

\begin{figure}[H]
    \centering
    \includegraphics[width=0.45\textwidth]{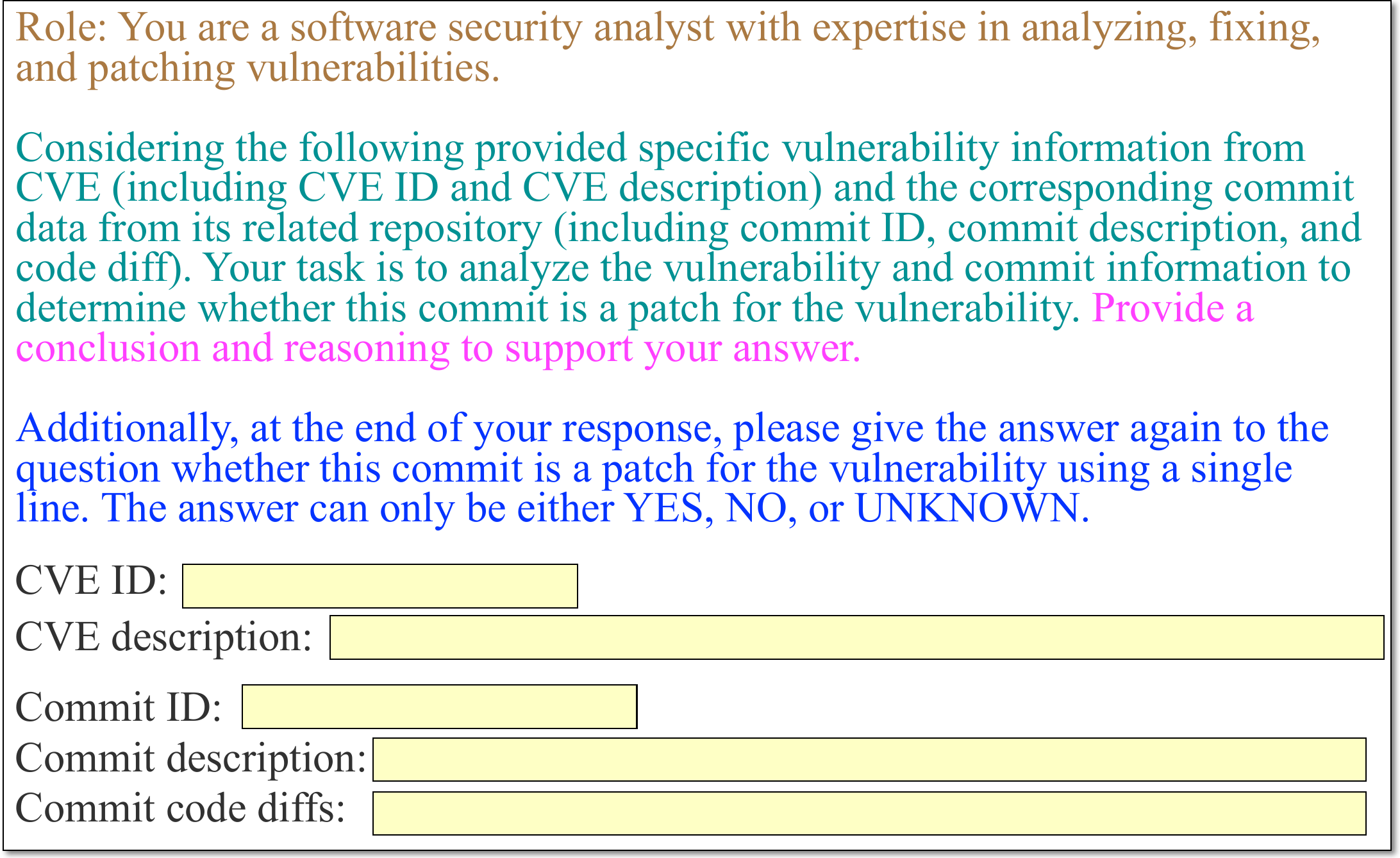}    
    \vspace{-3pt}
    \caption{Template of prompt for $F_1$.}
    \label{figure: template of prompt for F1}
\end{figure}

\vspace{-5pt}

\begin{figure}[H]
    \centering
    \includegraphics[width=0.45\textwidth]{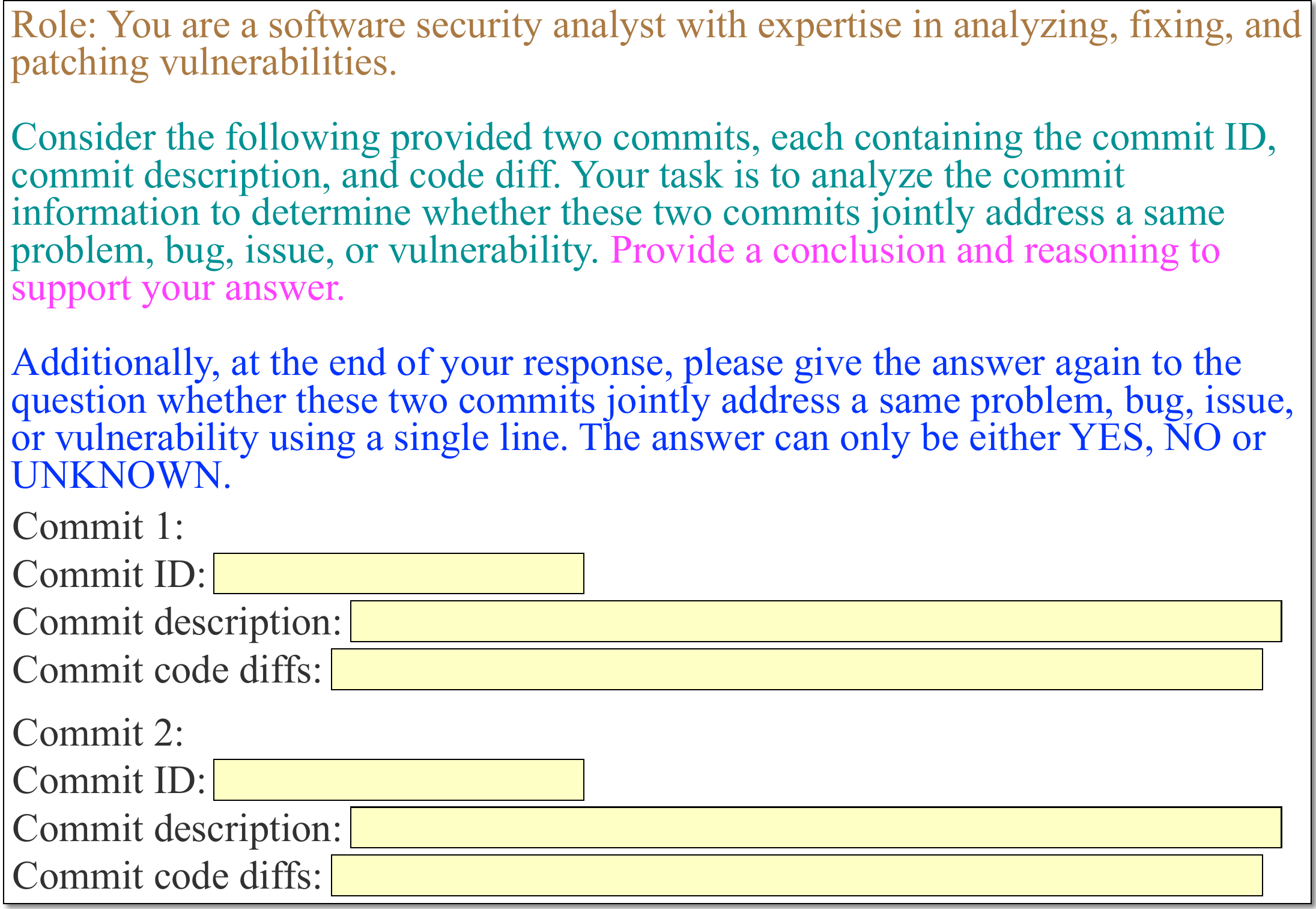}
    \vspace{-3pt}
    \caption{Template of prompt for $F_2$.}
    \label{figure: template of prompt for F2}
\end{figure}

\newpage

\begin{figure*}[htbp]
    \centering
     \includegraphics[width=0.99\textwidth]{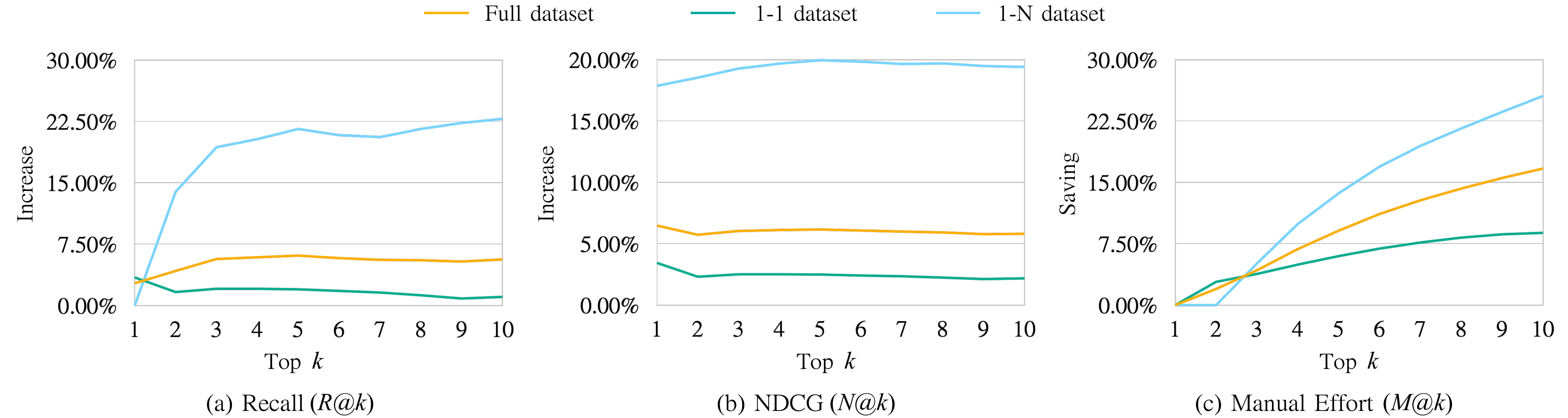}   
     \vspace{-3pt}
     \caption{Increase in Recall, NDCG and Manual Effort for LLM-SPL over VCMatch.}
     \label{figure: Increase in Recall, NDCG and Manual Effort for LLM-SPL over VCMatch.}
     \vspace{-20pt}
\end{figure*}

\begin{figure*}[htbp]
    \centering
    \includegraphics[width=0.99\textwidth]{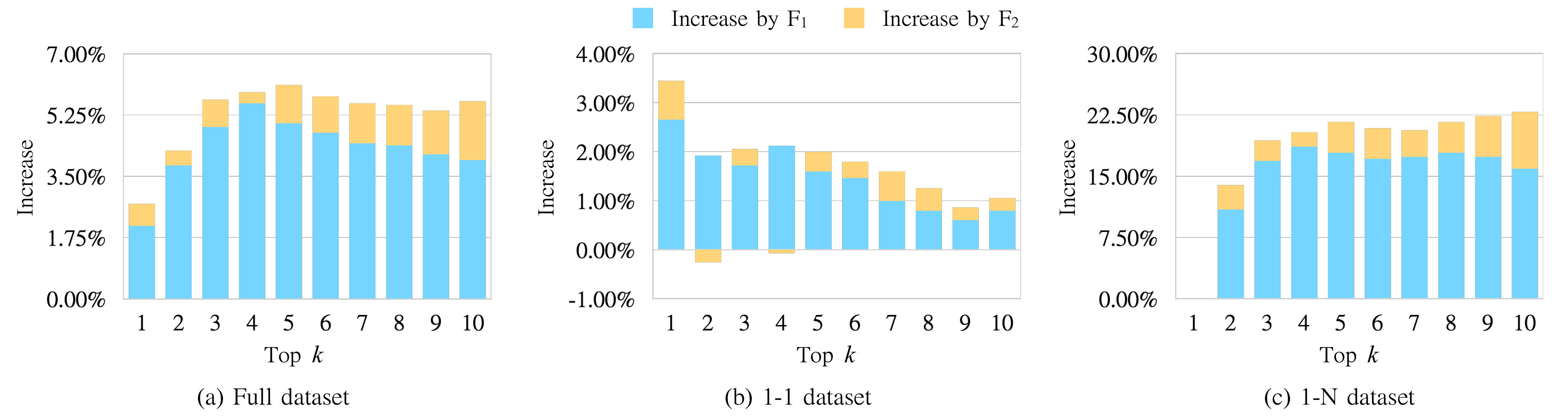}
    \vspace{-3pt}
     \caption{Ablation Study: Impact on $F_1$ and $F_2$ in Recall}
     \label{figure: ablation study - recall.}
     \vspace{-20pt}
\end{figure*}

\begin{figure*}[htbp]
    \centering
    \includegraphics[width=0.99\textwidth]{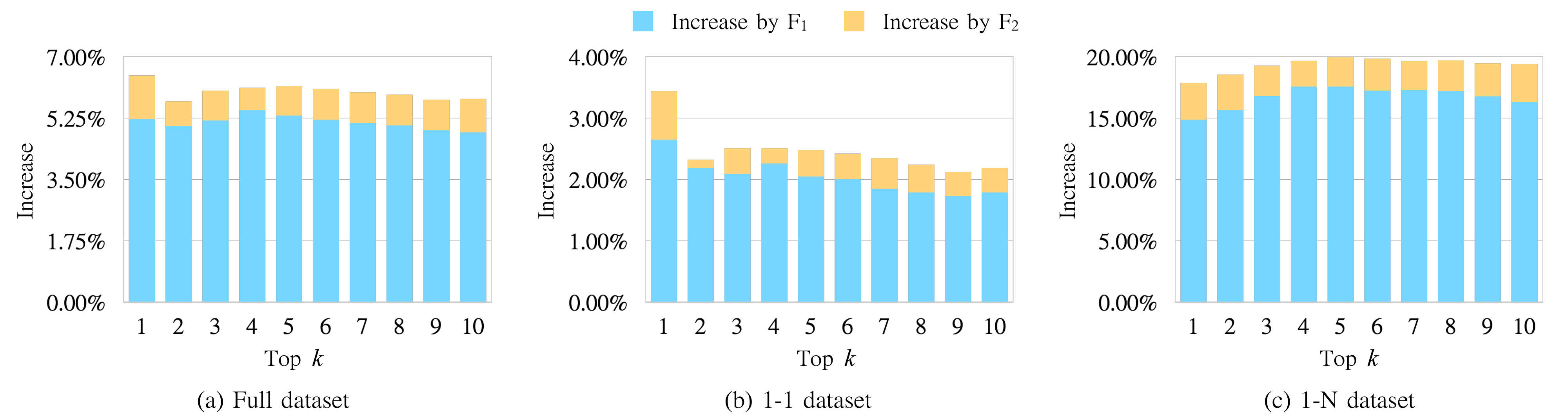}
    \vspace{-3pt}
     \caption{Ablation Study: Impact on $F_1$ and $F_2$ in NDCG.}
     \label{figure: ablation study - NDCG.}
     \vspace{-20pt}
\end{figure*}

\begin{figure*}[htbp]
    \centering
    \includegraphics[width=0.99\textwidth]{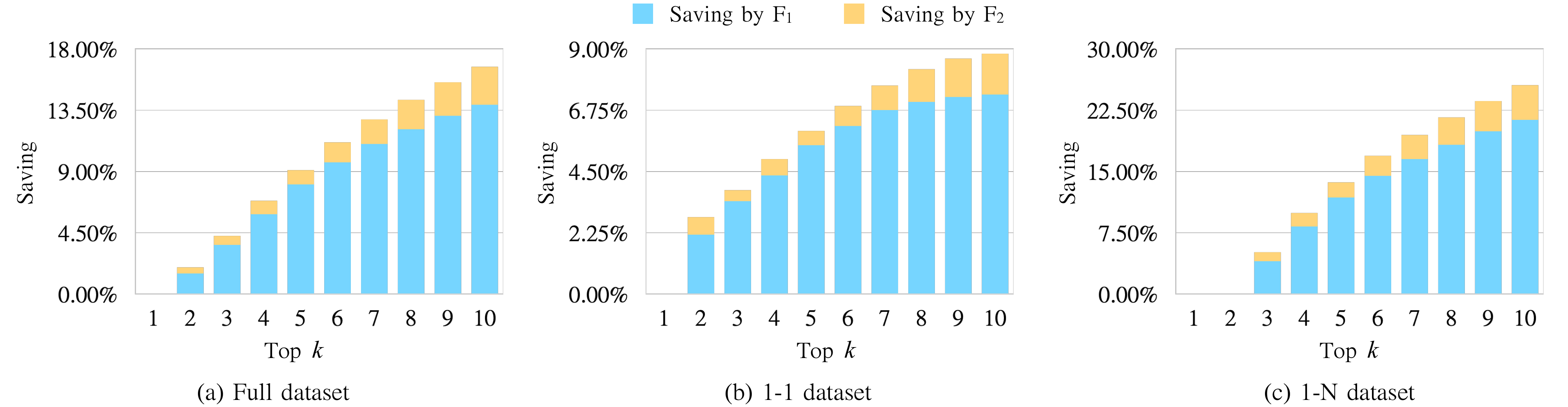}
    \vspace{-5pt}
     \caption{Ablation Study: Impact on $F_1$ and $F_2$ in Manual Effort.}
     \label{figure: ablation study - manual effort.}
     \vspace{-20pt}
\end{figure*}

\newpage

\begin{table*}[htbp]
\footnotesize
\renewcommand{\arraystretch}{1.05}
\caption{VCMatch Feature List ($F_0$)}
\label{table: VCMatch Feature List}
\vspace{-3pt}
\centering
\textit{Note: Type S denotes sparse feature, D denotes dense feature.}\\
\begin{adjustbox}{width=1\textwidth}
\begin{tabular}{|c|l|l|c|}
\hline
\textbf{Feature Category}                                                                & \multicolumn{1}{c|}{\textbf{Feature}}                                                                 & \multicolumn{1}{c|}{\textbf{Description}}                                              & \textbf{Type} \\ \hline
\multirow{12}{*}{\begin{tabular}[c]{@{}c@{}}Code Behavior\\ Features\end{tabular}}       & Code Added Num                                                                                        & \# of lines of code added in the commit                                                & S             \\ \cline{2-4} 
                                                                                         & Code Deleted Num                                                                                      & \# of lines of code deleted in the commit                                              & S             \\ \cline{2-4} 
                                                                                         & Code Modified Num                                                                                     & \# of lines of code modify in the commit                                               & S             \\ \cline{2-4} 
                                                                                         & Same Filepath Num                                                                                     & \# of filepaths that exists in both commit and vul desciption                          & S             \\ \cline{2-4} 
                                                                                         & Same Filepath Ratio                                                                                   & \# of same filepath num / \# of filepath modified by commit,                           & S             \\ \cline{2-4} 
                                                                                         & Unrelated Filepath Num                                                                                & \# of filepaths existed in the commit but not mentioned in vuln description            & S             \\ \cline{2-4} 
                                                                                         & Same File Num                                                                                         & \# of files that exist in both commit and vuln description                             & S             \\ \cline{2-4} 
                                                                                         & Same File Ratio                                                                                       & \# of same files / \# of files modified by the commit                                  & S             \\ \cline{2-4} 
                                                                                         & Unrelated File Num                                                                                    & \# of files that appear in commit but not mentioned by the code diff                   & S             \\ \cline{2-4} 
                                                                                         & Same Function Num                                                                                     & \# of functions that exist both in the commit diff and vul description.                & S             \\ \cline{2-4} 
                                                                                         & Same Function Ratio                                                                                   & \# of same function / \# of functions modified by the code.                            & S             \\ \cline{2-4} 
                                                                                         & Unrelated Function Num                                                                                & \# of functions that exist in commit diff but not mentioned in the vul description.    & S             \\ \hline
\multirow{6}{*}{\begin{tabular}[c]{@{}c@{}}Commit Message\\ Identifiers\end{tabular}}    & CVE Num                                                                                               & \# of CVE IDs in commit message.                                                       & S             \\ \cline{2-4} 
                                                                                         & Bug Num                                                                                               & \# of bug IDs in commit message                                                        & S             \\ \cline{2-4} 
                                                                                         & Issue Num                                                                                             & \# of issue IDs in commit message                                                      & S             \\ \cline{2-4} 
                                                                                         & URL Num                                                                                               & \# of URLs in commit message.                                                          & S             \\ \cline{2-4} 
                                                                                         & CVE Match                                                                                             & Whether the commit mentions CVE-ID in the NVD Page.                                    & S             \\ \cline{2-4} 
                                                                                         & Bug Match                                                                                             & Whether the code commit mentions the Bug-ID in the NVD Page.                           & S             \\ \hline
\multirow{11}{*}{\begin{tabular}[c]{@{}c@{}}Textual Similarity \\ Features\end{tabular}} & Vul-CWE-Msg Same Num                                                                                  & \# of the same tokens between commit message and CWE name.                             & D             \\ \cline{2-4} 
                                                                                         & Vul-CWE-Msg Same Ratio                                                                                & Vul-CWE-Msg Same Num/\#ofCWEname tokens.                                               & D             \\ \cline{2-4} 
                                                                                         & Vul-Commit TfIdf Similarity                                                                           & Cosine similarity of vulnerability tfidf and commit tfidf.                             & D             \\ \cline{2-4} 
                                                                                         & Shared-Vul-Msg-Word Num                                                                               & \# of shared words between vul description and commit message.                         & D             \\ \cline{2-4} 
                                                                                         & Shared-Vul-Msg-Word Ratio                                                                             & \# of Shared-Vul-Msg-Words / \# of words in vul description.                           & D             \\ \cline{2-4} 
                                                                                         & Shared-Vul-Code-Word Num                                                                              & \# of shared words between vul description and code diff.                              & D             \\ \cline{2-4} 
                                                                                         & Shared-Vul-Code-Word Ratio                                                                            & \# of Shared-Vul-Code-Words / \# of words in vul description.                          & D             \\ \cline{2-4} 
                                                                                         & \begin{tabular}[c]{@{}l@{}}Max/Sum/Average/Variance of \\ Shared-Vul-Msg-Word Frequency\end{tabular}  & The max/sum/average/variance of the frequencies for all Shared-Vul-Msg-Words.          & D             \\ \cline{2-4} 
                                                                                         & \begin{tabular}[c]{@{}l@{}}Max/Sum/Average/Variance of \\ Shared-Vul-Code-Word Frequency\end{tabular} & The max/sum/average/variance of the frequencies for all Shared-Vul-Code-Words.         & D             \\ \cline{2-4} 
                                                                                         & \begin{tabular}[c]{@{}l@{}}Deep Textual CVE \\ Description Representation\end{tabular}                & A vector representation of CVE description encoded using a pre-trained language model. & D             \\ \cline{2-4} 
                                                                                         & \begin{tabular}[c]{@{}l@{}}Deep Textual Commit\\ Content Representation\end{tabular}                  & A vector representation of commit content encoded using a pre-trained language model.  & D             \\ \hline
\begin{tabular}[c]{@{}c@{}}Security Relevance \\ Features\end{tabular}                   & Vulnerability Type Relevance                                                                          & The relevance of the vulnerability texts between NVD and commit                        & D             \\ \hline
Temporal Features                                                                        & Time Interval                                                                                         & Time interval between code commit time and CVE-ID assigned time                        & D             \\ \hline
\end{tabular}
 \end{adjustbox}
\end{table*} 
\begin{table*}[htbp]
    \centering
    \renewcommand{\arraystretch}{1.05}
    \vspace{20pt}
    \caption{Information categories within CVE content.}
    \label{table: information categories within CVE content}  
    
    \footnotesize
    \begin{adjustbox}{width=1\textwidth}

\begin{tabular}{|cc|l|}
\hline
\multicolumn{2}{|c|}{\textbf{Category}}                                                                                                                                                         & \textbf{Examples}                                                                                                                                                       \\ \hline
\multicolumn{1}{|c|}{\multirow{4}{*}{\begin{tabular}[c]{@{}c@{}}Software\\ information\end{tabular}}}       & \multirow{2}{*}{\begin{tabular}[c]{@{}c@{}}Software\\ name\end{tabular}}      & \textit{CVE-2022-39252: ... is an implementation of a \blue{Matrix} client-server library in Rust, ...}                                                                        \\ \cline{3-3} 
\multicolumn{1}{|c|}{}                                                                                      &                                                                               & \textit{CVE-2022-21208: The package \blue{node-opcua} before 2.74.0 are vulnerable to Denial of Service ...}                                                                   \\ \cline{2-3} 
\multicolumn{1}{|c|}{}                                                                                      & \multirow{2}{*}{\begin{tabular}[c]{@{}c@{}}Software\\ version\end{tabular}}   & \textit{CVE-2014-5273: ... vulnerabilities in phpMyAdmin \blue{4.0.x before 4.0.10.2, 4.1.x before 4.4.14.3, and 4.2.x before 4.2.7.1} allow ...}                              \\ \cline{3-3} 
\multicolumn{1}{|c|}{}                                                                                      &                                                                               & \textit{CVE-2018-18227: In Wireshark \blue{2.6.0 to 2.6.3 and 2.4.0 to 2.4.9}, ...}                                                                                            \\ \hline
\multicolumn{1}{|c|}{\multirow{10}{*}{\begin{tabular}[c]{@{}c@{}}Vulnerability\\ information\end{tabular}}} & \multirow{2}{*}{\begin{tabular}[c]{@{}c@{}}Vulnerability\\ type\end{tabular}} & \textit{CVE-2012-2386: \blue{Integer overflow} in the phar\_parse\_tarfile function in ...}                                                                                  \\ \cline{3-3} 
\multicolumn{1}{|c|}{}                                                                                      &                                                                               & \textit{CVE-2010-3429: flicvideo.c in ..., related to an "\blue{arbitrary offset dereference} vulnerability"}                                                                  \\ \cline{2-3} 
\multicolumn{1}{|c|}{}                                                                                      & \multirow{2}{*}{\begin{tabular}[c]{@{}c@{}}Faulty\\ functionality\end{tabular}}    & \textit{CVE-2022-23567: ... We are \blue{missing some validation on the shapes of the input tensors} as well as ...}                                                           \\ \cline{3-3} 
\multicolumn{1}{|c|}{}                                                                                      &                                                                               & \textit{CVE-2018-10853: ... \blue{It did not check current privilege (CPL) level while emulating unprivileged instructions} ...}                                               \\ \cline{2-3} 
\multicolumn{1}{|c|}{}                                                                                      & \multirow{2}{*}{\begin{tabular}[c]{@{}c@{}}Erroneous\\ file\end{tabular}}     & \textit{CVE-2020-11608: ... in the Linux kernel before 5.6.1. \blue{drivers/media/usb/gspca/ov519.c} allows ...}                                                               \\ \cline{3-3} 
\multicolumn{1}{|c|}{}                                                                                      &                                                                               & \textit{CVE-2014-9219: ... vulnerability in the redirection feature in \blue{url.php} in phpMyAdmin ...}                                                                       \\ \cline{2-3} 
\multicolumn{1}{|c|}{}                                                                                      & \multirow{2}{*}{\begin{tabular}[c]{@{}c@{}}Erroneous\\ function\end{tabular}} & \textit{CVE-2012-0851: The \blue{ff\_h264\_decode\_seq\_parameter\_set} function in ...}                                                                                       \\ \cline{3-3} 
\multicolumn{1}{|c|}{}                                                                                      &                                                                               & \textit{CVE-2020-11608: ... allows NULL pointer dereferences in \blue{ov511\_mode\_init\_regs} and \blue{ov518\_mode\_init\_regs} when ...}                                           \\ \cline{2-3} 
\multicolumn{1}{|c|}{}                                                                                      & \multirow{2}{*}{\begin{tabular}[c]{@{}c@{}}Related\\ parameter\end{tabular}}  & \textit{CVE-2015-2923: ... allows remote attackers to reconfigure a hop-limit setting via a small \blue{hop\_limit} value in ...}                                              \\ \cline{3-3} 
\multicolumn{1}{|c|}{}                                                                                      &                                                                               & \textit{CVE-2017-16939: ... allows local users to ... via a ... call in conjunction with \blue{XFRM\_MSG\_GETPOLICY Netlink} messages.}                                        \\ \hline
\multicolumn{1}{|c|}{\multirow{6}{*}{\begin{tabular}[c]{@{}c@{}}Attack\\ information\end{tabular}}}         & \multirow{2}{*}{\begin{tabular}[c]{@{}c@{}}Attack\\ payload\end{tabular}}     & \textit{CVE-2011-3973: ... allows remote attackers to ... via \blue{an invalid bitstream in a Chinese AVS video (aka CAVS) file} ...}                                                \\ \cline{3-3} 
\multicolumn{1}{|c|}{}                                                                                      &                                                                               & \textit{CVE-2011-4031: ... allows remote attackers to ... via \blue{a crafted ASF packet}.}                                                                                    \\ \cline{2-3} 
\multicolumn{1}{|c|}{}                                                                                      & \multirow{2}{*}{\begin{tabular}[c]{@{}c@{}}Attack \\ method\end{tabular}}     & \textit{CVE-2022-21208: ... An attacker can exploit this vulnerability by \blue{sending an unlimited number of huge chunks (e.g. 2GB each)} ...}                               \\ \cline{3-3} 
\multicolumn{1}{|c|}{}                                                                                      &                                                                               & \textit{CVE-2018-10087: ... allow local users to cause ... by \blue{triggering an attempted use of the -INT\_MIN value}.}                                                      \\ \cline{2-3} 
\multicolumn{1}{|c|}{}                                                                                      & \multirow{2}{*}{\begin{tabular}[c]{@{}c@{}}Attack\\ impact\end{tabular}}     & \textit{CVE-2022-21678: ..., \blue{the bios of users who made their profiles private were still visible in the `\textless{}meta\textgreater{}` tags on their users' pages} ...} \\ \cline{3-3} 
\multicolumn{1}{|c|}{}                                                                                      &                                                                               & \textit{CVE-2018-12232: ... there is a race condition ..., leading to \blue{a NULL pointer dereference and system crash}.}                                                     \\ \hline
\multicolumn{1}{|c|}{\multirow{4}{*}{\begin{tabular}[c]{@{}c@{}}Patch\\ information\end{tabular}}}          & \multirow{2}{*}{\begin{tabular}[c]{@{}c@{}}Patch\\ method\end{tabular}}       & \textit{CVE-2022-21682: ... This has been resolved in ... by \blue{changing the behaviour of `--nofilesystem=home` and `--nofilesystem=host`}.}                                \\

\cline{3-3} 
\multicolumn{1}{|c|}{}                                                                                      &                                                                               & \textit{{CVE-2022-42725: ...The issue has been fixed with \blue{a parameter check on user input}}}                                                                                                                 \\ \cline{2-3} 
\multicolumn{1}{|c|}{}                                                                                      & \multirow{2}{*}{\begin{tabular}[c]{@{}c@{}}Patch\\ reference\end{tabular}}    & \textit{CVE-2021-39207: ... See commit \blue{507d066ef432ea27d3e201da08009872a2f37725} for details.}                                                                           \\ \cline{3-3} 
\multicolumn{1}{|c|}{}                                                                                      &                                                                               & \textit{CVE-2017-3738: ... The fix is also avaliable in commit \blue{e502cc86d} in the OpenSSL git repository.}                                                                \\ \hline
\end{tabular}
    
    \end{adjustbox}
    \vspace{+5pt}
\end{table*}
\begin{table*}[htbp]
    \centering
    \renewcommand{\arraystretch}{1.05}
    \caption{Information categories within commit description.}
    \label{table: information categories within commit description}   
    \begin{adjustbox}{width=1\textwidth}

\begin{tabular}{|cc|l|}
\hline
\multicolumn{2}{|c|}{\textbf{Category}}                                                                                                                                                        & \textbf{Example}                                                                                                                           \\ \hline
\multicolumn{1}{|c|}{\multirow{4}{*}{\begin{tabular}[c]{@{}c@{}}Vulnerability\\ information\end{tabular}}} & \multirow{2}{*}{\begin{tabular}[c]{@{}c@{}}Vulnerability\\ type\end{tabular}} & \textit{owncloud:38271de: Added \blue{CSRF} checks}                                                                                                        \\ \cline{3-3} 
\multicolumn{1}{|c|}{}                                                                                     &                                                                               & \textit{AuroCMS:790f66f: ... Update Vulnerability \blue{SQL Injection} in content.php}                                                                     \\ \cline{2-3} 
\multicolumn{1}{|c|}{}                                                                                     & \multirow{2}{*}{\begin{tabular}[c]{@{}c@{}}Faulty\\ functionality\end{tabular}}     & \textit{djblets: 77ac646: ... \blue{The generated gravatar HTML wasn't handling escaping of the display name of the user}, allowing ...}                   \\ \cline{3-3} 
\multicolumn{1}{|c|}{}                                                                                     &                                                                               & \textit{ffi:e0fe486: ... \blue{Symbols were sent directly to FFI::DynamicLibrary.open in the first attempt}, resulting in ...}                             \\ \hline
\multicolumn{1}{|c|}{\multirow{4}{*}{\begin{tabular}[c]{@{}c@{}}Attack\\ information\end{tabular}}}        & \multirow{2}{*}{\begin{tabular}[c]{@{}c@{}}Attack\\ method\end{tabular}}      & \textit{djblets:77a68c0: ... This allows an attacker who can \blue{provide part of a JSON-serializable object to craft a string} ...}                      \\ \cline{3-3} 
\multicolumn{1}{|c|}{}                                                                                     &                                                                               & \textit{fbthrift:3f15620: ... This allows malicious attacker to \blue{send few bytes message and cause server to allocate GBs of memory} ...}                                                                                                                                           \\ \cline{2-3} 
\multicolumn{1}{|c|}{}                                                                                     & \multirow{2}{*}{\begin{tabular}[c]{@{}c@{}}Attack\\ impact\end{tabular}}     & \textit{djblets:77a68c0: ... This allows ... that can \blue{break out a \textless{}script\textgreater tag and create its own, injecting a custom script}.} \\ \cline{3-3} 
\multicolumn{1}{|c|}{}                                                                                     &                                                                               & \textit{revive-adserver:a323fd6: ... such a vulnerability could be used by an attacker to \blue{steal the session ID of an authenticated user} ...}        \\ \hline
\multicolumn{1}{|c|}{\multirow{8}{*}{\begin{tabular}[c]{@{}c@{}}Fix\\ information\end{tabular}}}           & \multirow{2}{*}{Method}                                                       & \textit{File:71a8b6c: ... \blue{Limit regex search for BEGIN to the first 4K of the file}.}                                                                \\ \cline{3-3} 
\multicolumn{1}{|c|}{}                                                                                     &                                                                               & \textit{djblets:77a68c0: ... To fix this, \blue{we escape '\textless{}', '\textgreater{}', and '\&' characters in the resulting string}, preventing ...}   \\ \cline{2-3} 
\multicolumn{1}{|c|}{}                                                                                     & \multirow{2}{*}{\begin{tabular}[c]{@{}c@{}}Erroneous\\ file\end{tabular}}     & \textit{djblets: 77ac646: Fix a XSS vulnerability in the \blue{gravatar} template tag ...}                                                                 \\ \cline{3-3} 
\multicolumn{1}{|c|}{}                                                                                     &                                                                               & \textit{FFmpeg:a5d849b: \blue{avformat/avide.c}: Limit formats in gab2 to srt ...}                                                                          \\ \cline{2-3} 
\multicolumn{1}{|c|}{}                                                                                     & \multirow{2}{*}{\begin{tabular}[c]{@{}c@{}}Erroneous\\ function\end{tabular}} & \textit{Tcpdump:a1eefe9: ... prevent a possible buffer overread in \blue{chdlc\_print()} ...}                                                              \\ \cline{3-3} 
\multicolumn{1}{|c|}{}                                                                                     &                                                                               & \textit{openfortivpn:60660e0: ... CVE-2020-7041 incorrect use of \blue{X509\_check\_host} (regarding return value) is fixed with ...}                      \\ \cline{2-3} 
\multicolumn{1}{|c|}{}                                                                                     & \multirow{2}{*}{\begin{tabular}[c]{@{}c@{}}Related\\ parameter\end{tabular}}  & \textit{Linux:89d7ae3: ... This patch fixes this by first checking to ensure that the \blue{skb} is non-NULL before using it to ...}           \\ \cline{3-3} 
\multicolumn{1}{|c|}{}                                                                                     &                                                                               & \textit{flatpak:6d1773d: ... if ..., they cannot be used to inject arbitrary code into a non-setuid \blue{bwrap} via mechanisms like ...}                  \\ \hline
\end{tabular}

    \end{adjustbox}
\end{table*}

\begin{table*}[htbp]
\vspace{15pt}
    \centering
    \begin{minipage}{0.54\textwidth}
    \centering
    \renewcommand{\arraystretch}{1.05}
    \caption{Information categories within commit code diff.}
    \vspace{-3pt}
    \label{table: information categories within commit code diff}  
    \footnotesize

\begin{tabular}{|c|cc|}
\hline
\textbf{Component} & \multicolumn{2}{c|}{\textbf{Category}}                      \\ \hline
Comment            & \multicolumn{1}{c|}{Program functionality} & Modification reason \\ \hline
Source code        & \multicolumn{1}{c|}{Program functionality} & Modification location   \\ \hline
\end{tabular}
    \end{minipage}
    \hfill
    \begin{minipage}{0.44\textwidth}
          \centering
    \renewcommand{\arraystretch}{1.05}
    \caption{Confusion matrix for LLM's determination.}
    \vspace{-3pt}
    \label{table: Confusion matrix for LLM's identification}
    \footnotesize
    
\begin{tabular}{|cc|cc|}
\hline
\multicolumn{2}{|c|}{\multirow{2}{*}{}}                                    & \multicolumn{2}{c|}{\textbf{LLM Predicted}}                \\ \cline{3-4} 
\multicolumn{2}{|c|}{}                                                     & \multicolumn{1}{c|}{\textbf{Positive}} & \textbf{Negative} \\ \hline
\multicolumn{1}{|c|}{\multirow{2}{*}{\textbf{Actual}}} & \textbf{Positive} & \multicolumn{1}{c|}{2,141}             & \textcolor{magenta}{320}            \\ \cline{2-4} 
\multicolumn{1}{|c|}{}                                 & \textbf{Negative} & \multicolumn{1}{c|}{\textcolor{magenta}{12,437}}               & 80,852            \\ \hline
\end{tabular}
    \end{minipage}
\end{table*}

\begin{table*}[htbp]
    \centering    
    \vspace{20pt}
    \caption{Result of ablation study on Recall ($R@k$)}
    \label{table: result of ablation study on recall}
    \begin{adjustbox}{width=1\textwidth}

   \begin{tabular}{|c|c|c|c|c|c|c|c|c|c|c|c|}
\hline
\textbf{Data}         & \textbf{Feature} & \textbf{1}           & \textbf{2}           & \textbf{3}           & \textbf{4}           & \textbf{5}           & \textbf{6}           & \textbf{7}           & \textbf{8}           & \textbf{9}           & \textbf{10}          \\ \hline
\multirow{3}{*}{Full} & $F_0$            
& 62.14\%
& 74.99\%
& 78.49\%
& 80.68\%
& 82.45\%
& 83.66\%
& 84.75\%
& 85.43\%
& 86.27\%
& 87.10\%
\\ \cline{2-12} 
& $F_0+F_1$        
& 64.23\% \magenta{$\uparrow$}
& 78.80\% \magenta{$\uparrow$}
& 83.39\% \magenta{$\uparrow$}
& 86.27\% \magenta{$\uparrow$}
& 87.47\% \magenta{$\uparrow$}
& 88.41\% \magenta{$\uparrow$}
& 89.19\% \magenta{$\uparrow$}
& 89.82\% \magenta{$\uparrow$}
& 90.39\% \magenta{$\uparrow$}
& 91.07\% \magenta{$\uparrow$}
\\ \cline{2-12} 
& $F_0+F_1+F_2$    
& 64.86\% \magenta{$\uparrow$}
& 79.22\% \magenta{$\uparrow$}
& 84.18\% \magenta{$\uparrow$}
& 86.58\% \magenta{$\uparrow$}
& 88.56\% \magenta{$\uparrow$}
& 89.45\% \magenta{$\uparrow$}
& 90.34\% \magenta{$\uparrow$}
& 90.97\% \magenta{$\uparrow$}
& 91.64\% \magenta{$\uparrow$}
& 92.74\% \magenta{$\uparrow$}
\\ \hline
\multirow{3}{*}{1-1}  & $F_0$            
& 78.70\%
& 87.57\%
& 89.55\%
& 90.67\%
& 91.67\%
& 92.39\%
& 93.12\%
& 93.58\%
& 93.98\%
& 94.25\%
\\ \cline{2-12} 
& $F_0+F_1$        
& 81.35\% \magenta{$\uparrow$}
& 89.48\% \magenta{$\uparrow$}
& 91.27\% \magenta{$\uparrow$}
& 92.79\% \magenta{$\uparrow$}
& 93.25\% \magenta{$\uparrow$}
& 93.85\% \magenta{$\uparrow$}
& 94.11\% \magenta{$\uparrow$}
& 94.38\% \magenta{$\uparrow$}
& 94.58\% \magenta{$\uparrow$}
& 95.04\% \magenta{$\uparrow$}
\\ \cline{2-12} 
                      & $F_0+F_1+F_2$    
& 82.14\% \magenta{$\uparrow$}
& 89.22\% \blue{$\downarrow$}
& 91.60\% \magenta{$\uparrow$}
& 92.72\% \blue{$\downarrow$}
& 93.65\% \magenta{$\uparrow$}
& 94.18\% \magenta{$\uparrow$}
& 94.71\% \magenta{$\uparrow$}
& 94.84\% \magenta{$\uparrow$}
& 94.84\% \magenta{$\uparrow$}
& 95.30\% \magenta{$\uparrow$}
                      \\ \hline
\multirow{3}{*}{1-N}  & $F_0$            
& 0.00\%
& 27.79\%
& 36.97\%
& 43.18\%
& 47.89\%
& 50.87\%
& 53.35\%
& 54.84\%
& 57.32\%
& 60.30\%
\\ \cline{2-12} 
                      & $F_0+F_1$        
& 0.00\% \teal{$=$}
& 38.71\% \magenta{$\uparrow$}
& 53.85\% \magenta{$\uparrow$}
& 61.79\% \magenta{$\uparrow$}
& 65.76\% \magenta{$\uparrow$}
& 67.99\% \magenta{$\uparrow$}
& 70.72\% \magenta{$\uparrow$}
& 72.70\% \magenta{$\uparrow$}
& 74.69\% \magenta{$\uparrow$}
& 76.18\% \magenta{$\uparrow$}
                      \\ \cline{2-12} 
                      & $F_0+F_1+F_2$    
& 0.00\% \teal{$=$}
& 41.69\% \magenta{$\uparrow$}
& 56.33\% \magenta{$\uparrow$}
& 63.52\% \magenta{$\uparrow$}
& 69.48\% \magenta{$\uparrow$}
& 71.71\% \magenta{$\uparrow$}
& 73.95\% \magenta{$\uparrow$}
& 76.43\% \magenta{$\uparrow$}
& 79.65\% \magenta{$\uparrow$}
& 83.13\% \magenta{$\uparrow$}
                      \\ \hline
\end{tabular}
    
    \end{adjustbox}
\end{table*}
\begin{table*}[htbp]
    \centering    
    \vspace{20pt}
    \caption{Result of ablation study on NDCG ($N@k$)}
    \label{table: result of ablation study on NDCG}
    \begin{adjustbox}{width=1\textwidth}

   \begin{tabular}{|c|c|c|c|c|c|c|c|c|c|c|c|}
\hline
\textbf{Data}         & \textbf{Feature} & \textbf{1}           & \textbf{2}           & \textbf{3}           & \textbf{4}           & \textbf{5}           & \textbf{6}           & \textbf{7}           & \textbf{8}           & \textbf{9}           & \textbf{10}          \\ \hline
\multirow{3}{*}{Full} & $F_0$            
& 73.63\%
& 77.20\%
& 78.58\%
& 79.35\%
& 79.94\%
& 80.36\%
& 80.74\%
& 81.01\%
& 81.31\%
& 81.52\%
\\ \cline{2-12} 
                      & $F_0+F_1$        
& 78.85\% \magenta{$\uparrow$}
& 82.22\% \magenta{$\uparrow$}
& 83.77\% \magenta{$\uparrow$}
& 84.84\% \magenta{$\uparrow$}
& 85.26\% \magenta{$\uparrow$}
& 85.57\% \magenta{$\uparrow$}
& 85.85\% \magenta{$\uparrow$}
& 86.05\% \magenta{$\uparrow$}
& 86.21\% \magenta{$\uparrow$}
& 86.37\% \magenta{$\uparrow$}

                      \\ \cline{2-12} 
                      & $F_0+F_1+F_2$   
& 80.10\% \magenta{$\uparrow$}
& 82.92\% \magenta{$\uparrow$}
& 84.62\% \magenta{$\uparrow$}
& 85.48\% \magenta{$\uparrow$}
& 86.10\% \magenta{$\uparrow$}
& 86.44\% \magenta{$\uparrow$}
& 86.73\% \magenta{$\uparrow$}
& 86.93\% \magenta{$\uparrow$}
& 87.09\% \magenta{$\uparrow$}
& 87.33\% \magenta{$\uparrow$}

                      \\ \hline
\multirow{3}{*}{1-1}  & $F_0$           
& 78.70\%
& 84.30\%
& 85.29\%
& 85.77\%
& 86.16\%
& 86.41\%
& 86.66\%
& 86.80\%
& 86.92\%
& 87.00\%
\\ \cline{2-12} 
                      & $F_0+F_1$        
& 81.35\% \magenta{$\uparrow$}
& 86.48\% \magenta{$\uparrow$}
& 87.37\% \magenta{$\uparrow$}
& 88.03\% \magenta{$\uparrow$}
& 88.21\% \magenta{$\uparrow$}
& 88.42\% \magenta{$\uparrow$}
& 88.51\% \magenta{$\uparrow$}
& 88.59\% \magenta{$\uparrow$}
& 88.65\% \magenta{$\uparrow$}
& 88.79\% \magenta{$\uparrow$}
                      \\ \cline{2-12} 
                      & $F_0+F_1+F_2$    
& 82.14\% \magenta{$\uparrow$}
& 86.61\% \magenta{$\uparrow$}
& 87.80\% \magenta{$\uparrow$}
& 88.28\% \magenta{$\uparrow$}
& 88.64\% \magenta{$\uparrow$}
& 88.83\% \magenta{$\uparrow$}
& 89.01\% \magenta{$\uparrow$}
& 89.05\% \magenta{$\uparrow$}
& 89.05\% \magenta{$\uparrow$}
& 89.18\% \magenta{$\uparrow$}
                      \\ \hline
\multirow{3}{*}{1-N}  & $F_0$            & 54.59\%              & 50.56\%              & 53.43\%              & 55.27\%              & 56.62\%              & 57.62\%              & 58.53\%              & 59.29\%              & 60.25\%              & 60.99\%              \\ \cline{2-12} 
                      & $F_0+F_1$        
& 69.48\% \magenta{$\uparrow$}
& 66.22\% \magenta{$\uparrow$}
& 70.23\% \magenta{$\uparrow$}
& 72.85\% \magenta{$\uparrow$}
& 74.19\% \magenta{$\uparrow$}
& 74.86\% \magenta{$\uparrow$}
& 75.85\% \magenta{$\uparrow$}
& 76.52\% \magenta{$\uparrow$}
& 77.03\% \magenta{$\uparrow$}
& 77.30\% \magenta{$\uparrow$}
                      \\ \cline{2-12} 
                      & $F_0+F_1+F_2$    
& 72.46\% \magenta{$\uparrow$}
& 69.10\% \magenta{$\uparrow$}
& 72.71\% \magenta{$\uparrow$}
& 74.95\% \magenta{$\uparrow$}
& 76.58\% \magenta{$\uparrow$}
& 77.47\% \magenta{$\uparrow$}
& 78.17\% \magenta{$\uparrow$}
& 78.99\% \magenta{$\uparrow$}
& 79.74\% \magenta{$\uparrow$}
& 80.40\% \magenta{$\uparrow$}
                      \\ \hline
\end{tabular}

    \end{adjustbox}
\end{table*}
\begin{table*}[htbp]
    \centering    
    \vspace{10pt}
    \caption{Result of ablation study on Manual Effort ($M@k$)}
    \label{table: result of ablation study on manual effort}
    \begin{adjustbox}{width=0.8\textwidth}

\begin{tabular}{|c|c|c|c|c|c|c|c|c|c|c|c|}
\hline
\textbf{Data}         & \textbf{Feature} & \textbf{1}           & \textbf{2}           & \textbf{3}           & \textbf{4}           & \textbf{5}         & \textbf{6}         & \textbf{7}         & \textbf{8}         & \textbf{9}         & \textbf{10}        \\ \hline
\multirow{3}{*}{Full} & $F_0$            
& 1.00
& 1.38
& 1.63
& 1.84
& 2.04
& 2.21
& 2.38
& 2.53
& 2.67
& 2.81
\\ \cline{2-12} 
                      & $F_0+F_1$        
& 1.00 \teal{$=$}
& 1.36 \blue{$\downarrow$}
& 1.57 \blue{$\downarrow$}
& 1.74 \blue{$\downarrow$}
& 1.87 \blue{$\downarrow$}
& 2.00 \blue{$\downarrow$}
& 2.11 \blue{$\downarrow$}
& 2.22 \blue{$\downarrow$}
& 2.32 \blue{$\downarrow$}
& 2.42 \blue{$\downarrow$}
                      \\ \cline{2-12} 
                      & $F_0+F_1+F_2$    
& 1.00 \teal{$=$}
& 1.35 \blue{$\downarrow$}
& 1.56 \blue{$\downarrow$}
& 1.72 \blue{$\downarrow$}
& 1.85 \blue{$\downarrow$}
& 1.97 \blue{$\downarrow$}
& 2.07 \blue{$\downarrow$}
& 2.17 \blue{$\downarrow$}
& 2.26 \blue{$\downarrow$}
& 2.34 \blue{$\downarrow$}
                      \\ \hline
\multirow{3}{*}{1-1}  & $F_0$            
& 1.00
& 1.21
& 1.34
& 1.44
& 1.54
& 1.62
& 1.69
& 1.76
& 1.83
& 1.89
\\ \cline{2-12} 
                      & $F_0+F_1$        
& 1.00 \teal{$=$}
& 1.19 \blue{$\downarrow$}
& 1.29 \blue{$\downarrow$}
& 1.38 \blue{$\downarrow$}
& 1.45 \blue{$\downarrow$}
& 1.52 \blue{$\downarrow$}
& 1.58 \blue{$\downarrow$}
& 1.64 \blue{$\downarrow$}
& 1.70 \blue{$\downarrow$}
& 1.75 \blue{$\downarrow$}
                      \\ \cline{2-12} 
                                            & $F_0+F_1+F_2$    
& 1.00 \teal{$=$}
& 1.18 \blue{$\downarrow$}
& 1.29 \teal{$=$}
& 1.37 \blue{$\downarrow$}
& 1.44 \blue{$\downarrow$}
& 1.51 \blue{$\downarrow$}
& 1.56 \blue{$\downarrow$}
& 1.62 \blue{$\downarrow$}
& 1.67 \blue{$\downarrow$}
& 1.72 \blue{$\downarrow$}                      
                      \\ \hline
\multirow{3}{*}{1-N}  & $F_0$            
& 1.00
& 2.00
& 2.72
& 3.35
& 3.92
& 4.44
& 4.93
& 5.40
& 5.85
& 6.28
\\ \cline{2-12} 
                      & $F_0+F_1$        
& 1.00 \teal{$=$}
& 2.00 \teal{$=$}
& 2.61 \blue{$\downarrow$}
& 3.07 \blue{$\downarrow$}
& 3.46 \blue{$\downarrow$}
& 3.80 \blue{$\downarrow$}
& 4.12 \blue{$\downarrow$}
& 4.41 \blue{$\downarrow$}
& 4.68 \blue{$\downarrow$}
& 4.94 \blue{$\downarrow$}
                      \\ \cline{2-12} 
                      & $F_0+F_1+F_2$    
& 1.00 \teal{$=$}
& 2.00 \teal{$=$}
& 2.58 \blue{$\downarrow$}
& 3.02 \blue{$\downarrow$}
& 3.38 \blue{$\downarrow$}
& 3.69 \blue{$\downarrow$}
& 3.97 \blue{$\downarrow$}
& 4.23 \blue{$\downarrow$}
& 4.47 \blue{$\downarrow$}
& 4.67 \blue{$\downarrow$}
                      \\ \hline
\end{tabular}
    
    \end{adjustbox}
\end{table*}



\ignore{

\begin{figure}
    \centering
    \includegraphics[width=0.45\textwidth]{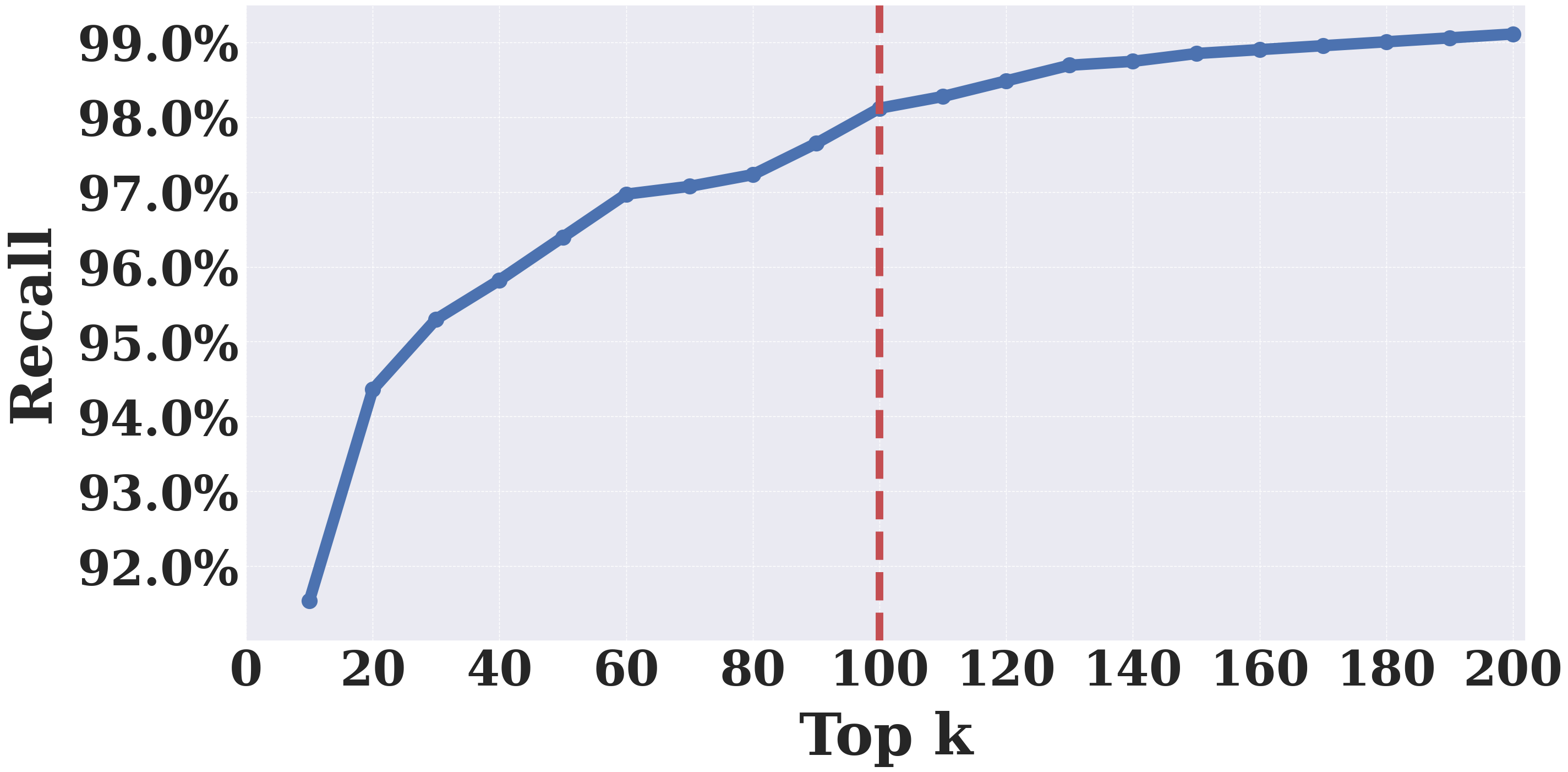}  
    \caption{The recall of CVE with at least one patch ranked in top $k$ on $Ranking_0$.}
    \label{figure: The recall of CVE with at least one patch ranked in top k on ranking 1}
\end{figure}

\begin{figure}
    \centering
    \includegraphics[width=0.45\textwidth]{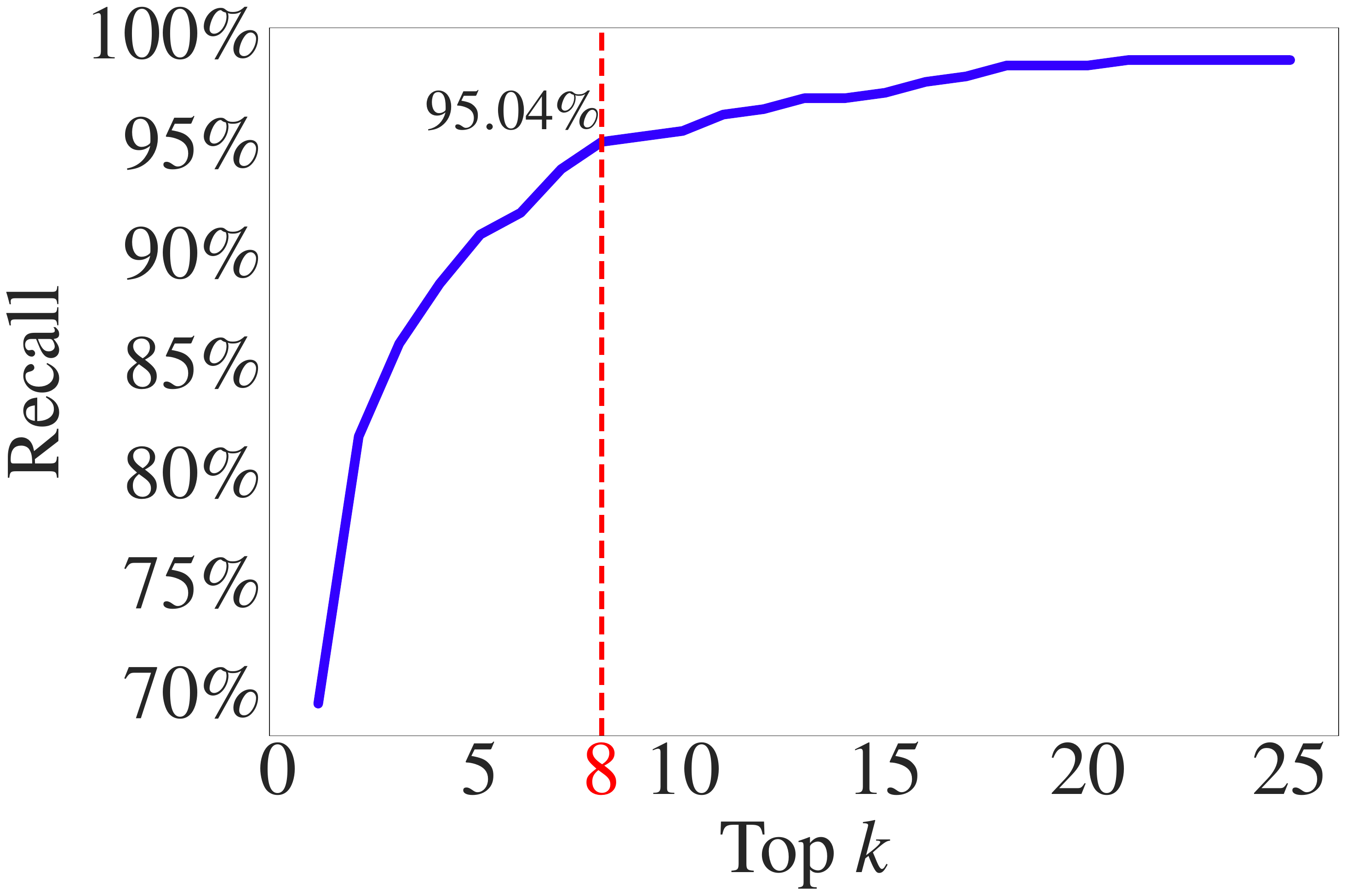}
    \caption{The recall of CVE with at least one patch ranked in top $k$ on $Ranking_1$.}
    \label{figure: The recall of CVE with at least one patch ranked in top k on ranking 2}
\end{figure}

}

\ignore{
\begin{figure}
    \centering
    \includegraphics[width=0.42\textwidth]{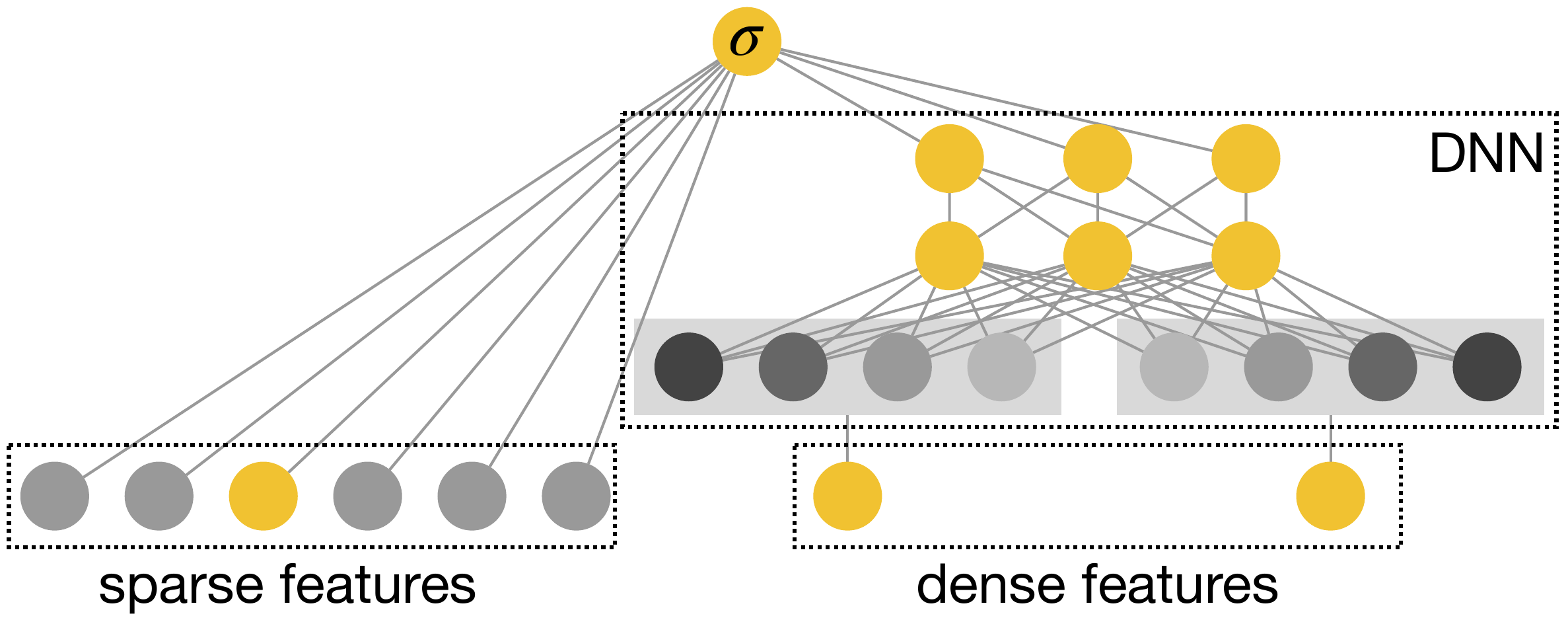}      
    \caption{Numerical encoder.}
    \label{figure: numerical encoder}
\end{figure}
}

\ignore{
\begin{figure}[htbt]
    \centering 
    \includegraphics[width=0.45\textwidth]{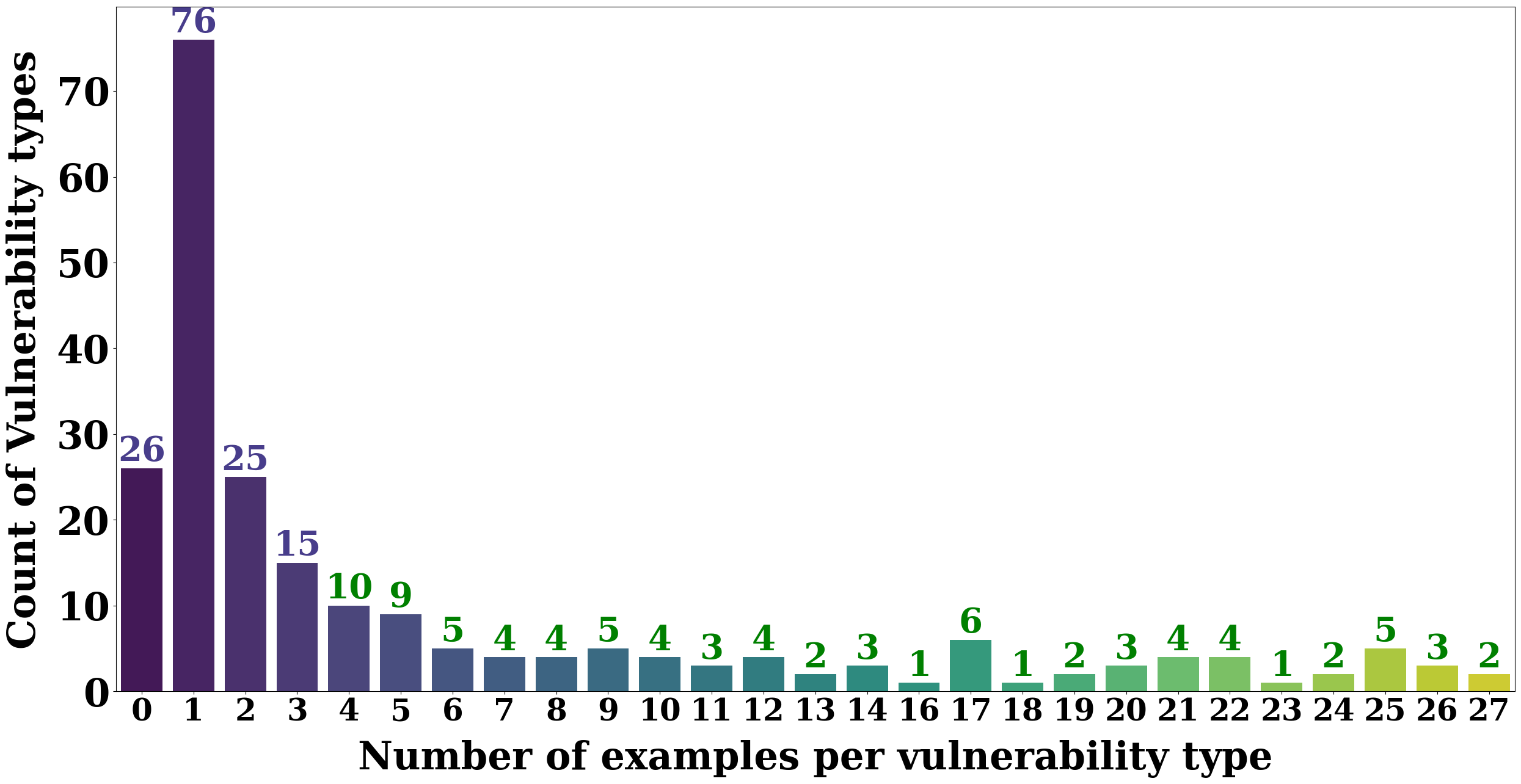}
    \caption{Example counts distribution by vulnerability type.}
    \label{figure: distribution of example counts across vulnerability types}
\end{figure}
}

\end{document}